\documentclass[fleqn,usenatbib,useAMS]{mnras}
\usepackage{tabu}
\usepackage{natbib}
\usepackage{graphicx}	
\usepackage{amsmath}	
\usepackage{amssymb}	
\usepackage{multicol}   
\usepackage{bm}		
\usepackage{pdflscape}	
\usepackage{float} 
\usepackage{lastpage}
\usepackage{color}
\usepackage{ulem}

\newcommand{\todo}{\ifmmode \text{\color{red}\Huge{\(\bullet\)}} \else {\color{red}{\Huge$\bullet$}}\fi}
\newcommand{\tido}{\ifmmode {{\color{red}\bullet}} \else {\color{red}$\bullet$}\fi}

\newcommand{\E        }[1]{\ifmmode 10^{#1} \else $10^{#1}$\fi}
\newcommand{\tE        }[1]{\ifmmode \times10^{#1} \else $\times10^{#1}$\fi}
\newcommand{\til}{\ifmmode \sim \else $\sim$\fi}
\renewcommand{\~} {\ifmmode \sim \else $\sim$\fi}

\newcommand{\pc}	{\ifmmode {\rm pc} \else pc\fi}
\newcommand{\kpc}	{\ifmmode {\rm kpc} \else kpc\fi}
\newcommand{\ld}	{\ifmmode {\rm l.d.} \else l.d.\fi}
\newcommand{\kms}	{\ifmmode {\rm km\,s}^{-1} \else km\,s$^{-1}$\fi}
\newcommand{\cc}	{\ifmmode {\rm cm}^{-3}    \else cm$^{-3}$\fi}
\newcommand{\cmii}	{\ifmmode {\rm cm}^{-2}    \else cm$^{-2}$\fi}
\newcommand{\ergs}	{\ifmmode {\rm erg\,s}^{-1} \else erg s$^{-1}$\fi}
\newcommand{\ergcms}	{\ifmmode {\rm erg\,cm}^{-2}\,{\rm s}^{-1} \else erg\,cm$^{-2}$\,s$^{-1}$\fi}
\newcommand{\ergcmsA}	{\ifmmode {\rm erg\,cm}^{-2}\,{\rm s}^{-1}\,{\rm\AA}^{-1}
\else erg\,cm$^{-2}$\,s$^{-1}$\,\AA$^{-1}$\fi}
\newcommand{  \ergcmsHz  }{\ifmmode{\rm erg\,cm}^{-2}\,{\rm s}^{-1}\,{\rm Hz}^{-1}
                       \else ergs\,cm$^{-2}$\,s$^{-1}$\,Hz$^{-1}$\fi}
\newcommand{\kev}	{\ifmmode {\rm keV} \else keV\fi}

\newcommand{\mic}	{\ifmmode {\rm \mu m} \else $\mu$m\fi}
\newcommand{\vFWHM}	{\ifmmode v_{\mbox{\tiny FWHM}} \else $v_{\mbox{\tiny FWHM}}$\fi}
\newcommand{\vBLR}	{\ifmmode v_{\mbox{\tiny BLR}} \else $v_{\mbox{\tiny BLR}}$\fi}
\newcommand{\sigBLR}	{\ifmmode \sigma_{\mbox{\tiny BLR}} \else $\sigma_{\mbox{\tiny BLR}}$\fi}
\newcommand{\vNLR}	{\ifmmode v_{\mbox{\tiny NLR}} \else $v_{\mbox{\tiny NLR}}$\fi}
\newcommand{\tauBLR}	{\ifmmode \tau_{\mbox{\tiny BLR}} \else $\tau_{\mbox{\tiny BLR}}$\fi}

\newcommand{\Hubble}	{\ifmmode {\rm km\,s}^{-1}\,{\rm Mpc}^{-1} \else km\,s$^{-1}$\,Mpc$^{-1}$\fi}
\newcommand{\NDunit}	{\ifmmode {\rm Mpc}^{-3} \else Mpc$^{-3}$\fi}
\newcommand{\LFunit}	{\ifmmode {\rm Mpc}^{-3}\,{\rm mag}^{-1} \else Mpc$^{-3}$\,mag$^{-1}$\fi}
\newcommand{\MFunit}	{\ifmmode {\rm Mpc}^{-3}\,{\rm dex}^{-1} \else Mpc$^{-3}$\,dex$^{-1}$\fi}

\newcommand{\Msun}{\ifmmode M_{\odot} \else $M_{\odot}$\fi}
\newcommand{\Lsun}{\ifmmode L_{\odot} \else $L_{\odot}$\fi}
\newcommand{\Zsun}{\ifmmode Z_{\odot} \else $Z_{\odot}$\fi}
\newcommand{\mpyr}{\ifmmode \Msun\,{\rm yr}^{-1} \else $\Msun\,{\rm yr}^{-1}$\fi}

\newcommand{\Msol}{\Msun}

\newcommand{\qnote}{\ifmmode q_{0} \else $q_{0}$\fi}
\newcommand{\Hnote}{\ifmmode H_{0} \else $H_{0}$\fi}
\newcommand{\hnote}{\ifmmode h_{0} \else $h_{0}$\fi}
\newcommand{\anote}{\ifmmode a_{0} \else $a_{0}$\fi}
\newcommand{\tnote}{\ifmmode t_{0} \else $t_{0}$\fi}

\def\gsim{\;\rlap{\lower 2.5pt \hbox{$\sim$}}\raise 1.5pt\hbox{$>$}\;}
\def\lsim{\;\rlap{\lower 2.5pt \hbox{$\sim$}}\raise 1.5pt\hbox{$<$}\;}

\newcommand{  \Halpha   }{\ifmmode {\rm H}\alpha \else H$\alpha$\fi}
\newcommand{  \halpha   }{\Halpha}
\newcommand{  \ha       }{\Halpha}
\newcommand{  \Hbeta    }{\ifmmode {\rm H}\beta \else H$\beta$\fi}
\newcommand{  \hbeta    }{\Hbeta}
\newcommand{  \hb       }{\Hbeta}
\newcommand{  \Hgamma   }{\ifmmode {\rm H}\gamma \else H$\gamma$\fi}
\newcommand{  \Hdelta   }{\ifmmode {\rm H}\delta \else H$\delta$\fi}
\newcommand{  \Lya      }{\ifmmode {\rm Ly}\alpha \else Ly$\alpha$\fi}
\newcommand{  \Lyb      }{\ifmmode {\rm Ly}\beta \else Ly$\beta$\fi}
\newcommand{  \Pa       }{\ifmmode {\rm P}\alpha \else P$\alpha$\fi}
\newcommand{  \Pb       }{\ifmmode {\rm P}\beta \else P$\beta$\fi}
\newcommand{  \Bra      }{\ifmmode {\rm Br}\alpha \else Br$\alpha$\fi}
\newcommand{  \Brg      }{\ifmmode {\rm Br}\gamma \else Br$\gamma$\fi}
\newcommand{  \hii      }{\ifmmode {\rm H}\,\textsc{ii} \else H\,\textsc{ii}\fi}
\newcommand{  \hei      }{\ifmmode {\rm He}\,\textsc{i} \else He\,\textsc{i}\fi}
\newcommand{  \heii     }{\ifmmode {\rm He}\,\textsc{ii} \else He\,\textsc{ii}\fi}
\newcommand{  \HeIIuv   }{\ifmmode {\rm He}\,\textsc{ii}\,\lambda1640 \else He\,\textsc{ii}\,$\lambda1640$\fi}
\newcommand{  \HeIIop   }{\ifmmode {\rm He}\,\textsc{ii}\,\lambda4686 \else He\,\textsc{ii}\,$\lambda4686$\fi}
\newcommand{  \CII	}{\ifmmode \left[{\rm C}\,\textsc{ii}\right]\,\lambda157.74\,\mu{\rm m} \else [C\,{\sc ii}]\ $\lambda157.74\,\mu{\rm m}$\fi}
\newcommand{  \cii	}{\ifmmode \left[{\rm C}\,\textsc{ii}\right] \else [C\,{\sc ii}]\fi}

\newcommand{  \ciii     }{\ifmmode {\rm C}\,\textsc{iii}\right] \else C\,\textsc{iii}]\fi}
\newcommand{  \CIII     }{\ifmmode {\rm C}\,\textsc{iii}\right]\,\lambda1909 \else C\,\textsc{iii}]\,$\lambda1909$\fi}
\newcommand{  \civ      }{\ifmmode {\rm C}\,\textsc{iv}  \else C\,\textsc{iv}\fi}
\newcommand{  \CIV      }{\ifmmode {\rm C}\,\textsc{iv}\,\lambda1549 \else C\,\textsc{iv}\,$\lambda1549$\fi}
\newcommand{  \NIIopt   }{\ifmmode \left[{\rm N}\,\textsc{ii}\right]\,\lambda6584 \else [N\,\textsc{ii}]\,$\lambda6584$\fi}
\newcommand{  \nii      }{\ifmmode \left[{\rm N}\,\textsc{ii}\right]  \else [N\,\textsc{ii}]\fi}
\newcommand{  \niii     }{\ifmmode {\rm N}\,\textsc{iii} \else N\,\textsc{iii}\fi}
\newcommand{  \NIII     }{\ifmmode {\rm N}\,\textsc{iii}\,\lambda4640 \else N\,\textsc{iii}\,$\lambda4640$\fi}
\newcommand{  \niv      }{\ifmmode {\rm N}\,\textsc{iv}  \else N\,\textsc{iv}\fi}
\newcommand{  \NIVuv    }{\ifmmode {\rm N}\,\textsc{iv}\,\lambda1486 \else N\,\textsc{iv}\,$\lambda1486$\fi}
\newcommand{  \nv       }{\ifmmode {\rm N}\,\textsc{v}   \else N\,\textsc{v}\fi}
\newcommand{\oi}{\ifmmode \left[{\rm O}\,\textsc{i}\right] \else [O\,{\sc i}]\fi}
\newcommand{\OI}{\ifmmode \left[{\rm O}\,\textsc{i}\right]\,\lambda6300 \else [O\,{\sc i}]$\,\lambda6300$\fi}
\newcommand{\oii}{\ifmmode \left[{\rm O}\,\textsc{ii}\right] \else [O\,{\sc ii}]\fi}
\newcommand{\OII}{\ifmmode \left[{\rm O}\,\textsc{ii}\right]\,\lambda3727 \else [O\,{\sc ii}]\,$\lambda3727$\fi}
\newcommand{\oiii}{\ifmmode \left[{\rm O}\,\textsc{iii}\right] \else [O\,{\sc iii}]\fi}
\newcommand{\OIII}{\ifmmode \left[{\rm O}\,\textsc{iii}\right]\,\lambda5007 \else [O\,{\sc iii}]\,$\lambda5007$\fi}
\newcommand{  \OIIIbf   }{\ifmmode {\rm O}\,\textsc{iii}\,\lambda3133 \else O\,\textsc{iii}\,$\lambda3133$\fi}
\newcommand{  \OIIIuv   }{\ifmmode {\rm O}\,\textsc{iii}\,\lambda1663 \else O\,\textsc{iii}\,$\lambda1663$\fi}
\newcommand{  \oiv      }{\ifmmode {\rm O}\,\textsc{iv}  \else O\,\textsc{iv}\fi}
\newcommand{  \OIVuv    }{\ifmmode {\rm O}\,\textsc{iv}\,\lambda1402  \else O\,\textsc{iv}\,$\lambda1402$\fi}
\newcommand{  \OIVIR    }{\ifmmode {\rm O}\,\textsc{iv}\,25.9\,\mu {\rm m} \else O\,\textsc{iv}\,$25.9\,\mu$m\fi}
\newcommand{  \ovi      }{\ifmmode {\rm O}\,\textsc{vi}   \else O\,\textsc{vi}\fi}
\newcommand{  \Ovi      }{\ifmmode {\rm O}\,\textsc{vi}\,\lambda1035 \else O\,\textsc{vi}\,$\lambda1035$\fi}
\newcommand{  \nei      }{\ifmmode {\rm Ne}\,\textsc{i}   \else Ne\,\textsc{i}\fi}
\newcommand{  \neii     }{\ifmmode {\rm Ne}\,\textsc{ii}  \else Ne\,\textsc{ii}\fi}
\newcommand{  \NeiiIR   }{\ifmmode {\rm Ne}\,\textsc{ii}\,12.8\,\mu {\rm m} \else Ne\,\textsc{ii}\,$12.8\,\mu$m\fi}
\newcommand{  \neiii    }{\ifmmode {\rm Ne}\,\textsc{iii} \else Ne\,\textsc{iii}\fi}
\newcommand{  \neiv     }{\ifmmode {\rm Ne}\,\textsc{iv}  \else Ne\,\textsc{iv}\fi}
\newcommand{  \nev      }{\ifmmode {\rm Ne}\,\textsc{v}   \else Ne\,\textsc{v}\fi}
\newcommand{  \NevIR    }{\ifmmode {\rm Ne}\,\textsc{v}\,24.3\,\mu {\rm m} \else Ne\,\textsc{v}\,$24.3\,\mu$m\fi}
\newcommand{  \nevi     }{\ifmmode {\rm Ne}\,\textsc{vi}  \else Ne\,\textsc{vi}\fi}
\newcommand{  \mgi      }{\ifmmode {\rm Mg}\,\textsc{i} \else Mg\,\textsc{i}\fi}
\newcommand{  \mgii     }{\ifmmode {\rm Mg}\,\textsc{ii} \else Mg\,\textsc{ii}\fi}
\newcommand{  \MgII     }{\ifmmode {\rm Mg}\,\textsc{ii}\,\lambda2798 \else Mg\,\textsc{ii}\,$\lambda2798$\fi}
\newcommand{  \sii      }{\ifmmode {\rm S}\,\textsc{ii} \else S\,\textsc{ii}\fi}
\newcommand{  \siii     }{\ifmmode {\rm S}\,\textsc{iii} \else S\,\textsc{iii}\fi}
\newcommand{  \siv      }{\ifmmode {\rm S}\,\textsc{iv} \else S\,\textsc{iv}\fi}
\newcommand{  \sili     }{\ifmmode {\rm Si}\,\textsc{i}   \else Si\,\textsc{i}\fi}
\newcommand{  \silii    }{\ifmmode {\rm Si}\,\textsc{ii}  \else Si\,\textsc{ii}\fi}
\newcommand{  \Siliv    }{\ifmmode {\rm Si}\,\textsc{iv}  \else Si\,\textsc{iv}\fi}
\newcommand{  \SilIVuv  }{\ifmmode {\rm Si}\,\textsc{iv}\,\lambda1400  \else Si\,\textsc{iv}\,$\lambda1400$\fi}
\newcommand{  \AlIII   }{\ifmmode {\rm Al}\,\textsc{iii}\,\lambda1857 \else Al\,\textsc{iii}\,$\lambda1857$\fi}
\newcommand{  \Aliii   }{\ifmmode {\rm Al}\,\textsc{iii} \else Al\,\textsc{iii}\fi}
\newcommand{  \caii     }{\ifmmode {\rm Ca}\,\textsc{ii} \else Ca\,\textsc{ii}\fi}
\newcommand{  \feii     }{\ifmmode {\rm Fe}\,\textsc{ii} \else Fe\,\textsc{ii}\fi}
\newcommand{  \feiii    }{\ifmmode {\rm Fe}\,\textsc{iii} \else Fe\,\textsc{iii}\fi}
\newcommand{  \Kalpha   }{\ifmmode {\rm K}\alpha \else K$\alpha$\fi}

\newcommand{ \Lhb   }{\ifmmode L_{\hb} \else $L_{\hb}$\fi}
\newcommand{ \Lha   }{\ifmmode L_{\ha} \else $L_{\ha}$\fi}
\newcommand{ \fwhb  }{\ifmmode {\rm FWHM}\left(\hb\right) \else FWHM(\hb)\fi}
\newcommand{\sighb  }{\ifmmode \sigma\left(\hb\right) \else $\sigma\left(\hb\right)$\fi}
\newcommand{ \ewhb  }{\ifmmode {\rm EW}\left(\hb\right) \else EW(\hb)\fi}
\newcommand{ \fwha  }{\ifmmode {\rm FWHM}\left(\ha\right) \else FWHM(\ha)\fi}
\newcommand{ \ewha  }{\ifmmode {\rm EW}\left(\ha\right) \else EW(\ha)\fi}
\newcommand{ \Lmg   }{\ifmmode L\left(\mgii\right) \else $L\left(\mgii\right)$\fi}
\newcommand{ \fwmg  }{\ifmmode {\rm FWHM}\left(\mgii\right) \else FWHM(\mgii)\fi}
\newcommand{ \Lciv  }{\ifmmode L\left(\civ\right) \else $L\left(\civ\right)$\fi}
\newcommand{ \fwciv }{\ifmmode {\rm FWHM}\left(\civ\right) \else FWHM(\civ)\fi}
\newcommand{ \fwhm  }{\ifmmode {\rm FWHM} \else FWHM\fi} 
\newcommand{ \voff  }{\ifmmode v_{\rm off} \else $v_{\rm off}$\fi} 
\newcommand{ \vmax  }{\ifmmode v_{\rm max} \else $v_{\rm max}$\fi} 

\newcommand{ \mumg  }{\ifmmode \mu\left(\mgii\right) \else $\mu\left(\mgii\right)$\fi}
\newcommand{ \fmg   }{\ifmmode f\left(\mgii\right) \else $f\left(\mgii\right)$\fi}
\newcommand{ \muciv }{\ifmmode \mu\left(\civ\right) \else $\mu\left(\civ\right)$\fi}
\newcommand{ \fciv  }{\ifmmode f\left(\civ\right) \else $f\left(\civ\right)$\fi}

\newcommand{  \auvo     }{\ifmmode \alpha_{\nu,{\rm UVO}} \else $\alpha_{\nu,{\rm UVO}}$\fi}
\newcommand{  \Ledd     }{\ifmmode L_{\rm Edd} \else $L_{\rm Edd}$\fi}
\newcommand{  \lamLlam  }{\ifmmode \lambda L_{\lambda} \else $\lambda L_{\lambda}$\fi}
\newcommand{  \lLl      }{\ifmmode \lambda L_{\lambda} \else $\lambda L_{\lambda}$\fi}
\newcommand{  \nuLnu    }{\ifmmode \nu L_{\nu} \else $\nu L_{\nu}$\fi}
\newcommand{  \nLn      }{\ifmmode \nu L_{\nu} \else $\nu L_{\nu}$\fi}
\newcommand{  \Luv      }{\ifmmode L_{1450} \else $L_{1450}$\fi}
\newcommand{  \Lop      }{\ifmmode L_{5100} \else $L_{5100}$\fi}
\newcommand{  \lLop     }{\ifmmode \log\left(\Lop/\ergs\right) \else $\log\left(\Lop/\ergs\right)$\fi}
\newcommand{  \Lthree   }{\ifmmode L_{3000} \else $L_{3000}$\fi}
\newcommand{  \lLthree  }{\ifmmode \log\left(\Lthree/\ergs\right) \else $\log\left(\Lthree/\ergs\right)$\fi}
\newcommand{  \Lsix      }{\ifmmode L_{6200} \else $L_{6200}$\fi}
\newcommand{  \lLisx     }{\ifmmode \log\left(\Lop/\ergs\right) \else $\log\left(\Lop/\ergs\right)$\fi}
\newcommand{  \Lxray    }{\ifmmode L_{\rm X} \else $L_{\rm X}$\fi}

\newcommand{  \Lsoft    }{\ifmmode L_{\rm 0.5-2} \else $L_{\rm 0.5-2}$\fi}

\newcommand{\Fthree}{\ifmmode F_{3000} \else $F_{3000}$\fi}
\newcommand{\fuv}{\ifmmode f_{\lambda}\left(1450{\rm \AA}\right) \else $f_{\lambda}\left(1450 {\rm \AA}\right)$\fi}
\newcommand{\fthree}{\ifmmode f_{\lambda}\left(3000{\rm \AA}\right) \else $f_{\lambda}\left(3000{\rm \AA}\right)$\fi}
\newcommand{\fH}{\ifmmode f_{\lambda}\left(1.65\micron\right) \else
$f_{\lambda}\left(1.65\micron\right)$\fi}

\newcommand{\fbol}{\ifmmode f_{\rm bol} \else $f_{\rm bol}$\fi}
\newcommand{\fbolwv}{\ifmmode f_{\rm bol}\left(\lambda\right) \else $f_{\rm bol}\left(\lambda\right)$\fi}
\newcommand{\fbolopt}{\ifmmode f_{\rm bol}\left(5100{\rm \AA}\right) \else $f_{\rm bol}\left(5100{\rm \AA}\right)$\fi}
\newcommand{\fbolthree}{\ifmmode f_{\rm bol}\left(3000{\rm \AA}\right) \else $f_{\rm bol}\left(3000{\rm \AA}\right)$\fi}
\newcommand{\fboluv}{\ifmmode f_{\rm bol}\left(1450{\rm \AA}\right) \else $f_{\rm bol}\left(1450{\rm \AA}\right)$\fi}

\newcommand{\fbolbat}{\ifmmode f_{\rm bol}\left(14-150\,\kev\right) \else $f_{\rm bol}\left(14-150\,\kev\right)$\fi}
\newcommand{\fbolhard}{\ifmmode f_{\rm bol}\left(2-10\,\kev\right) \else $f_{\rm bol}\left(2-10\,\kev\right)$\fi}

\newcommand{\fobs}{\ifmmode f_{\rm obs} \else $f_{\rm obs}$\fi}

\newcommand{  \mbh      }{\ifmmode M_{\rm BH} \else $M_{\rm BH}$\fi}
\newcommand{  \lmbh     }{\ifmmode \log\left(\mbh/\Msun\right) \else $\log\left(\mbh/\Msun\right)$\fi} 
\newcommand{  \lledd    }{\ifmmode L/L_{\rm Edd} \else $L/L_{\rm Edd}$\fi}
\newcommand{  \mmedd    }{\ifmmode \dot{m}/\dot{m}_{\rm \,Edd} \else $\dot{m}/\dot{m}_{\rm \,Edd}$\fi}
\newcommand{  \Lbol     }{\ifmmode L_{\rm bol} \else $L_{\rm bol}$\fi}
\newcommand{  \lbol     }{\ifmmode L_{\rm bol} \else $L_{\rm bol}$\fi}
\newcommand{  \lLbol    }{\ifmmode \log\left(\Lbol/\ergs\right) \else $\log\left(\Lbol/\ergs\right)$\fi} 
\newcommand{  \Lagn     }{\ifmmode L_{\rm AGN} \else $L_{\rm AGN}$\fi}
\newcommand{  \lagn     }{\ifmmode L_{\rm AGN} \else $L_{\rm AGN}$\fi}

\newcommand{  \tgrow     }{\ifmmode t_{\rm growth} \else $t_{\rm growth}$\fi}
\newcommand{  \tAD     }{\ifmmode t_{\rm acc} \else $t_{\rm acc}$\fi}
\newcommand{  \tacc    }{\ifmmode t_{\rm acc} \else $t_{\rm acc}$\fi}
\newcommand{  \tUni      }{\ifmmode t_{\rm Universe} \else $t_{\rm Universe}$\fi}

\newcommand{  \Mdotin	}{\ifmmode \dot{M}_{\rm infall} \else $\dot{M}_{\rm infall}$\fi}
\newcommand{  \Mdotbh	}{\ifmmode \dot{M}_{\rm BH} \else $\dot{M}_{\rm BH}$\fi}
\newcommand{  \Mdotad	}{\ifmmode \dot{M}_{\rm AD} \else $\dot{M}_{\rm AD}$\fi}
\newcommand{  \Mdotacc	}{\ifmmode \dot{M}_{\rm acc} \else $\dot{M}_{\rm acc}$\fi}
\newcommand{  \Mdotthin	}{\ifmmode \dot{M}_{\rm thin} \else $\dot{M}_{\rm thin}$\fi}
\newcommand{  \Mdotdisk	}{\ifmmode \dot{M}_{\rm disk} \else $\dot{M}_{\rm disk}$\fi}

\newcommand{  \Mindot	}{\ifmmode \dot{M}_{\rm infall} \else $\dot{M}_{\rm infall}$\fi}
\newcommand{  \Mbhdot	}{\ifmmode \dot{M}_{\rm BH} \else $\dot{M}_{\rm BH}$\fi}
\newcommand{  \Maddot	}{\ifmmode \dot{M}_{\rm AD} \else $\dot{M}_{\rm AD}$\fi}
\newcommand{  \Maccdot	}{\ifmmode \dot{M}_{\rm acc} \else $\dot{M}_{\rm acc}$\fi}
\newcommand{  \Mthdot	}{\ifmmode \dot{M}_{\rm thin} \else $\dot{M}_{\rm thin}$\fi}
\newcommand{  \Mdsdot	}{\ifmmode \dot{M}_{\rm disk} \else $\dot{M}_{\rm disk}$\fi}

\newcommand{  \as	}{\ifmmode a_{\rm *} \else $a_{\rm *}$\fi}
\newcommand{  \avec	}{\ifmmode \vec{a}_{\rm *} \else $\vec{a}_{\rm *}$\fi}
\newcommand{  \re	}{\ifmmode \eta      	 \else $\eta$\fi}
\newcommand{  \RISCO	}{\ifmmode R_{\rm ISCO}  \else $R_{\rm ISCO}$\fi}

\newcommand{  \mseed    }{\ifmmode M_{\rm seed} \else $M_{\rm seed}$\fi}
\newcommand{  \mbul     }{\ifmmode M_{\rm bulge} \else $M_{\rm bulge}$\fi} 
\newcommand{  \mstar    }{\ifmmode M_{*} \else $M_{*}$\fi} 
\newcommand{  \mgal     }{\ifmmode M_{*} \else $M_{*}$\fi} 
\newcommand{  \mhost    }{\ifmmode M_{\rm host} \else $M_{\rm host}$\fi}
\newcommand{  \mmsmall  }{\ifmmode M_{\rm BH}/M_{*} \else $M_{\rm BH}/M_{*}$\fi}
\newcommand{  \mmlarge  }{\ifmmode M_{*}/M_{\rm BH} \else $M_{*}/M_{\rm BH}$\fi}

\newcommand{  \mmdotlarge}{\ifmmode \dot{M}_*/\Mbhdot \else $\dot{M}_*/\Mbhdot$\fi}
\newcommand{  \mmdotsmall}{\ifmmode \Mbhdot/\dot{M}_* \else $\Mbhdot/\dot{M}_*$\fi}

\newcommand{  \mmwp     }{\ifmmode \left(M_{*}/M_{\rm BH}\right) \else $\left(M_{*}/M_{\rm BH}\right)$\fi}
\newcommand{  \ml       }{\ifmmode M_{*}/L_{*} \else $M_{*}/L_{*}$\fi}
\newcommand{  \mlwp     }{\ifmmode \left(M_{*}/L\right) \else $\left(M_{*}/L\right)$\fi}
\newcommand{  \mlk      }{\ifmmode \left(M_{*}/L_{K}\right) \else $\left(M_{*}/L_{K}\right)$\fi}
\newcommand{  \sigs     }{\ifmmode \sigma_{*} \else $\sigma_{*}$\fi}
\newcommand{  \Reff     }{\ifmmode R_{\rm e} \else $R_{\rm e}$\fi}
\newcommand{  \Rvir     }{\ifmmode R_{\rm vir} \else $R_{\rm vir}$\fi}
\newcommand{  \Rtwo     }{\ifmmode R_{200} \else $R_{200}$\fi}
\newcommand{  \Rfive    }{\ifmmode R_{500} \else $R_{500}$\fi}
\newcommand{  \Rgrp     }{\ifmmode R_{\rm grp} \else $R_{\rm grp}$\fi}
\newcommand{  \nser     }{\ifmmode n_{\rm s} \else $n_{\rm s}$\fi}
\newcommand{  \LSF      }{\ifmmode L_{\rm SF}  \else $L_{\rm SF}$\fi}
\newcommand{  \LFIR     }{\ifmmode L_{\rm FIR} \else $L_{\rm FIR}$\fi}
\newcommand{  \Lfir     }{\ifmmode L_{\rm FIR} \else $L_{\rm FIR}$\fi}
\newcommand{  \LTIR     }{\ifmmode L_{\rm TIR} \else $L_{\rm TIR}$\fi}
\newcommand{  \Ltir     }{\ifmmode L_{\rm TIR} \else $L_{\rm TIR}$\fi}

\newcommand{  \mdyn     }{\ifmmode M_{\rm dyn} \else $M_{\rm dyn}$\fi} 
\newcommand{  \mgas     }{\ifmmode M_{\rm gas} \else $M_{\rm gas}$\fi} 
\newcommand{  \mh       }{\ifmmode M_{\rm h} \else $M_{\rm h}$\fi}
\newcommand{  \mhalo    }{\ifmmode M_{\rm halo} \else $M_{\rm halo}$\fi}
\newcommand{  \sfr      }{\ifmmode {\rm SFR} \else SFR\fi}

\newcommand{ \Lcii     }{\ifmmode L_{\cii} \else $L_{\cii}$\fi}
\newcommand{ \fwcii  }{\ifmmode {\rm FWHM}\cii \else FWHM\cii\fi}

\newcommand{  \swift     }  {{\it Swift}}

\newcommand{\bj}{\ifmmode b_{\rm J} \else $b_{\rm J}$\fi}

\newcommand{\iab}{\ifmmode i_{\rm AB} \else $i_{\rm AB}$\fi}

\newcommand{\jab}{\ifmmode J_{\rm AB} \else $J_{\rm AB}$\fi}
\newcommand{\hab}{\ifmmode H_{\rm AB} \else $H_{\rm AB}$\fi}
\newcommand{\kab}{\ifmmode K_{\rm AB} \else $K_{\rm AB}$\fi}

\newcommand{\jveg}{\ifmmode J_{\rm Vega} \else $J_{\rm Vega}$\fi}
\newcommand{\hveg}{\ifmmode H_{\rm Vega} \else $H_{\rm Vega}$\fi}
\newcommand{\kveg}{\ifmmode K_{\rm Vega} \else $K_{\rm Vega}$\fi}

\def\arcmin{\hbox{$^\prime$}}
\def\arcsec{\hbox{$^{\prime\prime}$}}

\newcommand{  \Chisq    }{\ifmmode \chi^{2} \else $\chi^{2}$}
\newcommand{  \nelec    }{\ifmmode n_{e} \else $n_{e}$\fi}     
\newcommand{  \nh       }{\ifmmode n_{\rm H} \else $n_{\rm H}$\fi}     
\newcommand{  \Ncol     }{\ifmmode N_{\rm col} \else $N_{\rm col}$\fi} 
\newcommand{  \NH       }{\ifmmode N_{\rm H} \else $N_{\rm H}$\fi}     

\def\deg{\hbox{$^\circ$}}

\def\arcmin{\hbox{$^\prime$}}
\def\arcsec{\hbox{$^{\prime\prime}$}}
\def\ion#1#2{#1$\;${\small\rm\@Roman{#2}}\relax}

\usepackage[T1]{fontenc}
\usepackage{ae,aecompl}
\usepackage{newtxtext,newtxmath,amssymb}

\newcommand{  \lamEdd }{\ifmmode \lambda_{\rm Edd} \else $\lambda_{\rm Edd}$\fi}
\newcommand{  \Lrad   }{\ifmmode L_{\rm 1.4\,GHz} \else $L_{\rm 1.4\,GHz}$\fi}
\def\XS{XSHOOTER}
\newcommand{  \Lhard }{\ifmmode L_{2-10\,\kev} \else 
$L_{2-10\,\kev}$\fi}
\newcommand{  \Lbat }{\ifmmode L_{14-195\,\kev} \else 
$L_{14-195\,\kev}$\fi}

\title[BASS XIII: The most luminous obscured local AGN]{BAT AGN Spectroscopic Survey --  XIII. The nature of the most luminous obscured AGN in the low-redshift universe}

\author[R.~E.\ B{\"a}r et al.]{\Large\parbox{\textwidth}{Rudolf E. B\"{a}r$^{1}$\thanks{E-mail: baerr@phys.ethz.ch}, 
Benny Trakhtenbrot$^{2,3}$\thanks{E-mail: benny@astro.tau.ac.il},
Kyuseok Oh$^{4}$\thanks{JSPS fellow}, 
Michael J. Koss$^{5}$,
O. Ivy Wong$^{6}$, 
Claudio Ricci$^{7,8}$, 
Kevin Schawinski$^{1,9}$,
Anna K. Weigel$^{1,9}$, 
Lia F. Sartori$^{1}$, 
Kohei Ichikawa$^{10,11}$, 
Nathan J. Secrest$^{12}$,
Daniel Stern$^{13}$, 
Fabio Pacucci$^{14,15}$, 
Richard Mushotzky$^{16}$,
Meredith C. Powell$^{17}$,
Federica Ricci$^{18}$,
Eleonora Sani$^{19}$,
Krista L. Smith$^{20}$,
Fiona A. Harrison$^{21}$,
Isabella Lamperti$^{22}$, 
and C. Megan Urry$^{17}$}\\
~\\
$^{1}$Institute for Particle Physics and Astrophysics, Department of Physics, ETH Zurich, Wolfgang-Pauli-Strasse 27, CH-8093 Zurich, Switzerland\\
$^{2}$Department of Physics, ETH Zurich, Wolfgang-Pauli-Strasse 27, CH-8093 Zurich, Switzerland\\
$^{3}$School of Physics and Astronomy, Tel Aviv University, Tel Aviv 69978, Israel\\
$^{4}$Department of Astronomy, Kyoto University, Kyoto 606-8502, Japan\\
$^{5}$Eureka Scientific, 2452 Delmer Street Suite 100, Oakland, CA 94602, USA\\
$^{6}$International Centre for Radio Astronomy Research, The University of Western Australia M468, 35 Stirling Highway, Crawley, WA 6009, Australia\\
$^{7}$N\'ucleo de Astronom\'ia de la Facultad de Ingenier\'ia, Universidad Diego Portales, Av. Ej\'ercito Libertador 441, Santiago, Chile\\
$^{8}$Kavli Institute for Astronomy and Astrophysics, Peking University, Beijing 100871, China\\
$^{9}$Modulos AG, Technoparkstrasse 1, CH-8005 Zurich, Switzerland\\
$^{10}$Frontier Research Institute for Interdisciplinary Sciences, Tohoku University, Sendai 980-8578, Japan\\
$^{11}$Astronomical Institute, Tohoku University, Aramaki, Aoba-ku, Sendai, Miyagi 980-8578, Japan\\
$^{12}$U.S. Naval Observatory, 3450 Massachusetts Avenue NW, Washington, DC 20392, USA\\
$^{13}$Jet Propulsion Laboratory, California Institute of Technology, 4800 Oak Grove Drive, Mail Stop 169-221, Pasadena, CA 91109, USA\\
$^{14}$Kapteyn Astronomical Institute, Groningen, 9747 AD, Netherlands\\
$^{15}$BHI \& Clay Fellow | Harvard University \& Smithsonian Astrophysical Observatory, Cambridge, MA, USA\\
$^{16}$Department of Astronomy and Joint Space-Science Institute, University of Maryland, College Park, MD 20742, USA\\
$^{17}$Yale Center for Astronomy and Astrophysics, and Physics Department, Yale University, PO Box 2018120, New Haven, CT 06520-8120\\
$^{18}$Instituto de Astrof\'{i}sica, Facultad de F\'{i}sica, Pontificia Universidad Cat\'{o}lica de Chile, Casilla 306, Santiago 22, Chile\\
$^{19}$European Southern Observatory, Alonso de Cordova 3107, Casilla 19, Santiago 19001, Chile\\
$^{20}$Kavli Institute for Particle Astrophysics and Cosmology, Stanford University, SLAC National Accelerator Laboratory, Menlo Park, CA 94025, USA\\
$^{21}$Cahill Center for Astrophysics, 1216 E. California Blvd, California Institute of Technology, Pasadena, CA 91125, USA\\
$^{22}$Astrophysics Group, Department of Physics and Astronomy, University College London, 132 Hampstead Road, London NW1 2PS, UK
\vspace*{-0.75cm}
}

\date{\vspace*{-0.4cm}
Accepted 2019 August 12. Received 2019 July 25; in original form 2018 August 10.
}

\begin{document}
\label{firstpage}
\pagerange{\pageref{firstpage}--\pageref{lastpage}}
\maketitle

\begin{abstract}
We present a multi wavelength analysis of 28 of the most luminous low-redshift narrow-line, ultra-hard X-ray selected active galactic nuclei (AGN) drawn from the 70 month \swift/BAT all-sky survey, with bolometric luminosities of $\log(\Lbol/\ergs) \gtrsim 45.25$. 
The broad goal of our study is to determine whether these objects have any distinctive properties, potentially setting them aside from lower-luminosity obscured AGN in the local Universe.
Our analysis relies on the first data release of the BAT AGN Spectroscopic Survey (BASS/DR1) and on dedicated observations with the VLT, Palomar, and Keck observatories. 
We find that the vast majority of our sources agree with commonly used AGN selection criteria which are based on emission line ratios and on mid-infrared colours.
Our AGN are predominantly hosted in massive galaxies ($9.8 \lesssim \log [M_{*}/\Msol] \lesssim 11.7$); based on visual inspection of archival  optical images, they appear to be mostly ellipticals.
Otherwise, they do not have distinctive properties. 
Their radio luminosities, determined from publicly available survey data, show a large spread of almost 4 orders of magnitude -- much broader than what is found for lower X-ray luminosity obscured AGN in BASS. 
Moreover, our sample shows no preferred combination of black hole masses (\mbh) and/or Eddington ratio (\lamEdd), covering $7.5 \lesssim \log(\mbh/\Msol) \lesssim 10.3$ and $0.01 \lesssim \lamEdd \lesssim 1$.
Based on the distribution of our sources in the $\lamEdd-\NH$ plane, we conclude that our sample is consistent with a scenario where the amount of obscuring material along the line of sight is determined by radiation pressure exerted by the AGN on the dusty circumnuclear gas. 
\end{abstract}

\begin{keywords}
\vspace*{-0.1cm}
galaxies: active, galaxies: Seyfert 2, galaxies: nuclei, radio continuum: galaxies;
\end{keywords}

\section{Introduction}
\label{sec:intro}

The highest luminosity active galactic nuclei (AGN), with bolometric luminosities of $\Lbol > 10^{45}\,\ergs$, probe the epochs of maximal absolute accretion rates of the supermassive black holes (SMBHs) that power them, and naturally represent the consequences of the most extreme radiative outputs of such systems. 
Thus, they can provide key insights on a broad range of questions, ranging from accretion and jet-launching physics, through the interplay between the AGN output and circumnuclear material, and indeed the galaxy-scale mechanisms that drive extremely efficient SMBH growth.
In addition to unobscured, optically-selected quasars, which are commonly considered as representing the high-luminosity AGN regime, {\it obscured} high-luminosity AGN should also be considered, both to provide a more complete view of the AGN population, and because they offer unique opportunities to address some of the outstanding questions.

In terms of SMBH accretion demographics, the foremost question to address is whether the high accretion rates of the most luminous AGN (in terms of $\Mdotbh \propto \Lbol$) are driven by high-mass SMBHs accreting at moderate Eddington ratios (hereafter $\lamEdd \equiv \Lbol/L_{\rm Edd} \propto \Lbol/\mbh$), or by moderate-mass BHs with extremely high \lamEdd\ (or indeed super-Eddington accretion), or potentially a mixture of these two extreme cases.
As the distributions of \mbh\ and \lamEdd\ are observed to evolve with redshift \cite[e.g.,][]{TrakhtNetzer2012_Mg2,Schulze2015_BHMF,Trakhtenbrot2016_COSMOSFIRE_MBH},
forming a complete census of local AGN may provide a crucial benchmark for evolutionary studies of the cosmic growth of SMBHs.

The high intrinsic X-ray and/or UV-optical luminosities of highly luminous SMBHs should also be reflected in other spectral regimes and features. 
The intrinsically strong UV radiation, reprocessed by the narrow line region (NLR) and the cicumnuclear obscuring material, is expected to make the host galaxies of high-luminosity AGN easily distinguishable from other (inactive) galaxies in term of their narrow emission line ratios \cite[]{1981PASP...93....5B,Kewley2001_BPT} and mid-infrared colours \cite[e.g.,][]{2011ApJ...735..112J,2012ApJ...753...30S,Mateos_2012}.
Since in most accretion disc models the relative strength of the UV radiation is expected to depend on \mbh\ and/or \lamEdd\ \cite[see, e.g.,][]{Abramowicz2013,Capellupo2015_ADs,2016MNRAS.458.1839C}, these secondary AGN signatures may, again, depend on the basic properties of the SMBHs and accretion flows powering the most luminous AGN.
Furthermore, the high (hard-band) X-ray luminosities of some AGN may be associated with the observations of  significant core radio emission likely to originate from the central engine of the AGN, and thus high radio luminosities could be expected \cite[e.g.,][]{2016A&ARv..24...10T}. 
Therefore, a detailed analysis of the radio properties of high-luminosity, obscured AGN may provide additional insights into the strong non-thermal radiation linked to jets and associated radio lobes.

The most luminous obscured AGN may also serve to test different AGN unification models \cite[e.g,][]{Antonucci1993,UrryPadovani1995_rev}. 
Although the general, orientation-dominated unification model of AGN is well accepted, a number of studies point out some contradictions (see, e.g., \citealt{2014cosp...40E2695R,Netzer2015_torus_rev,2017MNRAS.464.2139A,2017ApJ...837..110V}, for a detailed discussion). 
The very existence of highly luminous obscured AGN can put strong constraints on, or indeed be in tension with, the ``receding torus'' model, put forward by \cite{1991MNRAS.252..586L} and used to interpret several AGN population studies \cite[e.g.,][]{2005MNRAS.360..565S,2015ApJS..219....1O}.
In this model, an increasing AGN luminosity would dictate a larger dust sublimation radius, and thus an inner edge of the torodial obscuring region (the ``torus'') that is further out from the accreting SMBH, leading to the observed decrease in the fraction of obscured AGN with increasing AGN luminosity.
An alternative scenario, developed in \cite{2006MNRAS.373L..16F}, \cite{Fabian2008_rad_press}, and \cite{2009MNRAS.394L..89F}, suggests instead that the distribution of cicrumnuclear obscuring material is dominated by the radiative pressure exerted by the AGN. 
This would mean that generally the fraction of mildly obscured (i.e., Compton-thin) AGN should critically depend on \lamEdd\, instead of on \Lbol\ alone. 
This would allow for highly luminous obscured AGN, provided that they are indeed powered by high-\mbh, moderate-\lamEdd\ SMBHs (with an additional dependence on column density).
This alternative ``radiative feedback driven unification'' scenario was recently shown to explain the obscuration and accretion rate properties of a large, highly complete sample of local AGN \citep{2017Natur.549..488R}.

High-luminosity AGN have been long suggested to be preferentially associated with major galaxy-galaxy mergers \citep[e.g,][]{Bahcall1997,2011ApJ...739...57K,2012ApJ...758L..39T,Glikman2015_red_QSO_mergers,2016IAUFM..29B.113H}. 
The recent study by \cite{2018MNRAS.476.2308W} showed that this can be (at least partially) explained by a combination of the well-known SMBH-host relations \citep{2013ARA&A..51..511K} and the higher probability of the most massive host galaxies to be associated with mergers (see also \citealt{2014ApJ...782....9H}). 
Moreover, there is evidence suggesting that the highest luminosity AGN should be found in hosts with high star formation rates (SFRs), marking periods of fast, ``co-evolutionary'' assembly of both BH and stellar mass \cite[e.g.,][]{Lutz2008_SF_AGN,Netzer2009_SF_AGN}.
Thanks to the obscured nature of the central AGN source, obscured high-luminosity AGN offer a unique opportunity to study these and other properties of the galaxies hosting the most vigorously accreting SMBHs.

To address all of these questions, one would require a large and complete sample of AGN, selected in a way that overcomes the obvious selection effects caused by circumnuclear obscuration, and preferably one that has a rich collection of ancillary multi-wavelength data, and in particular optical spectroscopy that allows measurements \mbh, and thus of \lamEdd\ (e.g., through stellar velocity  dispersion ($\sigs$) and the $\mbh-\sigs$ relation).
However, highly luminous obscured AGN are very rare \cite[e.g.,][]{2008AJ....136.2373R,2017MNRAS.468.3042M} and difficult to find, due to a combination of several well-established trends seen in the AGN population:
the steep decrease of the AGN luminosity function with increasing luminosity \citep{2008AJ....136.2373R}; 
the fact that the space density of more luminous AGN has peaked at higher redshifts \cite[e.g.,][]{2004MNRAS.349.1397C,2005A&A...441..417H,2006AJ....131.2766R,2013ApJ...773...14R,2015A&ARv..23....1B,2014ApJ...786..104U,2018PASJ...70S..34A}; 
and the decrease in the fraction of optically obscured or X-ray absorbed AGN with increasing luminosity (e.g., \citet{2015MNRAS.454.1202S,2017ApJ...841L..18M}, but see also \citet{Assef2015_HotDOGs}).\footnote{This is the observational consequence of the two scenarios linking AGN accretion and obscuration, discussed earlier in this introductory Section.}
At low redshifts, these requirements and limitations necessitate extremely wide-area AGN surveys with detailed spectroscopic follow-up.

Indeed, many studies pursued several observational approaches to construct statistical samples of high-luminosity type-2 AGN at different redshifts \cite[occasionally referred to as ``type-2 quasars''; e.g.,][]{2003AJ....126.2125Z,2004AJ....128.1002Z,2008AJ....136.2373R,2009ApJ...702.1098L,2013MNRAS.435.3306A}.
\citet{2003AJ....126.2125Z} used the optical spectroscopy of the Sloan Digital Sky Survey \citep[][assisted by wide-area radio and X-ray surveys]{2000AJ....120.1579Y} to identify a large sample of about 290 obscured quasar candidates at $0.3 < z < 0.8$. 
Follow-up multi-wavelength analyses of this sample \cite[e.g.,][]{2004AJ....128.1002Z} suggest that they are generally consistent with what is seen in the luminous, unobscured AGN population, including the fraction of radio-loud sources and a tendency towards high-SFR hosts, with a low fraction of mergers \citep{2006AJ....132.1496Z}.
This SF nature of the host galaxies was further investigated by \citet{2009ApJ...702.1098L} based on 9 sources, finding significant contributions from young stellar populations, broadly supporting the idea that intense SMBH growth may follow an epoch of fast host growth.
Although some follow-up studies of the SDSS type-2 quasars showed that   ionized gas outflows may be common in such systems \cite[e.g.,][]{Greene2011_Q2_outflows,Villar-Martin2011_Q2_outflows}, these may not necessarily strongly affect the host galaxies \cite[e.g.,][]{Villar-Martin2016_Q2_outflows}.
Later data releases of SDSS spectroscopy allow to extend the search for obscured high-luminosity AGN to larger samples, with over 2700 sources at $z<1$ \citep{2016MNRAS.462.1603Y}, as well as to higher redshifts, with over 140 candidates at $2 < z < 4.3$ \citep{2013MNRAS.435.3306A}.
These large SDSS-based samples suggested that many (and indeed, most) luminous obscured AGN would not be selected by commonly used mid-infrared (MIR) colour criteria, and that outflows of highly ionized gas are prevalent among such luminous AGN, thus providing further evidence for the possible impact of highly accreting SMBHs on their hosts.

Additional, complementary approaches for identifying large samples of high-luminosity obscured AGN focus on other spectral regimes, including X-rays, radio or MIR (see \citealt{HickoxAlexander2018_obs_AGN_rev} for a recent detailed review).
Several samples of broad-line but heavily reddened, luminous AGN (``red quasars'') were identified by combining survey data in the near-IR, mid-IR, X-rays and/or radio regimes \cite[e.g.,][]{Glikman2007_red_QSO_FIRST,Banerji2012_red_QSO_UKIDSS,Ross2015_red_QSO_BOSS,LaMassa2017_S82X_red_QSO}.
Such systems were suggested to trace a relatively short phase of growth of high-\mbh\ SMBHs \citep{Banerji2015_red_QSO_MBH}, preferentially associated with major galaxy mergers and/or intense host SF \citep{Glikman2015_red_QSO_mergers,Banerji2017_red_QSO_ALMA}.
It was recently suggested that this evolutionary phase may be indeed tracing the ``blow-out'' of dusty obscuring material implied by the radiative feedback scenario \citep{Glikman2017_red_QSO_note}.
Finally, a large fraction of MIR-selected extremely luminous, hot, dust-obscured galaxies \cite[``Hot-DOGs'';][and references therein]{Assef2015_HotDOGs} were also suggested to be powered by vigorously accreting SMBHs \cite[e.g.,][]{Stern2014_HotDOGs_Xray}.
The very existence of such systems, with extremely high bolometric luminosities and column densities ($\log[\Lbol/\ergs]\gtrsim 47$, $\log[\NH/\cmii]\gtrsim 23.5$; e.g., \citealt{Goulding2018_red_QSO_xray,Vito2018_HotDOGs_Xray}), challenges the ``receding torus'' model.
On the other hand, these high column densities mean that they can be accommodated within the radiative feedback scenario, despite their high accretion rates, $\lamEdd\sim1$ (see, e.g., \citealt{Wu2018_HotDOGs_MBH} for Hot-DOGs).

Most recently, \cite{KongHo2018_Q2} presented a systematic study of \mbh\ and \lamEdd\ in the  large, SDSS-based sample of obscured luminous AGN presented by \citet{2008AJ....136.2373R}, which spans bolometric luminosities in the range $45.5 \lesssim \log(\Lbol/\ergs) \lesssim 47.5$.
Relying on stellar velocity dispersion  measurements and the $\mbh-\sigs$ relation, they find that the sources in their sample have BH masses in the range $6.5 \lesssim \log(\mbh/\Msol) \lesssim 10$, and accrete at rates that correspond to $-2.9 \lesssim \log\lamEdd \lesssim 1.8$. 
At face value, the high accretion rates challenge the aforementioned radiative feedback scenario \cite[i.e.,][]{Fabian2008_rad_press,2017Natur.549..488R}.

The all-sky ultra-hard X-ray (14-195 keV) survey carried out by the BAT instrument onboard the \swift\ mission provides an optimal starting point for constructing a large sample of highly luminous, obscured AGN, in the local Universe.  
This is mainly due to the fact that the AGN emission in the ultra-hard X-ray band
is minimally affected by obscuring material along the line-of-sight.
Moreover, the BAT AGN Spectroscopic Survey \cite[BASS,][]{2017ApJ...850...74K,2017ApJS..233...17R} provides a large, rich, and ever-expanding set of ancillary multi-wavelength data, allowing the clear identification of optical counterparts (i.e., host galaxies), and reliable determination of redshifts, optical emission line properties, AGN sub-classifications, X-ray spectral properties, and -- crucially -- BH masses and accretion rates.

In this paper we investigate a sample of 28 of the most luminous obscured (type-2) AGN in the local Universe, selected from the 70-month catalogue of the \swift/BAT all-sky survey \citep{Baumgartner2013_SwiftBAT_70m} and further studies using the data obtained through the BASS project.
Our main objective is to determine whether as a group these sources can be set apart from the overall local AGN population as having some common characteristics (besides their luminosities). 
We present our sample of highly luminous, obscured AGN, and the optical spectroscopic observational data in Section~\ref{sec:data_sample}.
In Section~\ref{sec:analysis} we describe the data analysis, paying particular attention to several different characteristics of our sample.
We first examine how well our sample agrees with commonly used AGN selection methods, specifically strong (narrow) emission lines (Section \ref{subsec:emlines}) and mid-infrared colours (Section \ref{subsec:infraredl}).
In Section \ref{subsec:Hosts} we discuss the properties of the host galaxies, and particularly their morphology.
We then use available radio data to test the possible links between high accretion power and radio jet activity, and to place our sample in the context of the so-called fundamental plane of black hole activity (Section \ref{subsec:radio}). 
In Section \ref{sec:bhm} we investigate the black hole masses and accretion rates of our sample, and discuss them in the context of physically-motivated AGN unification models. 
Finally, we summarise our main findings in Section \ref{sec:conclusions}.
Throughout this paper, we use a standard $\Lambda$CDM cosmology with $\Omega_\Lambda = 0.7$, $\Omega_{\rm M} = 0.3$ and $H_0 = 70\,\kms\,{\rm Mpc}^{-1}$, consistent with observational measurements \citep{2011ApJS..192...18K}.

\section{Sample and Basic Observational Data}
\label{sec:data_sample}

\begin{figure}
\begin{center}
\includegraphics[trim={0 0 1cm 0},clip,width=0.49\textwidth]
{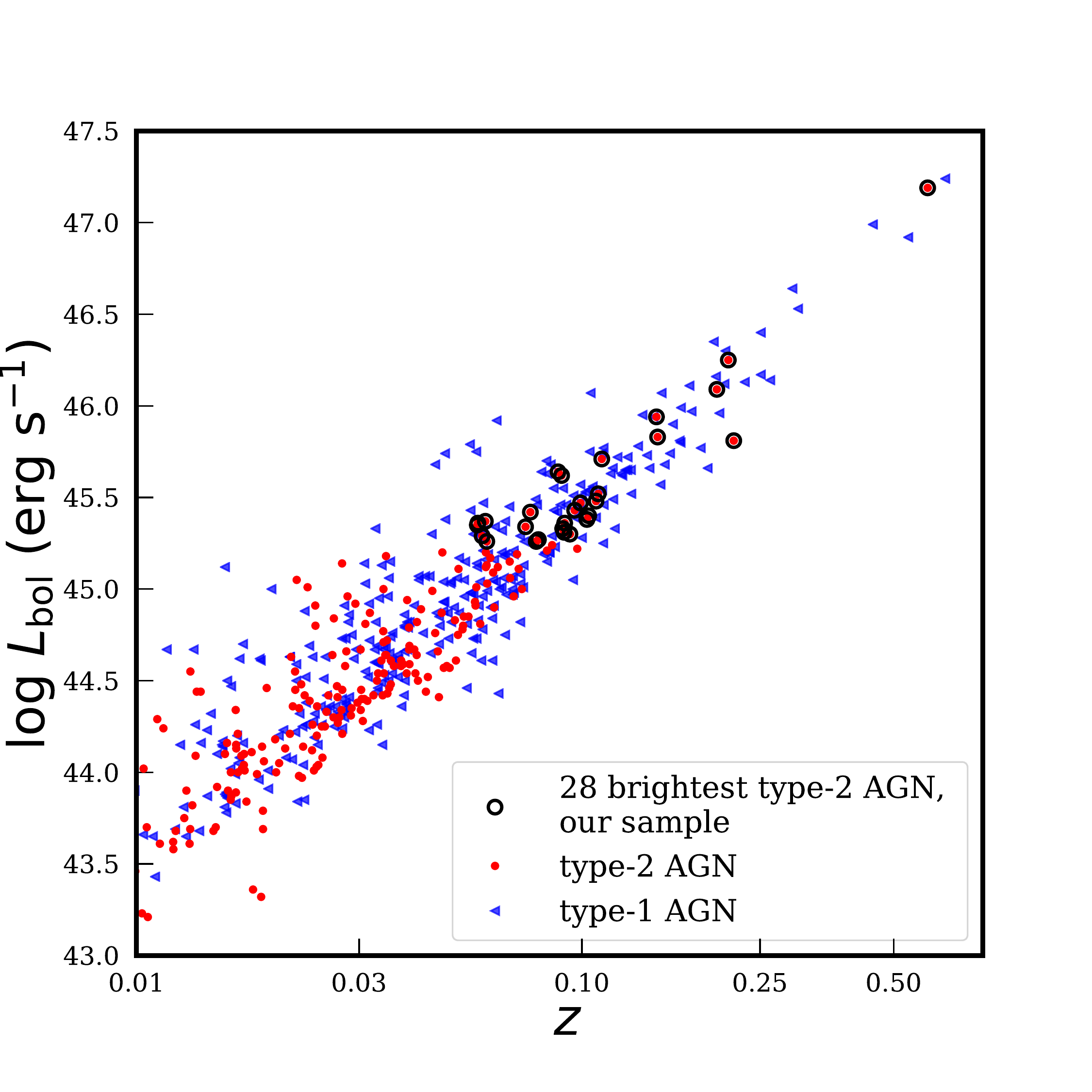}
\caption{
The luminosity-redshift plane for BASS/DR1 AGN. 
The black circles represent our sample of 28 of the most luminous type-2 AGN in BASS/DR1.
We have included in our sample the extremely luminous AGN at the centre of the Phoenix cluster, $z=0.597$ (2MASX J23444387$-$4243124; BAT ID 1204).
Blazars and sources within  6$^{\circ}$ of the galactic plane, which are included in  the BASS/DR1 population, are not shown.}
\label{fig:Lum_vs_z}
\end{center}
\end{figure}

\subsection{Sample selection and luminosity estimates}
\label{subsec:sample}

In this work we analyse the some of the most luminous type-2 AGN, not showing broad lines \cite[e.g.,][]{1981ApJ...249..462O}, from the 70-month \swift/BAT all-sky survey. 
We select a group of type-2 AGN based on X-ray luminosities and study their characteristics in the optical, infrared and radio regimes.
Our initial sample is based on the data collected with the BAT instrument \citep{2005SSRv..120..143B} on board the {\it Niel Gehrels Swift Observatory} \citep{2004ApJ...611.1005G} within its first 70 months of all-sky survey observations \citep{Baumgartner2013_SwiftBAT_70m}, as well as the data on the 836 AGN reported in the first data release of the BAT AGN Spectroscopic Survey \cite[BASS/DR1;][]{2017ApJ...850...74K}.
For general reference, we show in Fig.~\ref{fig:Lum_vs_z} the bolometric luminosities (\Lbol, based on X-ray emission; see below) vs. redshift of {\it all} type-1 and type-2 AGN of the BASS/DR1.

Throughout this work, we adopt the BASS/DR1 bolometric luminosities derived directly from the (observed; see below) ultra-hard X-ray emission at $14-195$ keV, by applying a constant bolometric correction of $f_{\rm bol}=8$, that is $\Lbol = 8 \times \Lbat$.
As explained in \citet{2017ApJ...850...74K}, this somewhat simplified correction was derived as follows: 
the 14-195 keV luminosity was first scaled down by factor of 2.67 as a way to obtain the (intrinsic) $2-10$ keV luminosity (\Lhard), following \citet{2009ApJ...700.1878R} which is, in turn, based on scaling the \cite{2004MNRAS.351..169M} templates to higher X-ray energies. 
Next, the median $2-10$ keV bolometric correction of the BAT sample \citep{2009MNRAS.399.1553V} was adopted, which resulted in a final bolometric correction of $f_{\rm bol}=8$.

Uncertainties on \Lbol\ are clearly dominated by systematics, given that the accretion-driven AGN SEDs span from the ultra-hard X-rays through the UV, to the optical (and perhaps beyond), and that the SED shape may depend on several physical properties of the accreting SMBH.
Indeed, several works have studied possible links between the bolometric corrections and AGN luminosity, Eddington ratio, and perhaps other properties \cite[see, e.g.,][and references therein]{2004MNRAS.351..169M,VasudevanFabian2007_BC,Jin2012_Xop_3_SEDs}.
Notwithstanding this range of possibilities, our experience within BASS \cite[e.g.,][]{2017MNRAS.464.1466O,2017Natur.549..488R,2017MNRAS.470..800T,Ichikawa2019} shows that these higher-order bolometric corrections have little effect on the conclusions that are drawn from the implied bolometric luminosities. 
We thus prefer to adopt the simple, fixed bolometric corrections. 
We verified that none of our main conclusions regarding our sample of luminous, obscured AGN would change if we adopt instead the alternative bolometric corrections. 
Here we only note that if we directly apply \Lhard-based prescription of \cite{2004MNRAS.351..169M} to the \Lhard\ measurements of our sample, the median difference between the resulting bolometric luminosities and the $\Lbat$-based ones would be $\lesssim0.3$ dex.
We finally stress that, throughout this work, we use the simpler, model-independent ``observed'' \Lbat\ measurements, as reported in BASS/DR1 \citep{2017ApJ...850...74K}, rather than the ``intrinsic'' ones tabulated in \cite{2017ApJS..233...17R}. 
For the sample studied here, the differences between these two sets of measurements are negligible: the mean and median differences are of about 0.03 dex.

We further constrain our sample to AGN classified as Seyfert 2 AGN, based on the information contained within BASS/DR1, and particularly on the prominence of their narrow optical emission lines (i.e., \hbeta, \OIII, \Halpha, \NIIopt). 
We note that \cite{2017ApJ...850...74K} includes (re-)classification of all the AGN in BASS/DR1, taking into account the best spectroscopic measurements available of broad and narrow optical emission lines.
We also stress that this selection does not directly involve the X-ray based classification of obscured or unobscured AGN (i.e., based on the line-of-sight column density, \NH; see below).
We next removed all beamed AGN (i.e., Blazars and BL Lacs), based on the 5th edition of the Roma BZCAT (\citealt{2015Ap&SS.357...75M}; see more details in \citealt{2017ApJ...850...74K}); and all AGN within 6$^{\circ}$ of the Galactic plane, due to the high levels of Galactic extinction and low optical spectroscopic completeness. 
We finally selected the 30 highest-\Lbol\ (i.e., highest \Lbat) AGN, with $45.25 <\log(\Lbol/\ergs) < 47.2$.
For two sources (BAT IDs 203 and 555) the BAT detections in the 70-month \swift/BAT catalogue were of particularly low significance (although the softer X-ray data leaves no doubt regarding the AGN and optical counterpart identification). 
We thus preferred to use instead the more reliable BAT flux measurements reported in the recently published 105-month catalogue \citep{2018ApJS..235....4O}.
As a result of our dedicated campaign aimed to complete the spectroscopy for the 30 sources (see Section~\ref{subsec:new_obs} below), we removed from our sample two sources (BAT IDs 811 and 303) which we re-classified as type 1.9 AGN.
We note that the highest luminosity source in our sample, BAT ID 1204 (2MASX J23444387$-$4243124, at $z=0.597$), is the AGN at the centre of the Phoenix cluster \cite[e.g.,][]{McDonald2012_Phoenix}.
It is an apparent outlier both in terms of its luminosity and location in the luminosity-redshift plane (see Fig.~\ref{fig:Lum_vs_z}). 
However, we have no clear indication for this source to be beamed or otherwise different from the other AGN that pass our selection criteria, and its spectral X-ray properties are consistent with those of an accretion-powered, non-beamed, obscured AGN \cite[e.g.,][]{2013ApJ...778...33U}.
We have thus decided to keep it in the sample.
Our final sample of extremely luminous type-2 AGN thus consists of 28 sources.

Given the high luminosities of our sources, one could suspect that their (ultra-)hard X-ray emission may be contaminated by emission from a jet component, particularly given the (resolved) radio emission detected in many of our sources (as discussed in detail in Section~\ref{subsec:radio}).
We have however verified that the jet contribution to \Lbat\ (and thus \Lbol) is negligible.
We first note again that we have explicitly excluded sources that were reported as Blazars (or beamed AGN).
Moreover, extended X-ray emission is more difficult to be significantly obscured (i.e., at the $\log[N_{\rm H}/\cmii]\geq22$ levels relevant to our AGN), and the jet emission is typically detected at energies below the photoelectric cutoff ($\lesssim 4-5\,\kev$). 
This, in turn, would increase the {\it scattered fraction} above the typical 1\% found for the entire BASS sample \citep{2017ApJS..233...17R}.
However, for all the sources of our sample the scattered fractions are $f_{\rm scat} \leq 7-8\%$, which implies that the contribution of extended jets to the X-ray spectra is sub-dominant, at most, and is typically well below 10\% of the X-ray emission in the BAT band. 
Several of our sources have published {\it Chandra} images that indeed resolve the jets, and confirm that the X-ray emission from these jets is much weaker than the central engine. 
Some examples for this are BAT IDs 57 and 209 (3C~033 and 3C~105, respectively; \citealt{2012A&A...545A.143B}), and BAT ID 118(3C~062; \citealt{2017MNRAS.470.2762M}).

\begin{figure*}
\centering
\begin{tabular}{cc}
    \includegraphics[width=0.49\textwidth]{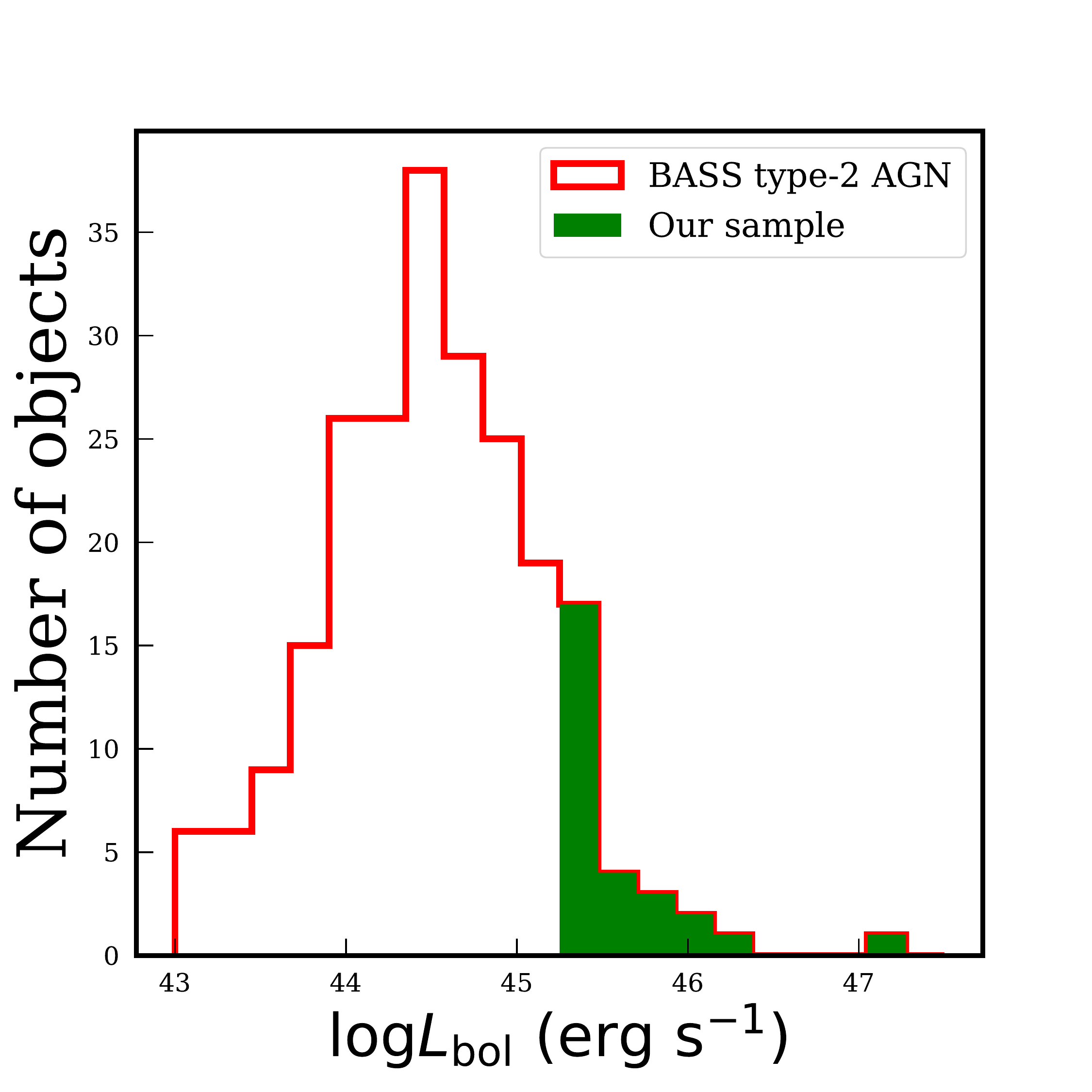} &
    \includegraphics[width=0.49\textwidth]{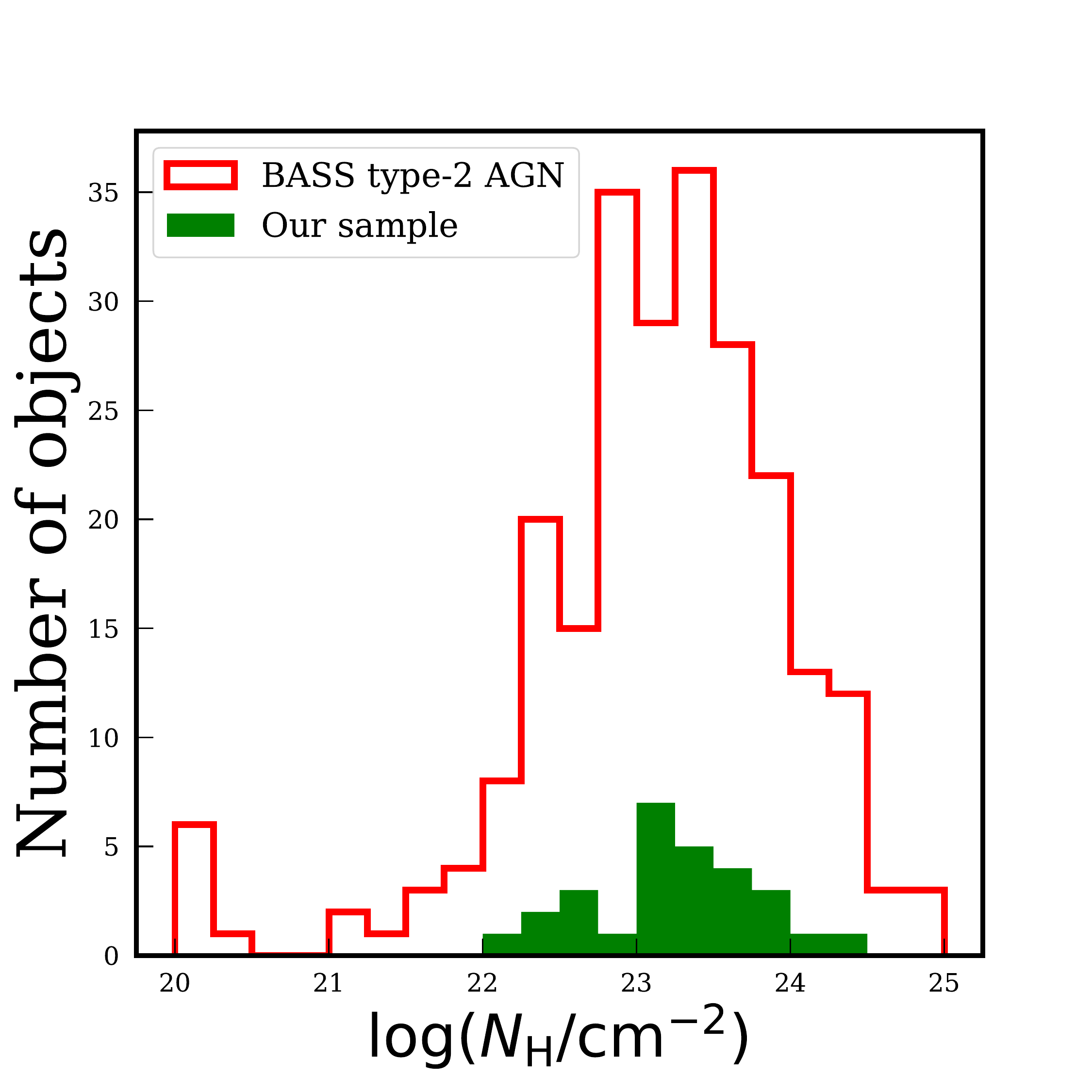} 
\end{tabular}
\caption{
Comparison of basic X-ray based properties of our sample of high luminosity \swift/BAT type-2 AGN to the general type-2 AGN population in BASS/DR1.
{\it Left:}
the distributions of (ultra-hard X-ray based) bolometric luminosities, $\log\Lbol$.
Our selected sample of 28 extremely luminous sources, with $45.3 \lesssim \log(\Lbol/\ergs) \lesssim  47.2$, represents the high-luminosity end of the BASS/DR1 type-2 AGN sample; some sources are not included in our sample, as we omit beamed AGN (blazars) and sources within  6$\rm ^o$ of the Galactic plane.
{\it Right:} the distributions of line-of-sight column densities, $\log\NH$. 
The sources of our sample are obscured with $\log(\NH/\cmii)  > 22$ and extend into the Compton thick range with  $\log(\NH/\cmii) \geq 24$. 
}
\label{fig:Lum_NH_dists}
\end{figure*}

In Fig.~\ref{fig:Lum_NH_dists} we show the distribution of X-ray based bolometric luminosities (\Lbol; left panel) and line-of-sight column densities (\NH, right panel) of our sample of 28 high luminosity type-2 AGN.
These are compared to the entire type-2 AGN population in BASS/DR1 (excluding Blazars and Galactic-plane sources).
This consistently-selected comparison sample of BASS/DR1 type-2 AGN is used through out this paper. 
The general BASS/DR1 type-2 population has bolometric luminosities in the range $43.0 < \log(\Lbol/\ergs)  < 47.2$, compared to $45.3 <  \log(\Lbol/\ergs) < 47.2$ for our sample (or $45.3 <  \log(\Lbol/\ergs) < 46.3$ if one excludes BAT ID 1204).
We note that the bolometric luminosities of our AGN overlap with the higher-luminosity sources of other samples of optically selected, luminous obscured AGN \cite[e.g.,][$\log(\Lbol/\ergs)\sim 44.6-47.0$]{2003AJ....126.2125Z}
All the sources in our sample are classified as obscured based on their X-ray SEDs, with $\log(\NH/\cmii) > 22$, and extend into the Compton-thick regime \cite[i.e., $\log(\NH/\cmii) \geq 24$;][]{2008ApJ...673...96A,2015ApJ...815L..13R,2016A&A...594A..73A,2017NatAs...1..679R,2018MNRAS.474.1225A}.

\subsection{New optical spectroscopy}
\label{subsec:new_obs} 

The main point of the present study is to investigate the most luminous obscured ultra-hard X-ray selected AGN in the local Universe in terms of their basic SMBH properties, including \mbh, \lamEdd, and multi-wavelength classification. 
While the BASS project is continuously gathering optical spectra, and thus reliable determinations of these properties for an ever-growing sample of local AGN, at the time of publication BASS/DR1 held such measurements for only a subset of our sample (10 of 28 sources).
We thus initiated a dedicated observational campaign to obtain reliable determinations of \mbh\ and \lamEdd\ for the remaining sources. 
Below we briefly describe these new observations and the related data reduction. 
Additional details on the observations are given in Table~\ref{tab:spec_obs}. 

Thirteen objects were observed with the VLT/\XS\ \citep{2011A&A...536A.105V} during 2017, as part of the upcoming BASS/DR2, through ESO programmes 098.A-0635 and 099.A-0403 (PI Kyuseok Oh). 
The \XS\ instrument uses three spectrograph arms, UVB, VIS, and NIR, covering approximately $3000-5600$ \AA, $5500-10,200$ \AA, and $10,200-24,800$ \AA, respectively. 
We observed our \XS\ targets through several cycles of ``AB/BA'' dithering patterns. 
In each dithering position (``A'' or ``B'') we had two consecutive exposures, each lasting 120s for the UVB and NIR arms, and 109s for VIS. 
These ``AB/BA'' dithering cycles were repeated for a number of times, according to the source brightness. 
Most typically, we repeated the ``AB/BA'' cycle twice, resulting in 960s of exposure in the UVB and NIR arms, and 872s in VIS.
The specific exposure times per each source are given in Table~\ref{tab:spec_obs}.
We used slits with widths of 1.6, 1.5, and 0.9\arcsec, resulting in spectral resolving powers of $R\simeq3300$, 5400, and 3890, (in the three arms, respectively).
The \XS\ data were reduced using the standard \textsc{reflex} pipeline \cite[v2.4.0;][]{2013A&A...559A..96F}.

\begin{table}
\caption{
Summary of new spectroscopic observations}
\begin{tabular}{lrlc}
\hline
\hline
{Observatory/} 		& {BAT} 	& {observation} & {total exposure}  \\
{instrument} & {ID}$^{a}$& {date [UT]} 	& {time [sec]$^{b}$}      \\
\hline
{VLT/\XS$^c$}	 & {   20} & {	02-Dec-2016	}  & (960, 872, 960) \\
{} 				 & {   32} & {	26-Nov-2016	}  & (960, 872, 960) \\
{} 				 & {  179} & {	21-Jan-2017	}  & (480, 436, 480) \\
{} 				 & {  200} & {	28-Jan-2017	}  & (480, 436, 480) \\
{} 				 & {  203} & {	09-Jan-2017	}  & (960, 872, 960) \\
{} 				 & {  209} & {	02-Dec-2016	}  & (960, 872, 960) \\
{} 				 & {  360} & {	02-Dec-2016	}  & (960, 872, 960) \\
{} 				 & {  442} & {	01-Feb-2017	}  & (480, 436, 480) \\
{} 				 & { 1072} & {	21-Mar-2017	}  & (480, 436, 480) \\
{}				 & { 1210} & {	01-Feb-2017	}  & (480, 436, 480) \\
\hline
{			} & {   57} & {	30-Sep-2017	}  & (960, 872, 960) \\
{			} & { 1204} & {	24-Jun-2017	}  & (1920, 1744, 1920)	\\
\hline
\hline
{Palomar/DBSP} & {   57} & { 02-Oct-2017	} & {1000} \\
{			 } & {  149} & { 02-Oct-2017	} & {1000} \\
{			 } & {  986} & { 31-Aug-2017	} & { 600} \\
{			 } & { 1051} & { 02-Oct-2017	} & {1000} \\
\hline
\hline
{Keck/LRIS} & {  353} & { 06-Mar-2018	} & {1000} \\
\hline
\end{tabular}
\flushleft 
\footnotesize{
Notes:\\ 
$^{a}$ See Table~\ref{tab:basic_info} for source names and basic information.\\
$^{b}$ For VLT/\XS\ observations, triplets of exposure times denote the total exposure times in the (UVB,VIS,NIR) arms.\\
$^{c}$ For VLT/\XS\ observations, we list separately the observations conducted as part of Periods 98 and 99 (Programmes 098.A-0635 and 099.A-0403, respectively).
}
\label{tab:spec_obs} 
\end{table}

Five additional objects were observed in August 2017 with the Double Spectrograph (DBSP) on the 200-inch Hale telescope at the Palomar observatory. 
These observations were part of the {\it NuSTAR} BAT snapshot programme, focusing mostly on (lower luminosity) type-2 AGN (PIs F.\ Harrison and D.\ Stern), or as part of a Yale follow-up program of BAT AGN towards BASS/DR2 (PI M.\ Powell). 
The Palomar observations were taken with the D55 dichroic and the 600/4000 and 316/7500 gratings using a 1.5\arcsec\ slit, providing spectral resolutions of 4.1 \AA\ and 6.0 \AA, respectively, and covering the wavelength range of 3700-10200 \AA. 
For wavelength and flux calibrations we used standard stars, which were observed at least once per night.
All newly-observed spectra were processed using standard tasks for fitting sky lines, cosmic ray removal, 1d spectral extraction, and flux calibration. As a result of these observations BASS ID 811 was discovered to have a prominent broad H$\alpha$ line and excluded from the study.

Two additional sources were observed in March 2018 with the Low Resolution Imaging Spectrometer (LRIS) on the Keck-I telescope at the W. M. Keck observatory. 
These data were taken as a part of a Yale-allocated time to observe high-redshift quasars (PI F. Pacucci). The LRIS observations covered the wavelength range 3200-10200 \AA\ and were taken with the 560 dichroic and the 600/4000 and 400/8500 gratings using a 1.0\arcsec\ slit. 
Flux calibration was obtained using standard star spectra, taken at the beginning and at the middle of the observing run. 
The spectra were processed using the standard pipeline for LRIS data provided by the Keck observatory. 
These observations provided a refined measurement of stellar velocity dispersion, for one high-luminosity type-2 AGN (BASS ID 353) and the detection of a prominent broad H$\alpha$ line in BASS ID 303 which was thus excluded from this study.

\subsection{Spectral measurements and \mbh\ estimation}
\label{subsec:ppxf_and_mbh}

\noindent
The continuum and the absorption features of the 1d extracted spectra were fitted using {\tt pPXF}  \citep{2004PASP..116..138C} to measure stellar kinematics and the central stellar velocity dispersion. 
More details regarding the {\tt pPXF} analysis are given in the BASS/DR1 paper \cite[][]{2017ApJ...850...74K}. 
Here we note that the typical measurement errors on \sigs\ are of about $10-20$ and $20-50$ \kms\ for the high- and acceptable-quality spectral fits, respectively (sources flagged as ``2'' and ``3'' in the ``\mbh\ Ref.'' column of Table~\ref{tab:basic_info}).

BH masses (\mbh) were derived in the same way as in BASS/DR1, relying on the measured velocity dispersions of the Ca H,K and Mg~{\textsc{i}} stellar absorption features, and adopting the relation $\log(\mbh/\Msol) = 4.38\times \log(\sigma_*/200\,\kms)+8.49$ given in \citet{2013ARA&A..51..511K}. 
We note that this \mbh\ prescription is practically indistinguishable from the one used in the recently published study of type-2 luminous SDSS AGN, by \cite{KongHo2018_Q2}.
Given the aforementioned uncertainties on \sigs, the uncertainties on our \mbh\ determinations are dominated by systematics.
\citet{2013ARA&A..51..511K} report an  intrinsic scatter of about 0.3 dex about their best-fit relation (that is, the scatter in \mbh\ at a given \sigs; see also, e.g, \citealt{Gultekin2009,McConnell2013_MM} for alternative determinations of the $\mbh-\sigs$ and the associated intrinsic scatter).
\mbh\ estimates like ours implicitly assume that AGN lie on the same $\mbh-\sigs$ relation as {\it in-active} galaxies (see, e.g., \citealt{Grier2013_sigs_PGs,Woo2013_RM_Msig}, but also \citealt{ReinesVolonteri2015_MM_AGN}).
The overall uncertainties in \mbh\ may thus be of order 0.5 dex.
Table \ref{tab:basic_info} presents our sample and some key properties, including luminosities and black hole masses.\\

\begin{table*}
\caption{Basic information about our sample of 28 luminous obscured AGN.} 
\begin{tabular}{rlccccclll}
\hline
{BAT} & {Name} & {$z$ $^{a}$} & {$\log\Lbat$ $^{\dagger}$} & {$\log\Lbol$ $^{b}$} & {$\log\mbh$ $^{\dagger}$} & {\mbh} & {Radio} & {MORX}  & {Optical}\\
{ID~~} & {of optical counterpart} & {} & {(\ergs)} & {(\ergs)} & {(\Msol)} &{ref.$^{c}$} & {morphology$^{d}$} & {class.$^{e}$}  & {image$^{f}$}\\
\hline
\\[-1em]
{  20} & {2MASX~J00343284$-$0424117 }  & {0.213} & {$45.35^{+0.12}_{-0.11}$} & {46.25} & {$ 9.89\pm0.11$} & {3} & {compact}  & {RX  }  &   {SDSS} \\[1ex]
{  32} & {ESP~39607                 }  & {0.201} & {$45.19^{+0.13}_{-0.12}$} & {46.09} & {$10.14\pm0.12$} & {3} & {compact} & {GRX }    & {\ldots} \\[1ex]
{  57} & {3C~033                    }  & {0.060} & {$44.38^{+0.07}_{-0.07}$} & {45.29} & {$ 8.75\pm0.08$} & {2} & {extended, FRII, double} & {AX} & {SDSS/PS}  \\[1ex]
{ 118} & {3C~062                    }  & {0.148} & {$44.92^{+0.16}_{-0.14}$} & {45.83} & {$ 8.43\pm0.24$} & {1} & {extended} & {K2X }   &  {PS} \\[1ex]
{ 149} & {2MASX~J02485937$+$2630391 }  & {0.058} & {$44.45^{+0.07}_{-0.07}$} & {45.35} & {$ 9.10\pm0.26$} & {1} & {compact}  &  {GRX }    &  {PS} \\ [1ex]
{ 179} & {PKS~0326$-$288            }  & {0.109} & {$44.61^{+0.14}_{-0.12}$} & {45.52} & {$ 8.49\pm0.13$} & {2} & {compact}  & {KRX  }   &     {PS}  \\[1ex]
{ 199} & {2MASX~J03561995$-$6251391 }  & {0.108} & {$44.58^{+0.12}_{-0.11}$} & {45.48} & {\ldots}         & { } & {\ldots}   & {ARX }    & {\ldots}  \\[1ex]
{ 200} & {2MASX~J03565655$-$4041453 }  & {0.075} & {$44.44^{+0.10}_{-0.09}$} & {45.34} & {$ 8.54\pm0.03$} & {2} & {\ldots}   & {GX  }    & {\ldots}  \\[1ex]
{ 203} & {SARS~059.28692$-$30.44439 }  & {0.094} & {$44.04^{+0.57}_{-0.25}$} & {44.94} & {$ 8.29\pm0.09$} & {2} & {compact}  & {GR  }    &  {PS}  \\[1ex]
{ 209} & {3C~105                    }  & {0.088} & {$44.74^{+0.08}_{-0.07}$} & {45.64} & {\ldots}         & { } & {extended, FRII, double} & {K2X} & {\ldots} \\[1ex]
{ 227} & {2MASX~J04332716$-$5843346 }  & {0.103} & {$44.48^{+0.17}_{-0.15}$} & {45.38} & {$ 8.50\pm0.31$} & {1} & {\ldots} & {GX  }   &{\ldots}    \\[1ex]
{ 238} & {PKS~0442$-$28             }  & {0.147} & {$45.04^{+0.10}_{-0.09}$} & {45.94} & {\ldots}         & { } & {extended,FRII} & {K2X }   & {PS}  \\[1ex]
{ 249} & {4C~$+$27.14               }  & {0.061} & {$44.36^{+0.09}_{-0.09}$} & {45.26} & {\ldots}         & { } & {compact} & {RX  }   &  {PS}   \\[1ex]
{ 353} & {2MASX~J06591070$+$2401400 } & {0.091} & {$44.41^{+0.19}_{-0.16}$} & {45.31} & {$ 8.30\pm0.11$} & {3} & {double, core-jet} & {GX  }    & {\ldots} \\[1ex]
{ 360} & {PKS~0707$-$35             } & {0.110} & {$44.81^{+0.09}_{-0.09}$} & {45.71} & {$ 8.97\pm0.29$} & {3} & {extended, double $\times$ 2} & {2X} & {\ldots} \\[1ex]
{ 406} & {2MASX~J08045299$-$0108476 } & {0.091} & {$44.46^{+0.19}_{-0.17}$} & {45.36} & {$ 7.47\pm0.20$} & {1} & {compact} & {\ldots}   & {SDSS} \\[1ex]
{ 442} & {2MASX~J09034285$-$7414170 } & {0.091} & {$44.43^{+0.16}_{-0.14}$} & {45.33} & {$ 8.45\pm0.32$} & {2} & {FRII} & {GRX }   & {\ldots} \\[1ex]
{ 555} & {SDSS~J113915.13$+$253557.9} & {0.219} & {$45.11^{+0.13}_{-0.10}$} & {46.01} & {$ 8.87\pm0.29$} & {1} & {compact} & {ARX}   & {SDSS/PS}   \\[1ex]
{ 591} & {B2~1204$+$34              } & {0.079} & {$44.36^{+0.14}_{-0.13}$} & {45.26} & {$ 8.55\pm0.23$} & {1} & {extended} & {KR2X}   & {SDSS/PS} \\[1ex]
{ 648} & {2MASX~J13000533$+$1632151 } & {0.080} & {$44.36^{+0.16}_{-0.14}$} & {45.27} & {$ 9.19\pm0.23$} & {1} & {compact} & {ARX }    &{SDSS/PS} \\[1ex]
{ 714} & {IGR~J14175$-$4641         } & {0.077} & {$44.51^{+0.10}_{-0.09}$} & {45.42} & {$ 8.80\pm0.27$} & {1} & {\ldots} & {KX  } & {\ldots}     \\[1ex]
{ 792} & {2MASX~J16052330$-$7253565 } & {0.090} & {$44.72^{+0.07}_{-0.07}$} & {45.62} & {$ 7.85\pm0.15$} & {2} & {\ldots} & {NRX }    & {\ldots}  \\[1ex]
{ 842} & {2MASX~J16531506$+$2349431 } & {0.104} & {$44.50^{+0.16}_{-0.14}$} & {45.40} & {$ 8.23\pm0.25$} & {1} & {compact} & {KRX}    &{SDSS/PS} \\[1ex]
{ 968} & {2MASX~J18212680$+$5955209 } & {0.099} & {$44.56^{+0.16}_{-0.14}$} & {45.47} & {\ldots}         & { } & {compact} & {X   }    & {PS}  \\[1ex]
{1051} & {3C~403                    } & {0.058} & {$44.46^{+0.07}_{-0.07}$} & {45.36} & {$ 9.15\pm0.23$} & {1} & {extended, FRII, double $\times$2} & {KX/2} & {PS} \\[1ex]
{1072} & {PKS~2014$-$55             } & {0.061} & {$44.46^{+0.07}_{-0.07}$} & {45.37} & {$ 9.18\pm0.09$} & {2} & {\ldots} & {KRX }   &{\ldots}   \\[1ex]
{1204} & {2MASX~J23444387$-$4243124 } & {0.597} & {$46.28^{+0.17}_{-0.15}$} & {47.19} & {$10.28\pm0.15$} & {3} & {\ldots} & {KRX }   &{\ldots}  \\[1ex]
{1210} & {PKS~2356$-$61             } & {0.096} & {$44.53^{+0.12}_{-0.11}$} & {45.43} & {$ 8.96\pm0.11$} & {2} & {\ldots} & {K2X}  &{\ldots}   \\[1ex]

\hline

\end{tabular}
\flushleft 
\footnotesize{
Notes:\\ 
$^{a}$ \OIII-based redshifts, mostly drawn from BASS/DR1 (rounded to 3rd decimal digit).\\
$^{b}$ Ultra-hard X-ray based bolometric luminosity, assuming $\Lbol = 8 \times \Lbat$.\\
$^{c}$ Reference for \mbh\ estimates: 
``1'' - measurements from BASS/DR1, \citealt{2017ApJ...850...74K}; 
``2'' - measurements from \XS\ and/or Palomar, good quality spectral fit; 
``3'' - measurements from \XS\ and/or Palomar with lower quality.\\
$^{d}$ Based on visual inspection of the NVSS radio emission contours.\\
$^{e}$ The MORX catalogue classification: ``G'' - galaxy; ``R'' - radio association; ``2'' - double radio lobes; ``X'' - X-ray association; ``K'' - type II object or AGN of unclear type; ``L'' - LINER.\\
$^{f}$ Source for the optical images: SDSS or PanSTARRS (``PS'').\\
$^{\dagger}$ Measurement errors are tabulated. See text for discussion of systematic uncertainties. 
} 
\label{tab:basic_info} 
\end{table*}

\section{Analysis -- in Search for Commonalities in Multi-Wavelength Data}
\label{sec:analysis}

\subsection{Multi-wavelength AGN selection criteria}
\label{subsec:mw_AGN_select}

The rich collection of ancillary multi-wavelength data available through the BASS project allows us to test the effectiveness of several widely used AGN selection criteria.
As our sample represents the most luminous AGN in the low-redshift Universe, the basic expectation is that essentially all of their emission would be dominated by AGN-related processes, and thus that they will all be classified as AGN when considering non-X-ray AGN  selection criteria.

\subsubsection{Optical emission line properties}
\label{subsec:emlines}

We rely on commonly-used strong emission line ratio diagnostic diagrams (so-called ``BPT diagrams'', following \citealt{1981PASP...93....5B}; see also  \citealt{1987ApJS...63..295V,2001ApJ...556..121K,2003MNRAS.341...33K,2007MNRAS.382.1415S} to test whether our sample of luminous type-2 AGN agrees with standard optical classification schemes.
We specifically focus on the \NIIopt/\halpha\ vs. \OIII/\hbeta\ line ratio diagnostics.
To obtain line flux measurements for our sources, we relied on BASS/DR1 and the newly observed spectra (part of the upcoming BASS/DR2), for which we followed an identical line-fitting scheme. 

\begin{figure}
\begin{center}
\includegraphics[width=0.49\textwidth]
{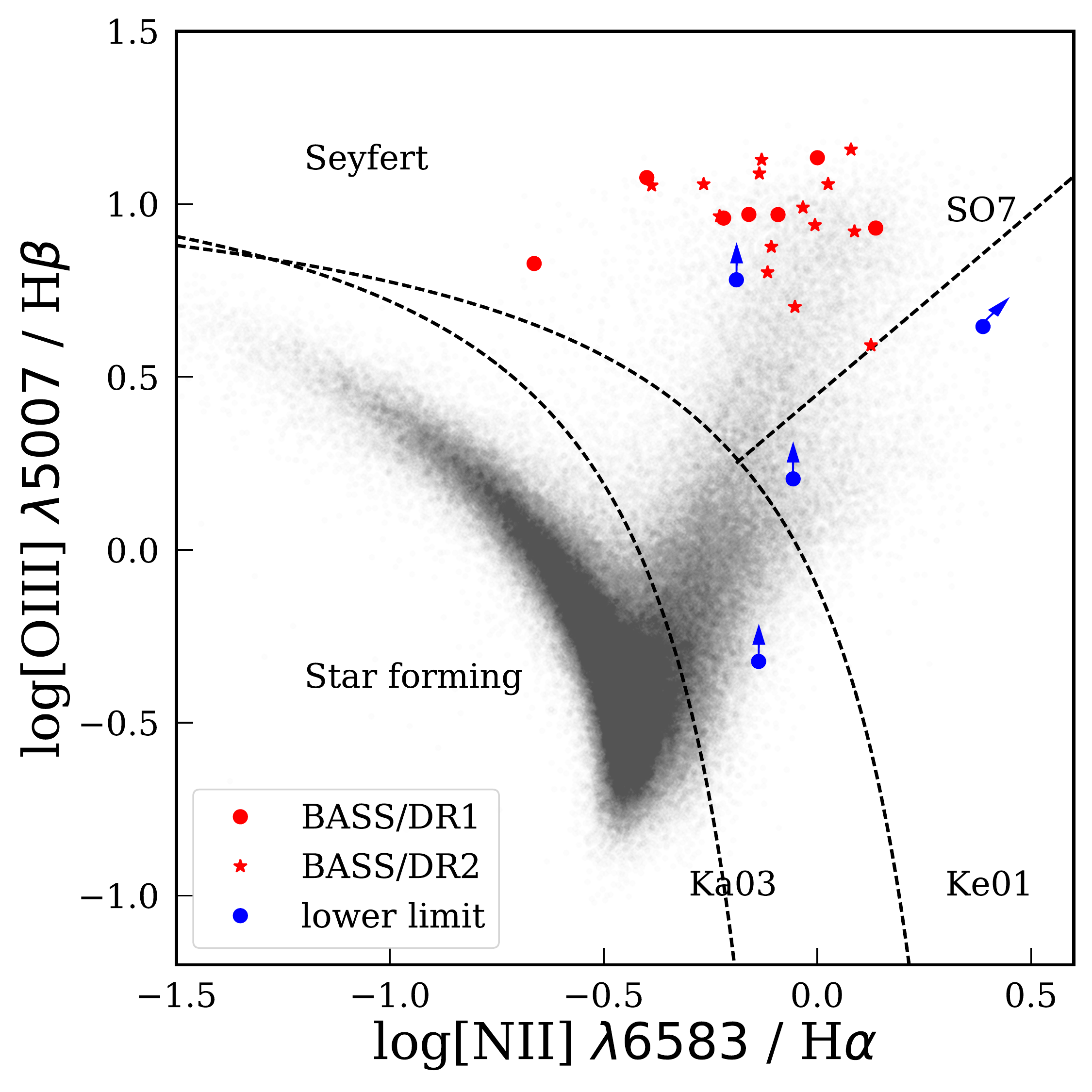}
\caption{Strong line ratio diagnostics (BPT) diagram for 25 sources of our sample.
The different symbols mark sources with spectral fitting performed either as part of BASS/DR1 or DR2, and either good or acceptable spectral fitting quality.
The dashed lines denote commonly-used demarcations between star-forming galaxies, Seyferts and LINERs, taken from 
\citet[``Ke01'']{Kewley2001_BPT},
\citet[``Ka03'']{2003MNRAS.341...33K}, 
and 
\citet[``S07'']{2007MNRAS.382.1415S}.
For comparison, the grey shaded area represents the total SDSS population for $0 < z < 0.1$. 
The vast majority of our sources are classified here as type-2 AGN (Seyferts), with relatively strong \oiii/\hb\ line ratios.
}
\label{fig:BPT}
\end{center}
\end{figure}

Figure~\ref{fig:BPT} shows the distribution of our sources in the \oiii/\hbeta\, vs.\, \nii/\halpha\  plane. 
We also show the classification criteria separating SF galaxies, AGN, and LINERs, following \citet{2001ApJ...556..121K}, \citet{2003MNRAS.341...33K} and \citet{2007MNRAS.382.1415S}. 
For context, the dense grey points represent the total SDSS galaxy population at $z < 0.1$ \citep{2009ApJS..182..543A,2011ApJS..195...13O}.
Of the 28 sources in our sample, 21 have robust measurements of all four emission lines, and are located well within the Seyfert region (see red symbols in Fig.~\ref{fig:BPT}).
For four additional sources we could only determine upper limits on either one, or both of the Balmer lines. Since the forbidden lines are robustly detected these four sources can still be placed on Fig.~\ref{fig:BPT}, as lower limits in either one or both axes (blue arrows). 
BAT ID 209 lacks both \ha\ and \hb\ measurements, and the lower limits place it within the LINER region, although it may still be consistent with a Seyfert classification.
BAT IDs 227 and 353 lack \hb\ measurements, and the lower limits place them in the LINER and Composite regions, respectively, although they may still be consistent with a Seyfert classification.
BAT ID 792 lacks an \hb\ measurement, but is found well within the Seyfert region.
Another source, BAT ID 118, has an optical spectrum (in BASS/DR1) that covers only the \ha+\nii\ spectral complex, and so cannot be shown in Fig.~\ref{fig:BPT}. 
However, since it has $\log(\nii/\ha) = 0.73$, it is most likely located in the Seyfert or LINER regions.
One other source, BAT ID 406, has only an \oiii\ measurement, and thus cannot be classified in terms of Fig.~\ref{fig:BPT}.
Finally, BAT ID 714 has an inadequate spectral fitting quality. 

We therefore conclude that the vast majority of our sources (25/28; 89\%) are clearly classified as AGN. 
Thus, our luminous narrow-line AGN, originally selected through their ultra-hard X-ray emission, would also have been classified as AGN based on their narrow, optical emission line ratios.
Moreover, most of our sources are located in the upper region of the BPT (\oiii/\hbeta\ vs. \nii/\ha) diagram, in the $\log(\oiii/\hb)\gtrsim0.8$ regime. 
This is in agreement with previous studies that have suggested that higher-luminosity AGN have higher \oiii/\hb\ line ratios \cite[e.g.,][]{Stern2013_BPT,2017MNRAS.464.1466O}.

\begin{figure}
\begin{center}
\includegraphics[width=0.49\textwidth]
{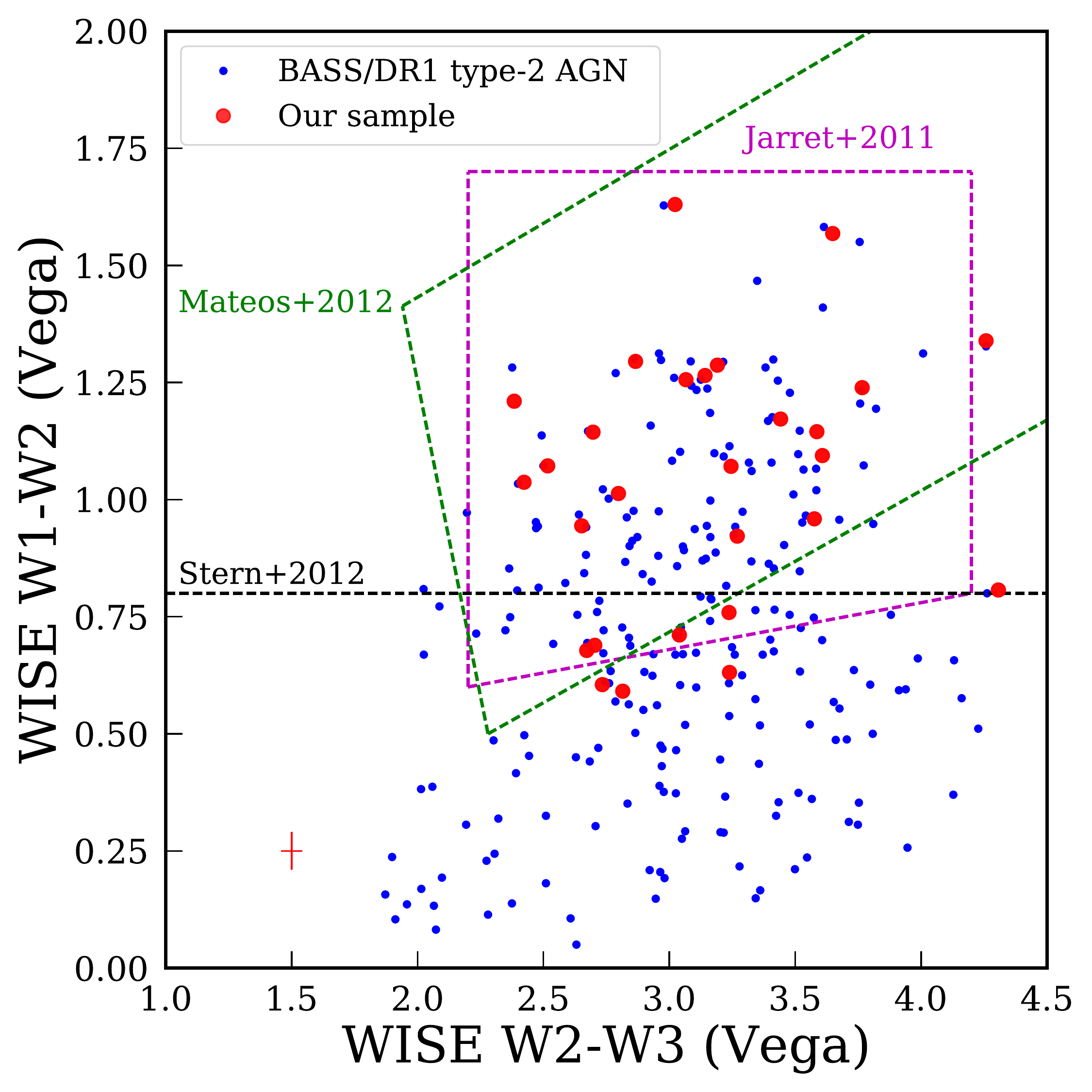}
\caption{Mid-infrared colour AGN selection criteria for our sample of luminous, obscured low-redshift AGN.
We show the $W1-W2$ ($3.4\mu m-4.6\mu m$) vs. $W2-W3$ ($4.6\mu m-12\mu m$) colour-colour plane, with our sample (large red symbols) compared with the general BASS/DR1 population of type-2 AGN (blue symbols). 
The error bars at the bottom-left corner illustrate the {\it maximal} uncertainties for our our sources.
All the data are based on the public all-sky survey carried out with {\it WISE}.
We also over-plot three commonly-used AGN selection criteria: the ``wedge'' by \citet{2011ApJ...735..112J} and the simple $W1-W2>0.8$ colour cut by \citet[][]{2012ApJ...753...30S}, as well as the selection region defined in \citet[][]{Mateos_2012}.
The vast majority of our sources (80\%) are classified as AGN by all criteria, similarly to previous studies based on these selection criteria \citep[e.g.,][]{2015ApJS..221...12S}.
} 
\label{fig:MIR_colors}
\end{center}
\end{figure}

\subsubsection{Mid-infrared colours}
\label{subsec:infraredl}

We next investigate whether our sources agree with commonly used mid-infrared (MIR) colour selection criteria for (luminous) AGN, which are driven by the radiation reprocessed by the dusty, obscuring toroidal circumnuclear structure \cite[see, e.g.,][]{2012ApJ...753...30S}.
Specifically, we use data available for all of our sources in the 3.4 $\mu$m, 4.6 $\mu$m, 12 $\mu$m and 22 $\mu$m wavelength bands ($W1-4$ hereafter), from the {\it Wide-field Infrared Survey Explorer} ({\it WISE}) all-sky survey \citep{Wright2010_WISE}.
In the {\it WISE} colours we use, the photometry of extended sources from the scaled 2MASS aperture photometry which introduces typical uncertainties of order 0.2-0.5 mag, depending on how well resolved the source is in different bands \citep{2012AJ....144...68J}.

In Fig.~\ref{fig:MIR_colors} we show the $W1-W2$ vs.\ $W2-W3$ colour-colour diagnostic diagram, with several different sets of AGN selection criteria. 

In magenta we show the AGN selection ``wedge'' proposed by \cite{2011ApJ...735..112J}, in black - the simpler $W1-W2 > 0.8$ AGN cut proposed by \cite{2012ApJ...753...30S}, whereas the green  presents another selection wedge, originally proposed by \cite{Mateos_2012} and further developed by \cite[][based on AGN drawn from the \swift/BAT 70 month catalogue]{2017ApJ...835...74I}.

As discussed in detail in the respective studies, the robustness of these selection criteria strongly depends on the AGN-related luminosity of the source. The MIR emission from lower luminosity sources could be affected by contamination from the host galaxies; therefore the AGN detection rate increases drastically for the higher luminosities. 
We would thus expect that our high-luminosity sources would show better agreement with the MIR selection criteria than the more general BASS AGN population. 
%
Indeed, about 80\% of our sample are classified as AGN by all criteria shown, while many of the lower luminosity type-2 BASS AGN do {\it not} agree with the selection criteria (i.e., the wedges or the simple $W1-W2$ colour line).

Specifically, out of the 204 lower-luminosity type 2 BASS/DR1 AGN, 135 (66\%) do {\it not} pass the \cite{2012ApJ...753...30S} $W1-W2$ cut.

The fact that some of our objects are not classified as AGN based on their MIR colours is in agreement with the findings of \citet{2015ApJS..221...12S} and the more recent analysis of the BASS/DR1 IR SEDs, by \citet{2017ApJ...835...74I}.

\subsection{Host galaxy morphologies and masses}
\label{subsec:Hosts}

\begin{figure*}
\begin{center}
\includegraphics[width=0.49\textwidth]{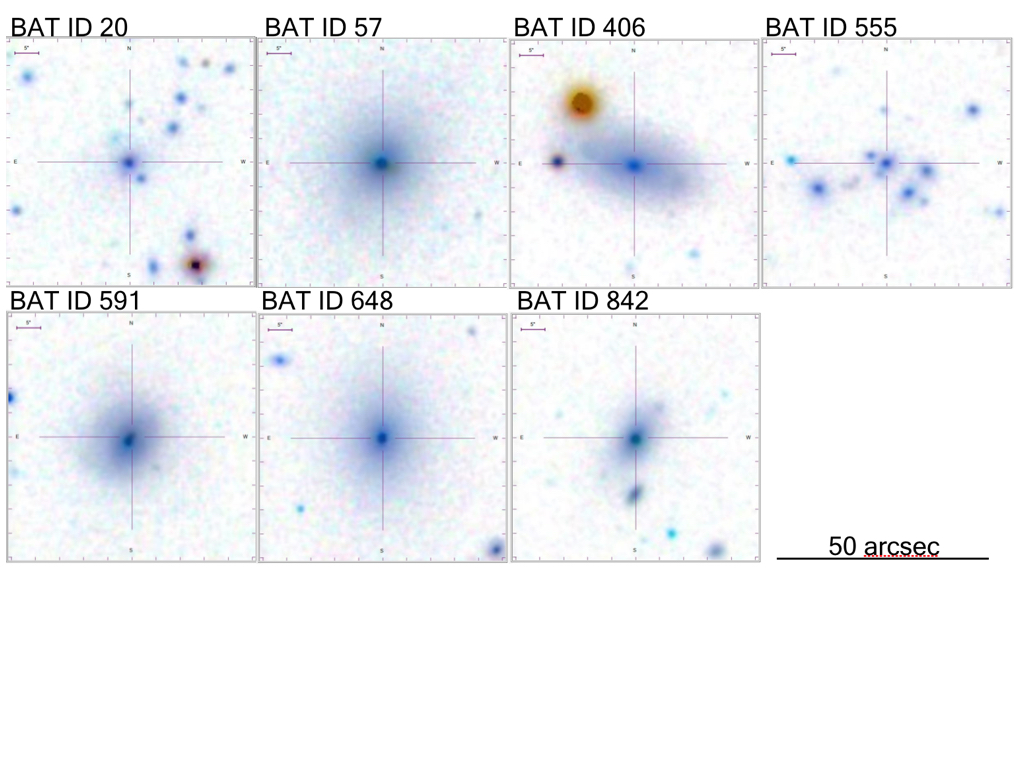}\hfill
\includegraphics[width=0.49\textwidth]{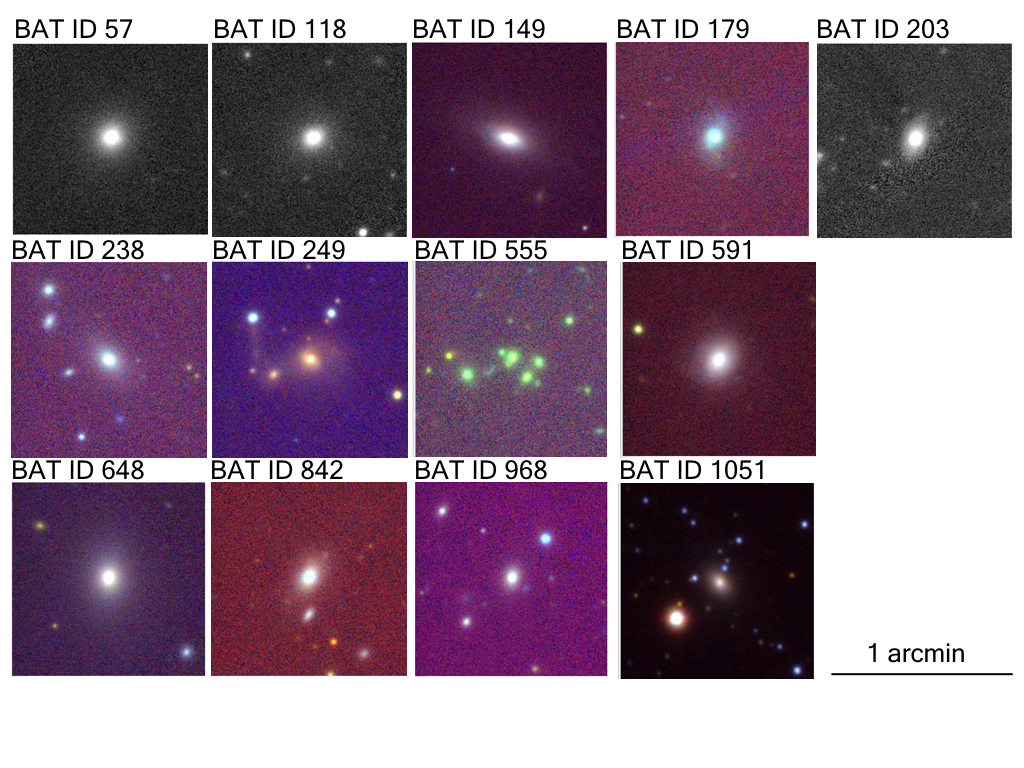}
\caption{
The host galaxies of the AGN in our sample.
{\it Left:} reverse $gri$ composite images of the host galaxies for seven of our sources, available from the SDSS. The images size is $50\arcsec \times 50\arcsec$. 
{\it Right:} $giy$ composite images of the host galaxies for 15 of our sources, available from PanSTARRS. The images size is $1\arcmin \times 1\arcmin$.  
Essentially all the AGN hosts in the two compilations are ellipticals, or clearly dominated by a bulge component.
} 
\label{fig:host_images}
\end{center}
\end{figure*}

We next investigate whether the host galaxies of our sample of luminous obscured AGN appear to have a common morphology.
For 15 of our 28 sources we have host galaxy images from the public databases of the Sloan Digital Sky Survey \cite[SDSS,][]{2018ApJS..235...42A} and/or the Panoramic Survey Telescope and Rapid Response System (PanSTARRS, \citealt{2016arXiv161205243F}).
Four sources have images from both surveys.
These images are shown in Fig.~\ref{fig:host_images}. 
Essentially all the hosts for which optical imaging is available appear to have elliptical morphologies. 
Although a few of them may arguably have some spiral-like features (e.g., BAT IDs 179, 249, and 406), the morphologies of these, too, are clearly dominated by a prominent (elliptical) bulge component.

Only three objects of our sample are classified in one of the Galaxy Zoo (GZ) catalogues \citep{2009MNRAS.396..818S,2011MNRAS.410..166L,2013MNRAS.435.2835W,2016MNRAS.461.3663H}, with some variation in classifications from one catalogue to another. 
The small number of GZ classification is not unexpected, given the number of sources with SDSS images and the fact that the GZ work is limited to $z<0.2$.
BAT ID 555 is identified as an elliptical galaxy in \citet{2011MNRAS.410..166L} and \citet{2013MNRAS.435.2835W}, whereas \citet{2016MNRAS.461.3663H} classifies it as having a spiral structure. 
BAT 648 appears only in \citet{2011MNRAS.410..166L} and is classified as elliptical. 
BAT 842 is listed in three catalogues, with \citet{2013MNRAS.435.2835W} and \citet{2016MNRAS.461.3663H} listing it as having an elliptical morphology while the morphology given in \citet{2011MNRAS.410..166L} is uncertain.

Given the redshift range of our sources, and the depth of the SDSS imaging data upon which the GZ classifications are based, it is not impossible that even those sources that are classified as ellipticals in fact harbour (faint) disc-like or spiral structures.

A detailed (parametric) morphological analysis of the hosts of our AGN would require deeper, higher-quality, multi-band imaging data. This, as well as a more elaborate investigation of the differences between the classifications in the various GZ catalogues and biases with redshift,
are well beyond the scope of the present study, which focuses on identifying the key common (or distinguishing) properties of luminous obscured AGN in the low-redshift Universe.

We briefly note here that the dominance of elliptical (and/or bulge-dominated) morphologies among the hosts of luminous, obscured AGN, was also reported by previous studies \cite[e.g.,][]{2006AJ....132.1496Z}. 
We further compare our findings to other studies below.

In order to determine whether the dominance of elliptical morphologies should be expected, we turn to assess the host galaxies stellar masses, \mstar, as galaxy mass is known to be closely related to morphology \cite[see, e.g.,][and references therein]{2017MNRAS.467.3934D}.
We follow two approaches to compare the morphologies of our AGN with the general population galaxies with comparable masses, as follows.

\begin{figure*}
\centering
\begin{tabular}{ccc}
\includegraphics[width=0.32\textwidth]
{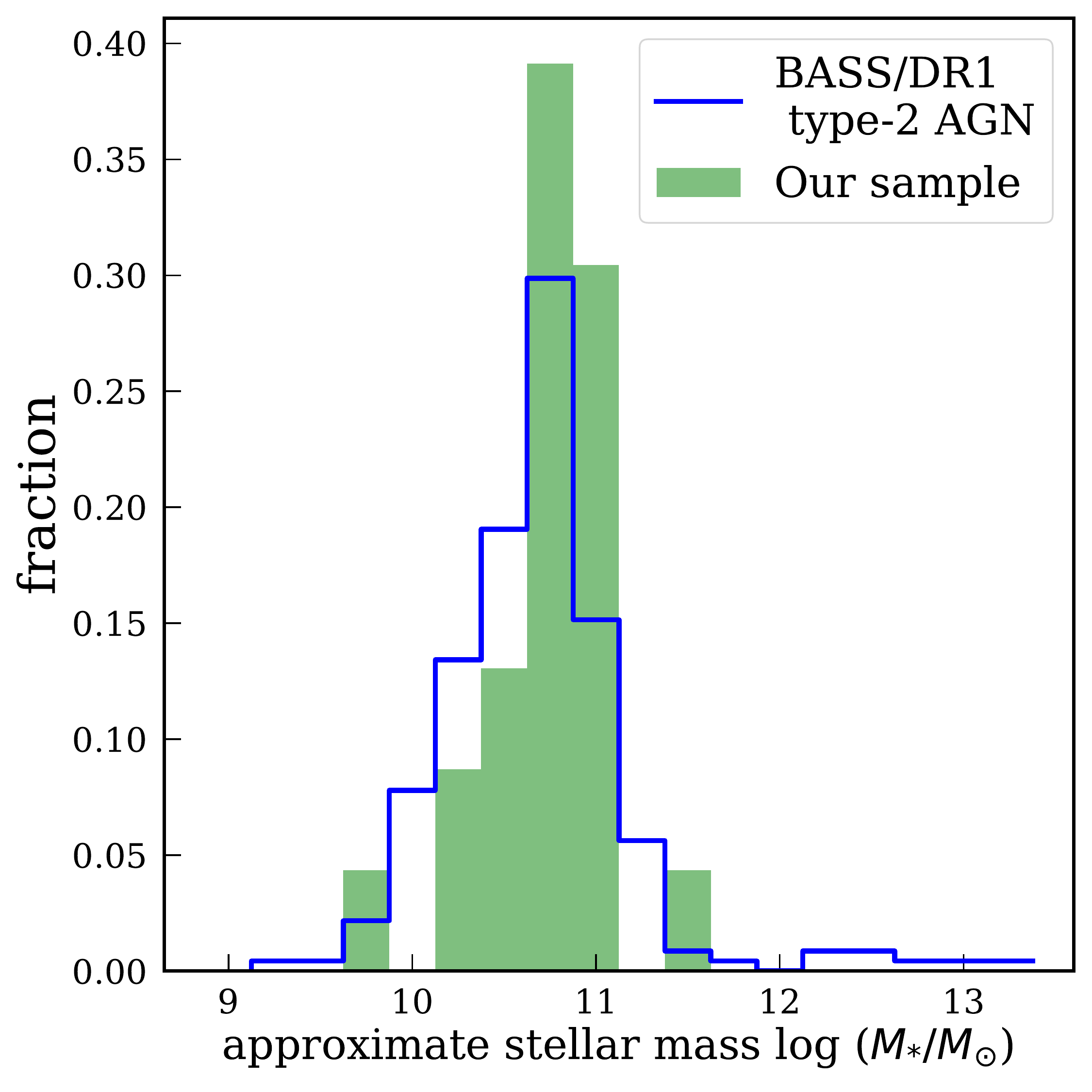}
\includegraphics[width=0.32\textwidth]
{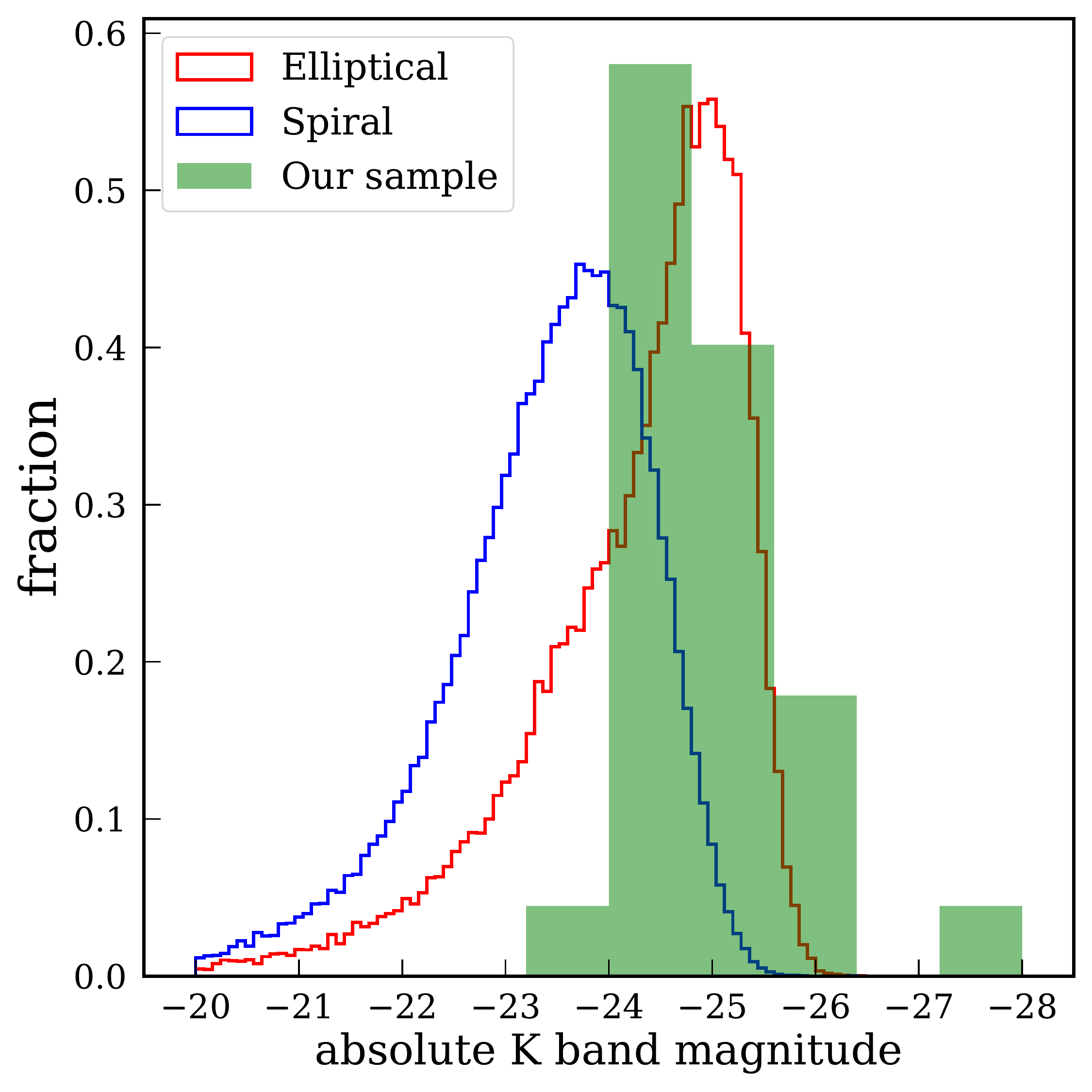}
\includegraphics[width=0.32\textwidth]
{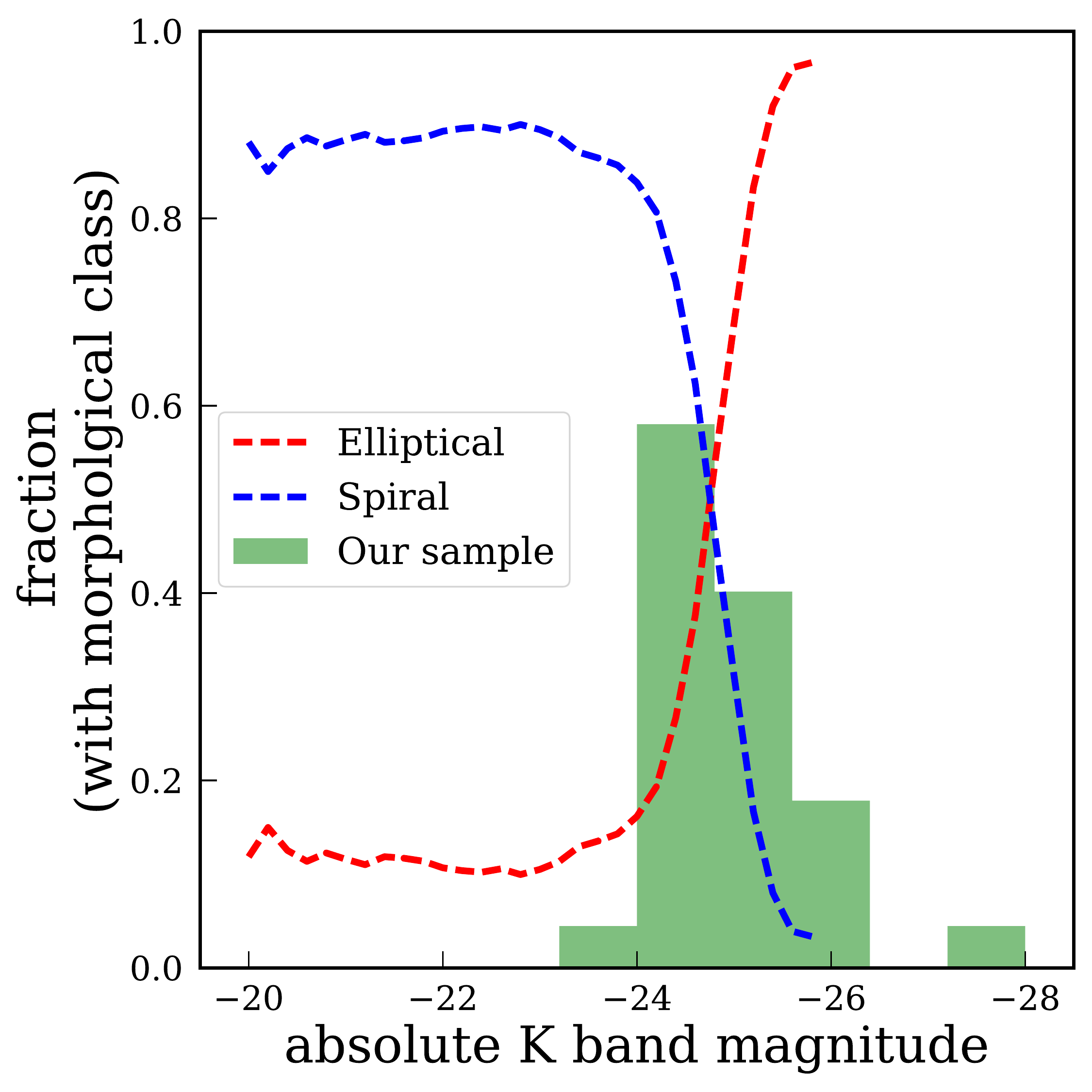}
\end{tabular}
\caption{
Stellar masses, absolute $K$-band magnitudes, and the expected morphological classification of the AGN hosts in our sample.
The left panel shows the distribution of stellar masses, \mstar, among the hosts of our luminous AGN sample (green bars), compared with that of the entire BASS/DR1 type-2 AGN population (blue). 
The host stellar masses of our luminous AGN extend over the entire range of BASS/DR1 type-2 AGN host masses, but are skewed towards the high-mass end.
{\it Centre:} the distribution of $K$-band absolute magnitudes ($M_K$) for the host galaxies of our AGN, compared to those of ellipticals and spirals drawn from a large SDSS-based sample  (see text for details).
{\it Right:} the fraction of spiral and elliptical galaxies, among galaxies in any of these two classes, as a function of $M_K$ (i.e., the two fractions always add up to 100 per-cent). 
The green bars illustrate again the $M_K$ distribution among our sources.
We would have expected to find {\it some} spirals among the hosts of our high-luminosity AGN; however, this is not seen in the available optical images (Fig.~ \ref{fig:host_images}).
} 
\label{fig:morphology}
\end{figure*}

We first use the host galaxy stellar masses derived for BASS AGN, as described in \citet{2018ApJ...858..110P}.
These were derived through careful, aperture-matched NIR/MIR multi-band photometry to all BASS AGN, 
modelled with the galaxy and AGN SED templates of \cite{Assef2010_SED_templates}.
The left panel of Fig.~\ref{fig:morphology} compares the \mstar\ distribution for our luminous type-2 AGN to that of the general type-2 AGN population in BASS/DR1.
The stellar masses of our sample extend over a wide range, $9.8 < \log(\mstar/\Msol) < 11.7$, which is however somewhat more concentrated towards the high-mass regime of the broader mass range of BASS type-2 AGN.
The median masses are indeed very similar, with $\log(\mstar/\Msol)=10.86$ for our sample vs. 10.79 for the general BASS type-2 AGN population. 
A formal Kolmogorov-Smirnov (KS) test confirms that the distributions of stellar masses of the two samples (BASS/DR1 type-2 AGN and our sample of luminous AGN) are indistinguishable ($P=0.15$).

Second, we use absolute $K$-band magnitudes ($M_K$) as a proxy of total galaxy stellar mass \cite[e.g.,][]{2003ApJS..149..289B,2013ApJ...764..151G,2016ASSL..418..431K}. 
The centre panel of Fig.~\ref{fig:morphology} shows the distribution of $M_K$ among our sample of luminous obscured AGN. 
For comparison, we use the distribution of $M_K$ for a large sample of galaxies with morphological classification, split into ``ellipticals'' and ``spirals'', drawn fro the SDSS.
This comparison sample is constructed from SDSS/DR7 \citep{2009ApJS..182..543A} through cross-matching with the New York Value-Added Galaxy Catalog (NYU VAGC; \citealt{2005AJ....129.2562B, 2008ApJS..175..297A,2008ApJ...674.1217P}), and 2MASS \citep{2006AJ....131.1163S}, using a 1\arcsec\ separation. 
We used morphological classifications made available through the Galaxy Zoo 1 data release \cite[GZ1;][]{2011MNRAS.410..166L}, focusing on galaxies with a debiased vote fraction that exceeds a threshold of 0.8. 
This provides a comparison sample of 197,551 galaxies with a robust morphological classification, of which 151,163 are classified as ``spirals'' and 46,388 as ``ellipticals''.
The normalized distributions of $M_K$ for this SDSS-2MASS-GZ1 based comparison sample are shown in centre panel of Fig.~\ref{fig:morphology}, while the right panel shows the fraction of galaxies of each class as a function of $M_K$ (among galaxies with a robust morphological classification).
For the most luminous, most massive galaxies (lowest magnitudes, $M_K\lesssim-26$) we can expect a very high fraction of ellipticals; however for galaxies with $-25 < M_K < -23$ we should expect a non-negligible fraction of spirals, which increases towards lower luminosities (and masses).
Indeed, $\sim$30 per-cent of SDSS-2MASS-GZ1 spirals have $M_K < -24$ (compared with $\sim$68 per-cent of ellipticals).

A straightforward KS-test to compare the $M_K$ distribution of our sample with the SDSS-2MASS-GZ1 comparison samples indicates that they differ, with great statistical significance ($P \approx 10^{-11} $), from the spiral galaxies. 
Conversely, the $M_K$ distribution of our sample does {\it not} differ, statistically, from that of the SDSS-2MASS-GZ1 ellipticals ($P=0.07$). 
These simple tests thus suggest that the tendency of our sample towards elliptical morphologies may be driven by higher luminosities and masses. 

However, the right panel of Fig.~\ref{fig:morphology} clearly shows that our sample covers the range where one would expect a significant fraction of spirals, in fact covering the region where spirals are just as common as ellipticals (i.e., the fraction of spirals is $\sim50$ per-cent in the SDSS-2MASS-GZ1 reference sample).

As our sample covers the host luminosity range mentioned above, we further quantified the expected fraction of ellipticals using the aforementioned SDSS-2MASS-GZ1 cross-matched sample.
For each of our luminous AGN, we constructed a corresponding comparison sample of galaxies with similar $M_K$ and redshift, defined to lie within $\Delta M_K = \pm0.5$ and $\Delta z=\pm0.0005$.
In three cases we had to slightly adjust these ranges in order to include at least 50 objects in each of our test samples. 
We did not include galaxies that lack a consensus morphological classification in GZ1.
We also note that, since the large cross-matched SDSS-2MASS-GZ1 galaxy sample is restricted to $z<0.2$, we could not construct such per-source comparison samples for four of our objects.
The detailed list of elliptical-to-total fractions found for the control samples matched to our AGN is given in Table~\ref{tab:morph_test}.
While the full range of elliptical fractions is broad, $6-97\%$, the typical fractions are broadly in good agreement with the overall distributions of high-luminosity galaxies: the median (average) elliptical fraction among the matched control samples is 69\% (63\%, respectively), compared to the essentially 100\% ellipticals among our luminous AGN. 
Moreover, two thirds of the matched control samples have a majority of ellipticals (i.e., for 18 of 24 AGN have $P(E)>50\%$), while one third have a majority of spirals.

We finally note that those AGN among our sample that lack optical host images, and thus host classifications, do not bias our tests: their $K$-bad luminosities, of $-25.9 \leq  M_K \leq -24.4$, cover the core of the $M_K$ distribution of the entire sample (and included in the histograms shown in Fig.~\ref{fig:morphology}). 
They also do not show particularly high or low $P(E)$ in Table~\ref{tab:morph_test}, thus indicating that these are statistically expected to be mostly, but not solely, ellipticals.

In summary, all our tests indicate that one should have expected to see {\it some} spirals among the hosts of our AGN (i.e., roughly one third of sources), even when considering their rather high ($K-$band) luminosities and/or stellar masses.
We thus conclude that the dominance of elliptical (or bulge-dominated) galaxies among the hosts of our luminous AGN is unlikely to be solely driven by their (high) luminosities and/or stellar masses.
We caution, however, that the modest size of our sample, and the type of imaging data used here, limit our ability to draw stronger conclusions regarding the host galaxies.

Several previous studies have found that (optically selected) AGN in the local Universe generally tend to be in spiral host galaxies \cite[e.g.,][]{2003AJ....126.1750M,2009MNRAS.400.1803W,2017MNRAS.466.4917D}. 
Moreover, \citet{2010ApJ...716L.125K} show that a high fraction of \swift/BAT AGN are found in galaxy mergers, and this may be even more pronounced for obscured systems \cite[see also][]{2016ApJ...825...85K,Ricci2017_mergers,2018Natur.563..214K}.
\citet{2011ApJ...739...57K} analysed the host galaxy morphologies of the 185 AGN selected in the shallower, 22-month \swift/BAT ultra hard X-ray all-sky survey, and found that they are predominantly host in massive spirals. 
This difference may possibly be explained by the fact that our sample has much higher luminosities than the typical luminosity of the AGN studied by \cite{2011ApJ...739...57K}.  Specifically, \cite{2011ApJ...739...57K} focused on $z<0.05$ AGN with $\log (L_{\rm BAT}/\ergs) \sim 42-44$, compared with $z>0.05$ and $\log (L_{\rm BAT}/\ergs) \gtrsim 44.5$ for our sample (see Fig.~\ref{fig:Lum_vs_z}.

Thus, previous studies of lower-luminosity BAT AGN further emphasise that high AGN luminosity could be linked to predominantly elliptical (or bulge-dominated) host galaxy morphologies. 
Indeed, the intrinsic AGN luminosities of our sample are even higher those of PG quasars and (FIR-selected) ultra luminous IR galaxies (ULIRGs), which tend to show elliptical morphologies \citep{2009ApJ...701..587V}.

We finally note that our highly luminous AGN appear to span the range of host luminosities, and stellar masses, that are associated with the transition of the galaxy population from (star forming) spirals to (quiescent) ellipticals (see right panel of Fig.~\ref{fig:morphology} and, e.g., \citealt{Bell2003_LF_MF,Baldry2004_galaxy_bimod,Moffett2016_SMF_morph,Weigel2016_SMF}, and references therein).

If confirmed through the analysis of higher-quality, multi-band data,
this may lend some (indirect and non-causal) support to the popular idea that intense SMBH growth is somehow linked to such dramatic galactic transformations \cite[see, e.g.,][for a recent review]{Harrison2017_BH_SF_review}.

\begin{table}
\caption{Host galaxy morphology and comparison samples test.}
\begin{tabular}{|rlc|rlc|}
\hline
{BAT}  & {Morph.} $^{a}$ & {$P(E)$} $^{b}$  & {BAT}  & {Morph.}  $^{a}$ & {$P(E)$} $^{b}$  \\
{ID~~} & {}              & {test}           & {ID~~} & {}               & {test}\\
\hline	
    { 57} & {E} & {64\%} & { 360} & $-$ & {91\%	}\\
	{118} & {E} & {59\%} & { 406} & {E} & {42\%	}\\
	{149} & {E} & {91\%} & { 442} & $-$ & {68\%	}\\
	{179} & {E} & {18\%} & { 591} & {E} & {54\%	}\\
	{199} & {E} & {97\%} & { 648} & {E} & {95\%	}\\
	{200} & $-$ & {71\%} & { 714} & $-$ & {91\%	}\\
	{203} & {E} & {24\%} & { 792} & $-$ & {84\%	}\\
	{209} & $-$ & {21\%} & { 842} & {E} & {17\%	}\\
	{227} & $-$ & {70\%} & { 968} & {E} & {~6\%	}\\
	{238} & {E} & {67\%} & {1051} & {E} & {73\%	}\\
	{249} & {E} & {86\%} & {1072} & $-$ & {80\%	}\\
	{353} & $-$ & {88\%} & {1210} & $-$ & {58\%}\\
\hline 
\end{tabular}
\flushleft 
\footnotesize{
Notes:\\
$^{a}$ Morphology of the host galaxies of our AGN sample, based on SDSS and/or PanSTARRS images, with ``E'' indicating an elliptical.\\
$^{b}$ The fraction of ellipticals in the $M_K-$ and redshift-matched test samples constructed for each object. The median fraction among all objects is 70\%.
} 
\label{tab:morph_test} 
\end{table}

\subsection{Radio properties}
\label{subsec:radio}

We next investigate the radio properties of our luminous obscured AGN. 
From the large number of observable phenomena known in the radio regime \cite[e.g.,][]{2014ARA&A..52..589H}, we focus on simple observed, phenomenological attributes of our luminous obscured AGN: their radio luminosities and related radio loudness; the rough shape of their radio SED; the identification of radio lobes; and the so-called fundamental plane of Black Hole activity.

\subsubsection{Survey data used}

Two commonly used, wide-area, public radio surveys at 1.4 GHz can be considered as data sources for a large all-sky survey like BASS (and thus our BASS-based sample). 
The National Radio Astronomy Observatory (NRAO) Very Large Array (VLA) Sky Survey \cite[NVSS][]{1998AJ....115.1693C}, which reaches flux densities of $S_\nu \approx 2.5$ mJy, and the deeper Faint Images of the Radio Sky at Twenty centimetres survey \cite[FIRST;][]{1995ApJ...450..559B}, which reaches $S_\nu \approx 0.75$ mJy. 
The NVSS has a spatial resolution (i.e., synthesized beam size) of 45\arcsec\ whereas FIRST has a much better resolution of $\sim$5\arcsec. 
However, the higher resolution of FIRST has the disadvantage of potentially underestimating (or, indeed, missing) emission from more extended sources \citep{2005MNRAS.362....9B}. 
On the other hand the, low spatial resolution of the NVSS may lead to misclassification of radio sources as ``compact'' (or, rather, unresolved).
The advantage of NVSS is that the beam is sufficiently large so that, for the vast majority (${\sim}$99\%) of radio sources, the radio emission would be contained within a single beam, which would thus capture all their (spatially) integrated flux, except for a few extremely extended sources \citep{2005MNRAS.362....9B}.

Of our sample of 28 luminous type-2 AGN, 
19 sources are located within the NVSS footprint ($\delta \gtrsim -40\deg$), 
and 14 sources are associated with robustly detected radio sources, identified by cross-matching our sample with the NVSS catalogue through the \texttt{VizieR} service \cite[30\arcsec\ search radius; see][]{1998AJ....115.1693C}.

Of the five remaining sources, three are well known radio sources: BAT IDs 57, 360, and 1051 (3C 033, PKS 0707-35, and 3C 403, respectively; see Table~\ref{tab:basic_info}). 
For these three sources, the NVSS catalogue has separate entries for the radio lobes, and so we instead adopt 1.4 GHz measurements from the literature (\citealt{1992ApJS...79..331W} for BAT IDs 57 and 1051; \citealt{2010A&A...511A..53V} for ID 360).
For the two last sources (BAT IDs 406 and 968), we used the NVSS Flux Server\footnote{\url{https://www.cv.nrao.edu/nvss/NVSSPoint.shtml}} to obtain flux density measurements, by querying the locations of the optical counterparts of the AGN.\footnote{We treat the non-integrated flux densities (i.e., mJy/beam) provided by the NVSS Flux Server as integrated flux densities (i.e., mJy), which is a reasonable choice given the large beam of the NVSS. 
Indeed, for the 14 sources that are found in the NVSS catalogue, the median difference between the types of measurements is -0.03 dex and the standard deviation is $<$0.1 dex.} 
Of the 19 sources within the NVSS footprint, five sources are also detected by FIRST (there are no sources that are detected solely by FIRST).
However, in order to have consistent data for the analysis of our sample, we relied exclusively on the NVSS data (i.e., in cases where both surveys have robust detections). 

We have also constructed a larger sample of {\it all} type-2 AGN in BASS which are associated with catalogued NVSS sources, using again the \texttt{VizieR} service to cross-match with the \cite{1998AJ....115.1693C} catalogue (within 30\arcsec), and yielding 146 type-2, non-galactic-plane, non-beamed AGN.\footnote{For this specific comparison sample, in order to robustly avoid radio-loud (mildly obscured) quasars, we excluded any sources that were reported as ``Sy1-1.9'' in the more recent 105 month \swift/BAT catalogue \cite[][]{2018ApJS..235....4O}.}

We stress that this comparison sample contains, by construction, the 14 aforementioned sources from our sample of high X-ray luminosity AGN that are associated with NVSS sources (and thus 132 other, lower-\Lbat\ type-2 BASS/DR1 sources).
%
In a further step we used the The Million Optical - Radio/X-ray Associations (MORX) Catalog \citep{2016PASA...33...52F}, to obtain classifications of our (radio-detected) AGN, and specifically identification of resolved radio lobes.

\subsubsection{Radio luminosity}
\label{subsec:radio_lum}

The relationship of radio luminosity to X-ray luminosity has developed into an important tool for the analysis of AGN \citep{2013MNRAS.431.2471B,2004A&A...414..895F,2015MNRAS.447.1289P,2016MNRAS.460.1588W}.
We derive monochromatic radio luminosities at rest-frame 1.4 GHz (i.e., $\nu L_\nu [1.4 {\rm GHz}]$, or \Lrad\ hereafter), from the aforementioned NVSS measurements and assuming a spectral index of $\alpha_\nu = -0.7$ (see below).

\begin{figure*}
\begin{center}
\includegraphics[width=0.49\textwidth]
{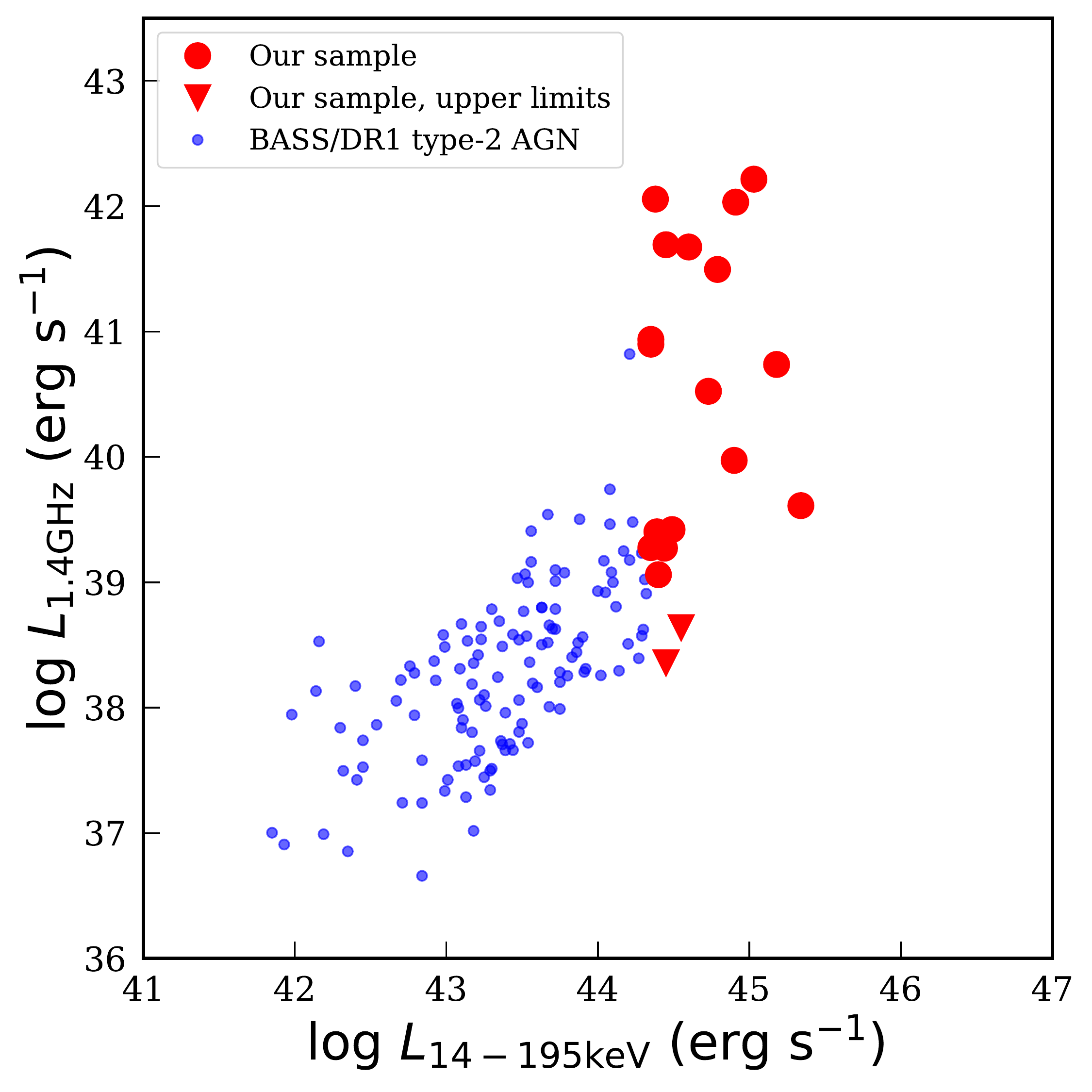}\hfill
\includegraphics[width=0.49\textwidth]
{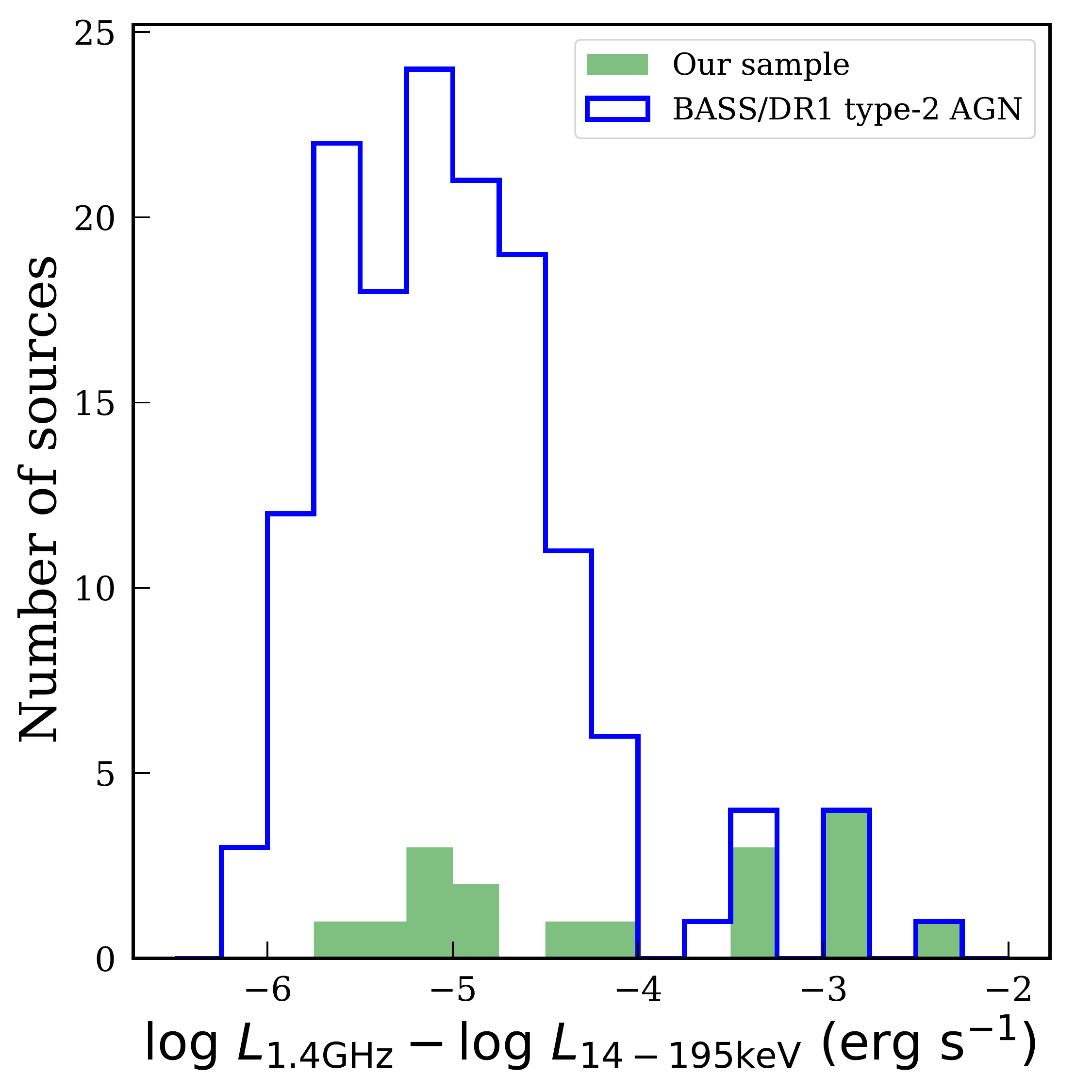}
\caption{Comparing radio and ultra-hard X-ray luminosities for our luminous obscured (type-2) BASS AGN.
{\it Left:} 
the radio luminosity, $\nu L_\nu({\rm 1.4\, GHz})$ plotted against the ultra-hard X-ray luminosity, \Lbat. 
The blue dots represent all type-2 BASS/DR1 AGN for which NVSS data are available, while
the red triangles highlight the 19 sources that belong to our sample of high X-ray luminosity obscured AGN (see text for more details on these samples). 
While our X-ray luminous obscured AGN dominate the high {\it radio} luminosity regime, their radio luminosities in fact show a large scatter, which extends over almost 4 orders of magnitude over a rather limited range in \Lbat.
{\it Right:} distributions of radio luminosity relative to X-ray luminosity, $\log L_{\rm 1.4 GHz} - \log\Lbat$. 
Here the blue line refers to the total population of BASS/DR1 type-2 AGN for which NVSS data are available, while the green bars represent our sample of the highest X-ray luminosity sources.
Our sample has a much larger scatter in relative radio luminosity than what is found for the general type-2 AGN population in BASS ($\sim$1.0 vs. $\sim$0.5-0.6 dex), thus clearly confirming  the apparently large scatter seen in the left panel.
} 
\label{fig:radio_bat}
\end{center}
\end{figure*}

In the left panel of Figure~\ref{fig:radio_bat} we show the \Lrad\ vs.\Lbat\ plane for our 19 high X-ray luminosity sources with relevant radio data (red circles for detections and red triangles for upper limits), and for the larger sample of BASS/DR1 type-2 AGN for which NVSS radio data is available (146 sources, including our 19 AGN; blue points). 
The radio luminosities of our sources cover a huge range, extending over 3.5 orders of magnitude, and clearly extending towards higher radio luminosities over a rather limited range in X-ray luminosities.
To further emphasize this, in the right panel of Fig.~\ref{fig:radio_bat} we show the distributions of the radio luminosities relative to ultra-hard X-ray luminosities, that is $\Delta \log L_{\rm R/X} \equiv \log \Lrad - \log L_{14-195\,\kev}$. 

As the right panel of Fig.~\ref{fig:radio_bat} shows, 
our sample covers the same range in $\Delta \log L_{\rm R/X}$ as does the general population of (NVSS detected) type-2 AGN in BASS, $-6 \lesssim \Delta \log L_{\rm R/X} \lesssim -3$.
However, while the general population is well concentrated around $\Delta \log L_{\rm R/X} \approx -5$, our highly X-ray luminous AGN are much more uniformly distributed, and constitute the vast majority of BASS type-2 AGN at high radio-to-X-ray luminosity ratios (i.e., $\Delta \log L_{\rm R/X} > -4$).
Indeed, the standard deviation of the relative radio luminosity of BASS/DR1 type-2 AGN is $\sigma(\Delta \log L_{\rm R/X}) \simeq 0.6$ dex (or 0.5 dex excluding our 19 AGN), whereas the standard deviation for our sample exceeds 1 dex. 
We conclude that the large range of jet-related radio luminosities seen in our sample exceeds what is expected simply from scaling the accretion-related (ultra-hard) X-ray luminosities.

One possible explanation for the huge range in \Lrad, for a rather limited range in \Lbat, may be given by the diverse morphologies of the radio emission, since extended radio structures do not necessarily trace the concurrent, small-scale physics related to the accretion flow, as the X-ray luminosities do.
The radio morphological classifications of our sources are listed in Table~\ref{tab:basic_info}. 
The classification into ``compact'' and ``extended'' sources is based on visual inspection of the NVSS flux contour maps, while the identification of sources with radio lobes is based on the MORX catalogue \citep{2016PASA...33...52F}. 
Twelve of our sources are compact and six sources are extended; furthermore, seven sources have radio lobes. 
A closer inspection of the ``compact'' and ``extended'' subsets shows these two subsets overlap in \Lrad\ and present only a mildly narrower range than our overall sample (of 19 sources).
The radio luminosities of the 11 compact sources with robust radio detections span over 3.3 dex, $ 38.4 \lesssim \log(\Lrad/\ergs) \lesssim 41.7$, and have a median of $\log(\Lrad/\ergs) = 39.4$ and a standard deviation of 1 dex. 
The seven extended sources span 1.7 dex, $ 40.5 \lesssim \log(\Lrad/\ergs) \lesssim 42.2$, with a median of $\log(\Lrad/\ergs) = 41.7$ and a standard deviation of  $\gtrsim$0.6 dex.

The observed radio luminosities may be used to classify our sources as ``radio-loud'' or ``radio-quiet'' -- a commonly used phenomenological characteristic of AGN 
(see, e.g., \citealt{1974MNRAS.167P..31F,1993MNRAS.263..461P,UrryPadovani1995_rev,1989AJ.....98.1195K,2004MNRAS.351...70B,2007ApJ...658..815S,2016ApJ...831..168K}, 
but also contradicting evidence in, e.g., \citealt{2009AJ....137...42R,2012A&A...545A..66B,2013MNRAS.429.1970B}, as well as the recent critique of this approach by \citealt{2017NatAs...1E.194P}).
For obscured, type-2 AGN, where the UV-optical emission is dominated by the stellar content of the host galaxy, the only sensible radio-loudness measure (apart from relative to X-rays, which we discussed above) is based on a simple cut in radio luminosity.
Indeed, \cite{2016ApJ...831..168K} define radio loudness as spectral luminosity $L_{\rm 6 GHz} > 10^{23.2}\,{\rm W\, Hz}^{-1}$. 

To assess the radio loudness of our sources (i.e., from their observed 1.4 GHz measurements), we calculated the 6.0 GHz flux densities assuming again a power-law radio SED, $S_\nu \propto \nu^{\alpha_\nu}$ with a spectral index $\alpha_\nu = -0.7$.
We note that in reality, each source is expected to have a different spectral slope, driven by the nature and properties of the dominant radio emission mechanism (i.e., synchrotron vs. bremsstrahlung; see, e.g., \citealt{1993ApJ...407..549K}), and/or the age of the radio-emitting jet (see, e.g., the discussion in the recent works of \citealt{2015ApJ...809..168C} and \citealt{2018ApJ...859...23N}; and also \citealt{1991ApJ...383..554C,1992MNRAS.257..545L,1998AJ....116.2953A,2011MNRAS.416.1135R} and \citealt{2012A&A...548A..75M} for additional specific examples). 
Here we use the simplistic $\alpha_\nu=-0.7$ assumption as a practical choice to derive \Lrad\ for our sources in a way that is consistent with many previous studies.
Using the derived 6.0 GHz luminosities and the definition above, 11 sources of our sample are radio loud. 
This corresponds to $57.4^{+10.9} _{-10.5}\%$ of the sources for which we have NVSS data (i.e., 19 high X-ray luminosity AGN).\footnote{Fractions and uncertainties are calculated through the inverse beta distribution, using the the 50-, 18- and 84-th percentiles.\label{foot:frac_calc}}
The larger sample of type-2 AGN in BASS/DR1 has only two additional sources that would qualify as radio-loud based on this definition (i.e., 2 of the 127 lower-\Lbat\ AGN), and the total fraction of such radio loud sources among NVSS-detected BASS/DR1 type-2 AGN is thus $9.3^{+2.6}_{-2.2} \%$ (13/146 sources{\color{blue}; see footnote~\ref{foot:frac_calc}}).
The difference in fractions is highly significant -- a formal Fisher's exact test results in $P < 10^{-5}$.

We thus conclude that our extremely high X-ray luminosity AGN show a vast range in radio luminosities (almost 4 dex), and a higher fraction of extremely radio-luminous (radio-loud) sources, compared to the general population of BASS/DR1 AGN.
This large range in radio luminosities is unlikely to be driven {\it solely} by the (diverse) morphologies of the radio-emitting regions in our sample, and instead may be predominantly driven by a diversity of evolutionary stages and/or time-scales \cite[e.g.,][]{1995ApJ...453L..13K}.

\subsubsection{Double radio lobes}
\label{subsec:other_rad_props}

We finally examine the occurrence of double lobed radio sources among our sample of high X-ray luminosity obscured AGN.
Such radio emission is driven by pairs of jets that interact with matter in the host galaxies and (large-scale) environments of AGN, and is extremely rare \cite[e.g.,][]{2006AJ....131..666D}.  
Of all the 264 type-2 AGN in BASS/DR1, 176 (67.7\%) are classified as ``Radio'' in the MORX catalogue, and only nine (3.4\%) have double radio lobes.
In Table~\ref{tab:radio_lobes} we list the basic information regarding these double radio lobes.\footnote{Note that Table~\ref{tab:radio_lobes} extends beyond our sample of 28 highly luminous type-2 BASS AGN.}
We note that the source BAT ID 1051 (3C~403) has two sets of double radio lobes \cite[see, e.g.,][]{2005ApJ...622..149K}, which are both listed in Table~\ref{tab:radio_lobes}.
Seven of these nine sources belong to our sample of extremely luminous type-2 AGN within BASS/DR1, that is 25\% of the highly luminous sources have double lobes, and constitute $\sim$78\% of all type-2 BASS with such features.
Conversely, we note that BAT ID 1092 {\it does} have double radio lobes, although it has a luminosity of $\log(\Lbat/\ergs) = 43.33$, which is one order of magnitude below the luminosity of the sources in our sample.

We therefore conclude that the occurrence rate of double radio lobes is significantly higher among our highly X-ray luminous sources than among the general population of BASS/DR1 obscured AGN population (a formal Fisher's exact test results in $P < 10^{-3}$).
With only $25\%$ of our sources having such radio features, however, it is still far from being a {\it defining} characteristic of highly X-ray luminous obscured AGN.

\begin{table*}
\caption{Type-2 AGN from BASS/DR1 with double radio lobes.}
\resizebox{\textwidth}{!}
{\begin{tabular}{|lc|lrrrr|lrrrr|}
\hline
{BAT}  &{$\log\Lbat$} &  {Lobe 1}& {$F_\nu$~~} &{$\Delta F_\nu$~} & {a$^{\ast}$}& {b$^{\dagger}$}& {Lobe 2}& {$F_\nu$~~} &{$\Delta F_\nu$~}& {a$^{\ast}$}&  {b$^{\dagger}$}  \\
{ID$^{a}$} & (\ergs) & {Radio ID} $^{b}$ & {(mJy)} & {(mJy)} & {(\arcsec)} & {(\arcsec)} & {Radio ID} $^{b}$&     {(mJy)}& {(mJy)} &{(\arcsec)}& {(\arcsec)}  \\
\hline
\textbf{ 118} & {44.92} & {NVSS~J021539.1$-$125933 } & {1915.8} & { 73.7} & { 21.2} & {16.2}  & {NVSS~J021535.3$-$125929 } & { 2764.1} & { 97.4} & { 15.1} & {17.7}\\
\textbf{ 209} & {44.74} & {NVSS~J040722.2$+$034117 } & {1221.6} & { 38.6} & {110.9} & {33.0}  & {NVSS~J040711.1$+$034342 } & { 1074.7} & { 98.0} & {127.3} & {29.8}\\
\textbf{ 238} & {45.04} & {NVSS~J044438.5$-$281012 } & {3474.2} & {113.0} & { 41.7} & {18.8}  & {NVSS~J044436.2$-$280922 } & { 3262.9} & {108.0} & { 35.2} & {19.4}\\
\textbf{ 360} & {44.80} & {NVSS~J070917.7$-$360217 } & { 255.3} & {  8.3} & { 85.6} & {47.5}  & {NVSS~J070910.3$-$360020 } & {  262.6} & {  8.6} & {262.6} & {45.0}\\
{ 426}        & {44.21} & {FIRST~J084002.7$+$294914} & { 159.6} & {0.153} & { 5.44} & {339.0} & {FIRST~J084001.6$+$294845} & {  163.5} & {0.153} & { 30.4} & {20.4}\\
\textbf{ 591} & {44.36} & {FIRST~J120734.0$+$335227} & { 172.4} & {0.134} & { 11.24} & {4.3}  & {FIRST~J120731.2$+$335256} & {   60.6} & {0.135} & {  3.0} & {0.0}\\
\textbf{1051$^{c}$} & {44.35} & {NVSS~J195218.2$+$023029 } & {2467.6} & { 85.1} & { 45.7} & {28.0}  & {NVSS~J195213.0$+$023026 } & { 2741.0} & {101.7} & { 27.9} & {20.8}\\
\textbf{1051$^{c}$} & {44.46} & {NVSS~J195217.7$+$022920 } & { 378.8} & { 12.3} & { 43.2} & {13.9}  & {NVSS~J195211.2$+$023113 } & {  458.0} & { 16.2} & { 50.2} & {15.0}\\
{1092}        & {43.33} & {SUMSS~J205206.9$-$570357} & { 198.5} & {  9.8} & {110.7} & {53.9}  & {SUMSS~J205202.0$-$570406} & { 1972.0} & { 59.2} & { 54.3} & {48.2}\\
\textbf{1210} & {44.53} & {SUMSS~J235855.5$-$605428} & {9498.0} & {303.6} & {100.7} & {90.4}  & {SUMSS~J235909.9$-$605548} & {12376.0} & {408.5} & { 98.1} & {74.1}\\
\hline
\end{tabular}}
\flushleft 
\footnotesize{
Notes:\\
$^{a}$ Objects with BAT IDs in boldface are part of our sample of high-luminosity type-2 AGN.\\
$^{b}$ Classification and data from the MORX catalogue.\\
$^{c}$ BAT ID 1051 (4C~403) is known to have two sets of double radio lobes \cite[see, .e.g.,][]{2005ApJ...622..149K}.\\
$^{\ast}$ Semi-major axis.\\
$^{\dagger}$ Semi-minor axis.
}
\label{tab:radio_lobes}
\end{table*}

\begin{figure} 
\begin{center}
\includegraphics[width=0.48\textwidth]{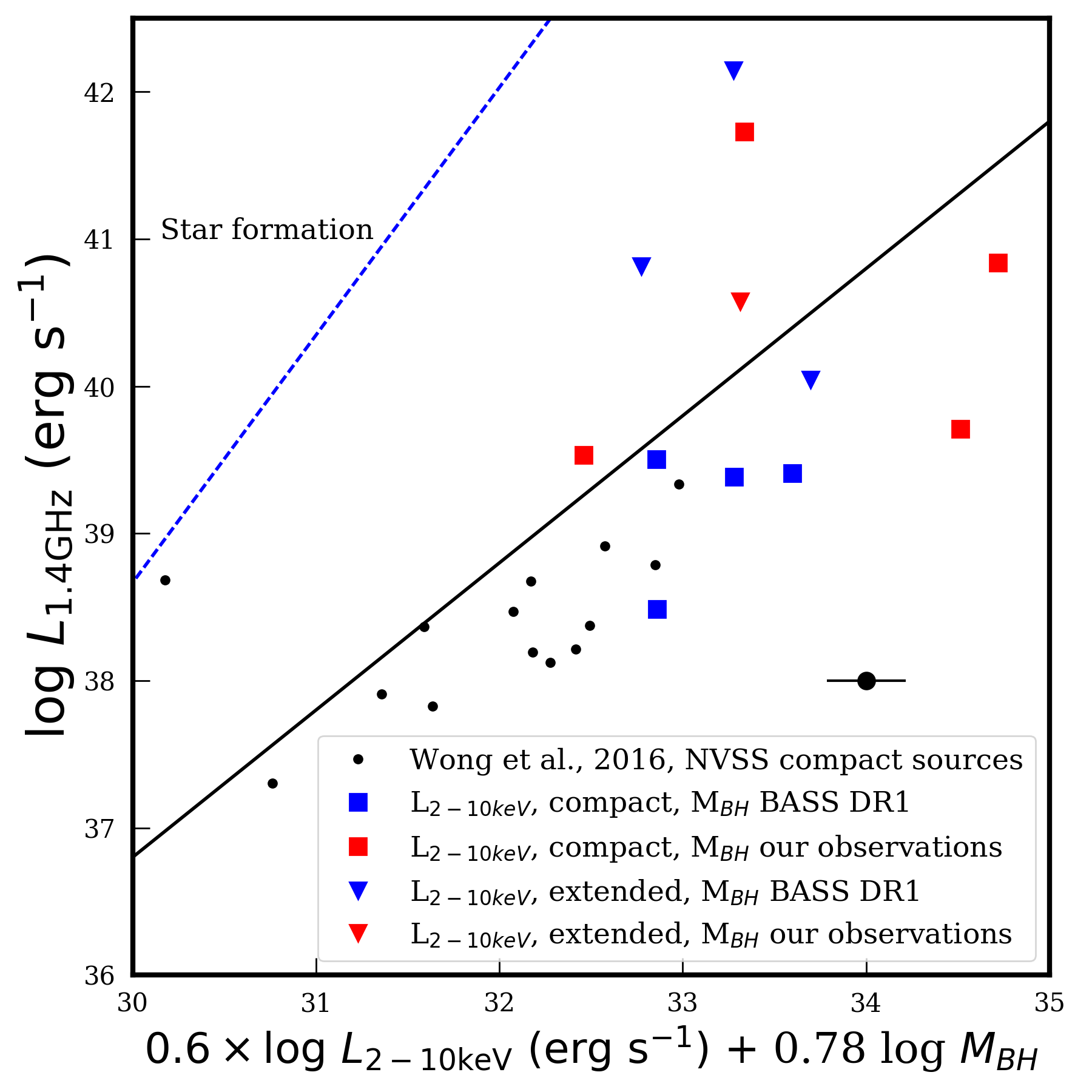}
\caption{
The ``fundamental plane of BH activity'' for our sample, in the context of other BAT AGN studies \citep{2016MNRAS.460.1588W}.
The black solid line represents the fundamental plane, adapted from \protect\cite{2003MNRAS.345.1057M}.
The thick blue dashed line shows the expected relationship if the sole origin of the observed radio and X-ray luminosities were from star formation (that is, not related to the BH mass in the x-axis; \protect\citealt{2003ApJ...586..794B,2014MNRAS.437.1698M}). 
The black points represent NVSS compact sources from \protect\citet{2016MNRAS.460.1588W}.
Large symbols represent sources from our sample of 28 of the most X-ray luminous obscured AGN in BASS/DR1, with different symbols tracing radio morphological classes and different colours tracing the source for \mbh\ estimates, as indicated in the legend.
The values for the intrinsic \Lhard\ are taken from the catalogue of X-ray properties of the \swift/BAT 70-month survey associated with BASS/DR1 \protect\cite[][]{2017ApJS..233...17R}.
The single error bar represents the mean measurement-related errors on the combination of \mbh\ and \Lhard\ (propagating errors in quadrature). 
Our sources are scattered above and below the fundamental plane and do not show any common characteristics as a group. 
The positions of the compact sources are predominantly below the fundamental plane, those of the extended sources above the fundamental plane, which is not unexpected (see text for discussion). 
}
\label{fig:Fundplane}
\end{center}
\end{figure}

\subsubsection{The fundamental plane of black hole activity}
\label{subsec:fund}

Several lines of study tried to put forward a way to unify accreting BHs across the mass range -- from stellar-mass BHs through SMBHs. 
This is inspired by scale-independent (and to some extent accretion model independent) relations between the BH mass (\mbh), accretion rate (\Mdotbh), and the emergent radio emission from jets (i.e., the radio luminosity, $L_{\rm R}$). 
Since \citet{2003MNRAS.343L..59H} suggested a strong correlation between these quantities, many studies have investigated what has become to be known as the ``fundamental plane of BH activity'' - a plane in the $\log\mbh - \log L_{\rm R} - \log L_{\rm X}$ three-dimensional space that indeed links BHs of all mass scales \citep{2003MNRAS.345.1057M,2004A&A...414..895F,2009ApJ...706..404G,2012MNRAS.419..267P,2013MNRAS.429.1970B,2009ApJ...703.1034Y,2016MNRAS.455.2551N,2018ApJ...863..117D,2018MNRAS.474.1342M}. 
The data available for our sample of high X-ray luminosity type-2 BASS AGN allows us to test the relevance of the fundamental plane for a well-defined sub-region of the (3-dimensional) parameter space, and specifically to test whether the broad range of radio luminosities (Section~\ref{subsec:radio_lum} above) can be accounted for by the \mbh\ dependence of the fundamental plane.

In Fig.~\ref{fig:Fundplane} we show the radio / X-ray fundamental plane, as proposed by \cite{2003MNRAS.345.1057M}, and further adjusted to follow the definition given in \citet[black solid line]{2016MNRAS.460.1588W}. 
We also show the expected relationship provided the sole origin of the observed radio and X-ray luminosities is from star formation \cite[and effectively ignoring any \mbh\ dependence; dashed blue line, following][]{2003ApJ...586..794B,2014MNRAS.437.1698M}. 
For reference, we plot NVSS measurements for compact sources, taken from \citet[][black points]{2016MNRAS.460.1588W}. 
The sources in our sample of luminous type-2 BASS AGN are presented according to their radio morphological classification (i.e., Table \ref{tab:basic_info}) and the source for the optical spectra used for their \mbh\ measurements (see figure legend).
Here we choose to use intrinsic $2-10\,\kev$ luminosities, \Lhard, taken from the respective BASS catalogue \cite[Table~C9 therein]{2017ApJS..233...17R} as tracing $L_{\rm X}$.
For radio luminosities we use the aforementioned NVSS-based measurements of \Lrad.

As can be seen in Fig.~\ref{fig:Fundplane}, our AGN show a considerable scatter with the Fundamental Plane, and do not occupy any specific region in it. 
The scatter of our sources (i.e., standard deviation of their residuals with regard to the Fundamental Plane) exceeds 3 dex, which is much larger than the intrinsic scatter derived for the Fundamental Plane \cite[$<1$ dex; see, e.g.,][]{Kording2006_FP,2012MNRAS.419..267P}.
Moreover, most of the extended sources are found above the fundamental plane, whereas the compact sources are found somewhat below it.
This apparent dichotomy is somewhat be expected, as the fundamental plane establishes correlations only between tracers of nuclear emission \citep{1979ApJ...232...34B,1995A&A...293..665F,2004A&A...414..895F}.
The reference compact NVSS sources from \citet{2016MNRAS.460.1588W} are found in a relatively narrow band below the fundamental plane (with a single exception).
We note that, although the reference NVSS compact sources taken from \cite{2016MNRAS.460.1588W} were also based on an X-ray bright sample of ultra-hard X-ray selected AGN \cite[drawn from an earlier \swift/BAT catalogue;][]{2011ApJ...739...57K}, the host stellar masses of those sources are typically lower than the masses of our sources (average $\log[\mstar/\Msol] = 10.27$ vs. $10.63$). Accordingly, the BH masses are also somewhat lower.

Even in the case that our highly X-ray luminous, obscured AGN do not occupy a particular region of the Fundamental plane, they could have been located close to, and broadly along the solid black line, or instead show a common offset from it.
This is clearly not the case as they scatter above and below the fundamental plane. 
In principle, one possible explanation for the large scatter could be that some of the sources are beamed \citep{2003MNRAS.345.1057M}. 
However, we recall that we excluded beamed sources in the selection of our sample (Section \ref{subsec:sample}).
We also verified that our sample does not include any robustly detected $\gamma$-ray sources.\footnote{Although \citet{2013ApJS..206...17M} lists one of our sources, BAT ID 249, as a $\gamma$-ray candidate, we have found no match in the Fermi Large Area Telescope Third Source Catalogue \citep{2015ApJS..218...23A}.}

We conclude that the high X-ray luminosity, obscured AGN in our sample do not occupy any characteristic region of the fundamental plane of BH activity, and specifically that the \mbh\ dependence of the plane cannot fully account for the large range on radio luminosities seen in our sample.

\subsubsection{Summary of radio analysis}

To summarise our radio analysis, the highly X-ray luminous obscured AGN in our sample span a wide range in essentially all the radio properties we have examined, and are not necessarily concentrated in specific regions of radio-related parameter space. 
In particular, we found that:
\begin{itemize}

\item The radio luminosities of our sources span a large range, of almost 4 orders of magnitude, over a rather limited range in ultra-hard X-ray luminosity. 

\item Based on the NVSS contours (Table \ref{tab:basic_info}) 12 sources are compact, six sources are extended, and the remaining objects have different qualifications or the contours are not available. 

\item A high fraction of sources (7 of 28; or 25\%) have double radio lobes. These constitute the vast majority of all BASS/DR1 type-2 AGN with such radio features ($\sim$75\%).

\item 
A significantly higher fraction of luminous X-ray AGN have radio lobes, compared to the total BASS/DR1 type-2 AGN population (${\sim}25\%$ vs. ${\sim}3\%$).

\item 
A significantly higher fraction of luminous X-ray AGN are classified as radio loud, based on a simple radio luminosity threshold, compared to the total type-2 BASS/DR1 AGN population (${\sim}57 \%$ vs. ${\sim}9\%$).

\item Our sources scatter widely both above and below the fundamental plane of BH activity and, again, do not show any distinctive properties. 
This can be partially explained by the fact, that our sample includes both extended and compact sources.

\end{itemize}
We finally note that our finding of a very high occurrence rate of strong radio emission from our sample of X-ray luminous AGN agrees with the finding of a high fraction of elliptical (or bulge-dominated), massive host galaxies (Section~\ref{subsec:Hosts}), and the well-established links between the two phenomena in large samples of radio-emitting AGN (see, e.g., \citealt{2005MNRAS.362...25B,2014ARA&A..52..589H}, and references therein).

Moreover, the high occurrence rate of strong radio emission (and of double radio lobes) could be indicative of a strong link between intense SMBH growth, as probed by the exceptionally high ultra-hard X-ray luminosities, and efficient jet launching, as probed by the radio data.
Such close links between accretion disc power and radio jet power were revealed in several previous AGN studies \cite[see, e.g.,][and references therein]{Ghisellini2014_jets_disks}.
Given that our sources trace the top of the distribution of AGN luminosities, one intriguing (though speculative) possibility is that our sources are powered by highly-spinning SMBHs, which would in turn result both in a high radiative efficiency ($\eta\equiv \Lbol / \dot{M} c^2$) and in an enhanced jet production efficiency and/or jet power (through mechanisms reminiscent of, e.g., \citealt{BlandfordZnajek1977}; see also \cite{2018arXiv181206025B} for a recent review).

\section{Accretion demographics}
\label{sec:bhm}

The mass-normalized accretion rate, or Eddington ratio of an AGN ($\lamEdd\equiv \Lbol/L_{\rm Edd} \propto \Lbol/\mbh$) is one of its key characteristics, as it yields important information about the small scale accretion process and the mass build-up of the SMBH. 
As our sample is selected to include the most luminous AGN in the low-redshift Universe (as probed through BASS/DR1), a central question is what drives the extremely high luminosities of these sources.
One could expect them to have either extremely high \mbh\ (with modest \lamEdd); to have extremely high \lamEdd\ (with modest \mbh); or indeed a combination of extremely high \mbh\ and \lamEdd. 
Each of these scenarios may be linked to different stages in the growth of the SMBHs in question, or at the very least to different stages in the specific accretion episode that powers our AGN.
As mentioned in Section  \ref{subsec:new_obs}, we obtain \mbh, and thus \lamEdd, for our sources either directly from the BASS/DR1 catalogue  \citep{2017ApJ...850...74K}, or by applying similar spectral analysis procedures to our newly acquired optical spectra of luminous type-2 AGN.

\begin{figure}
\begin{center}
\includegraphics[width=0.49\textwidth]{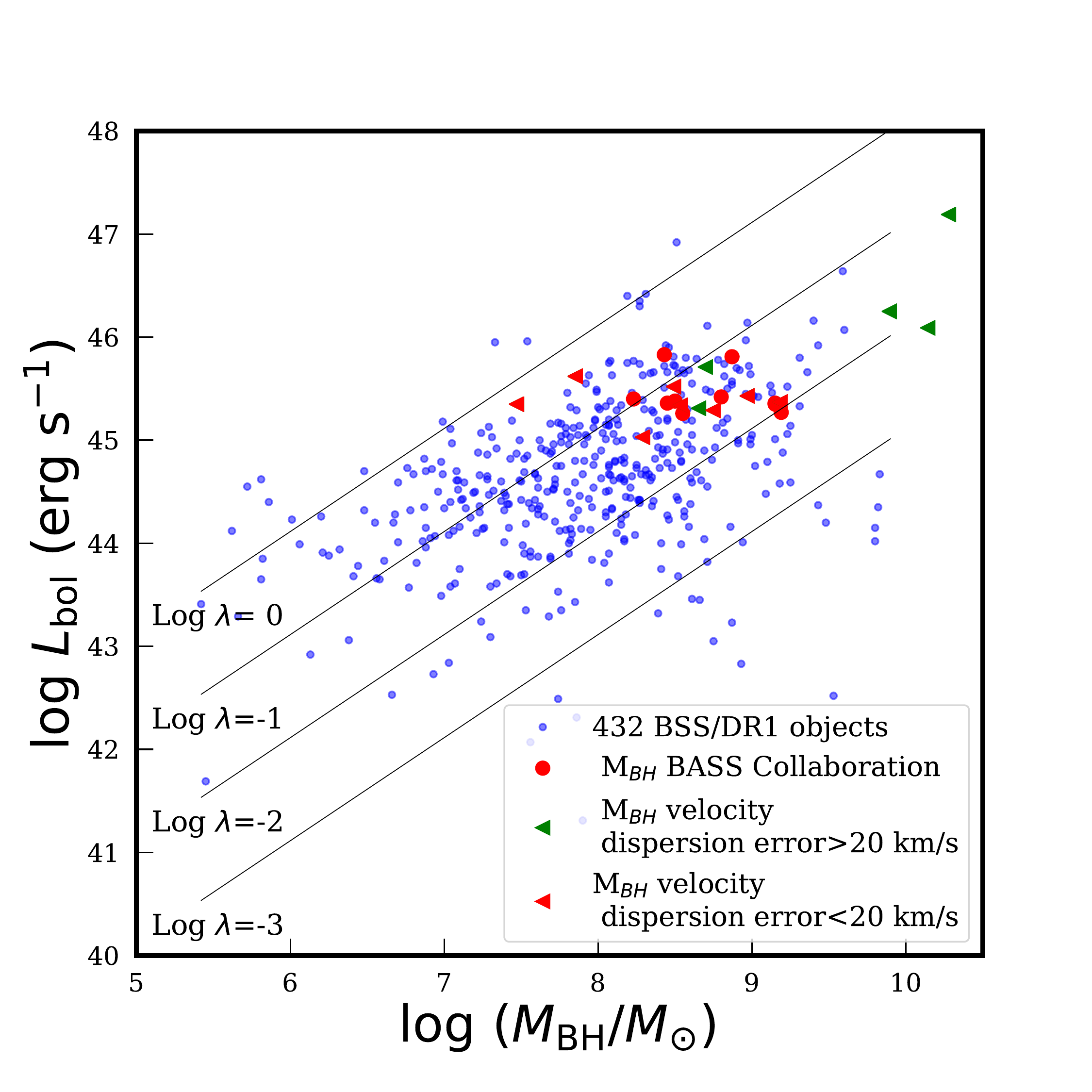}
\caption{Log  $\Lbol$ vs.\ $\log\mbh$ for our sample compared to the AGN of the BASS/DR1 (blue dots). The black lines represent constant Eddington ratios  $\lambda$. 
Our sample lies at the high mass end between $7.2 < \log(\mbh/\Msol) < 10.2$ and at Eddington ratios $-2.2 < \log\lamEdd  < -0.2$. Contrary of what one might intuitively expect, our sample of extremely luminous objects are not concentrated either in the low mass/ high Eddington ratio range or in the high mass/low Eddington ratio range, but are spread out across the whole high mass and Eddington ratio range.} 
\label{fig:Eddratio}
\end{center}
\end{figure}

In Fig.~\ref{fig:Eddratio} we show the $\Lbol-\mbh$ plane for our objects, compared with the general BASS/DR1 AGN population.
Different symbols mark sources that differ in the source of optical spectra used for their \mbh\ measurements, and in the quality of their stellar velocity dispersion measurements (which is used to derive \mbh; see figure legend and caption). 
The black lines trace several constant Eddington ratios.

Fig.~\ref{fig:Eddratio} shows that our sources are neither very massive, nor have particularly high Eddington ratios. 
Their BH masses are in the range $7.5 \lesssim \log(\mbh/\Msol) \lesssim 9.3$ if we consider only high-quality \mbh\ determinations (i.e., velocity dispersion errors $\Delta\sigs \leq 20\,\kms$).
This range is consistent with the peak of the active BH mass function of luminous AGN in the low-redshift Universe, as traced by {\it unobscured} sources \cite[i.e., broad-line quasars; see, e.g.,][]{VestergaardOsmer2009_BHMF,Schulze2010_BHMF,KellyShen2013_BHMF}.
If we also include less reliable \mbh\ estimates ($\Delta\sigs >  20\,\kms$), the range extends to $\log(\mbh/\Msol) \simeq 10.3$.
Notwithstanding the uncertainties related to this latter high-mass end, we note that it is consistent with the 
highest masses seen in {\it inactive} SMBHs in the local Universe \cite[e.g.,][]{McConnell2011_MBH_10,McConnell2012_BCGs}; 
the most massive SMBHs observed out to $z\sim6$ \cite[e.g.,][]{Shemmer2004,Wang2015_z5_hiM,Wu2015_z6_nature}
and perhaps the highest BH masses that could be observed as (radiatively efficient) accreting systems \citep{Inayoshi2016_maxMBH,King2016_maxMBH,Pacucci2017_maxMBH_feedback}.

\begin{figure}
\begin{center}
\includegraphics[width=0.48\textwidth]{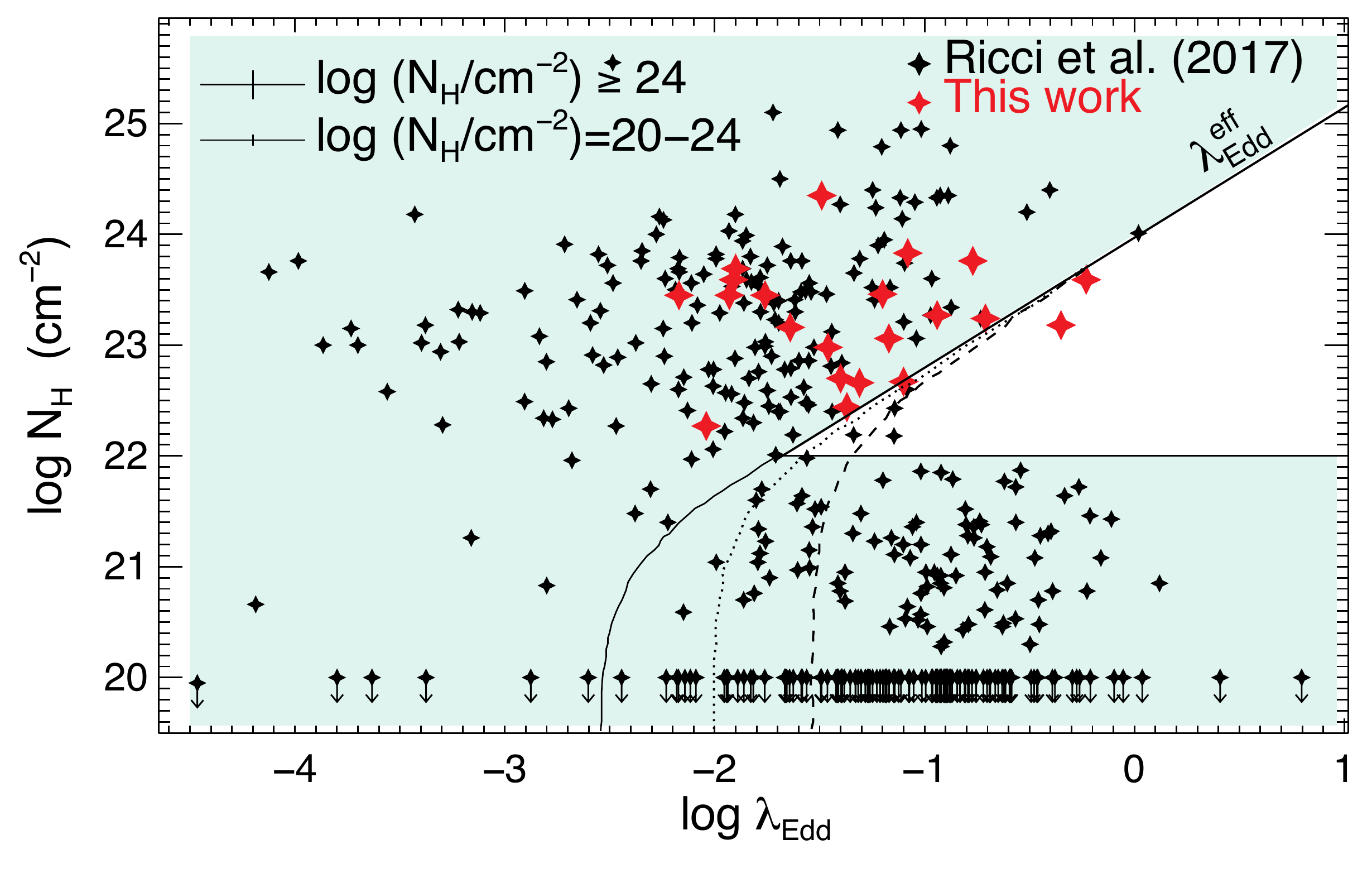}
\caption{Eddington ratio - column density diagram reproduced from \citet{2017Natur.549..488R}. 
The diagram shows in black  392 AGN from \citet{2017ApJ...850...74K} for which the black hole mass and the Eddington ratio have been determined. The green shaded area delineates the region where the Eddington ratio is below the Eddington ratio for dusty gas, the white area reflects the region where radiation pressure would push out the obscuring material from the torus. Our sample of highly luminous type-2 AGN is shown in red. Their positions are outside the wedge with two exceptions. The positions of the two sources, which are just inside the wedge are not conclusive given  the systematic uncertainties of \lamEdd. Our sample is therefore consistent with a radiation driven unification model.} 
\label{fig:Lum_NH_dists_LLEdd}
\end{center}
\end{figure}

The Eddington ratios of our sources lie below the Eddington limit and in the range $-2.2 \lesssim \log \lamEdd \lesssim -0.2$, with a median log $\lamEdd=-1.3$.
This would not change dramatically if one would use the (intrinsic) $2-10$ \kev luminosities, and the  \Lhard-based bolometric corrections of \cite{2004MNRAS.351..169M}. Such choices would result in  $\log\lamEdd$ ranging  $-2.4$ to $-0.2$, and a median of $-1.1$.
We note that the distribution of \lamEdd\ among our highly luminous sources does {\it not} extend to the lower end of the \lamEdd\ distribution seen for BASS/DR1 AGN (i.e., our sources not reaching below $\log\lamEdd \approx -2.2$, while the BASS AGN reach $\log\lamEdd \approx -4$; see Figs.~\ref{fig:Eddratio} and \ref{fig:Lum_NH_dists_LLEdd}). 
This is expected, given the high \Lbol\ cut used to select our sources.
Specifically, even an exceptionally massive BH, with $\log(\mbh/\Msol)=10$ would need to have $\log\lamEdd \gtrsim -3$ to reach $\log(\Lbol/\ergs)\gtrsim 45.3$.

The range of \lamEdd\ for our sources is, again, consistent with what is seen in large samples of (luminous) broad-line quasars, reaching $z\sim2$ \cite[and beyond; e.g.,][]{TrakhtNetzer2012_Mg2,KellyShen2013_BHMF,Schulze2015_BHMF,Trakhtenbrot2016_COSMOSFIRE_MBH}.
The recent study by \citet{2017ApJ...845..134W} used a forward modelling approach to predict the AGN luminosity function from the galaxy stellar mass function. 
They show that the observations are consistent with a mass independent Eddington ratio distribution function (ERDF), which is further assumed to take a broken power-law shape with a break at $\log \lamEdd^{*}$.
For X-ray selected AGN, drawn from earlier (shallower) all-sky ultra-hard X-ray surveys using \swift/BAT, this ERDF has a break at $\log \lamEdd^{*} \simeq -1.8$ \cite[see also][]{Caplar2015_coeval}).
Other studies \cite[e.g.,][]{Schulze2010_BHMF,Schulze2015_BHMF} promoted alternative ERDF shapes, which resulted in $\log \lamEdd^{*} \simeq (-1)-(-0.5)$.
In any case, the range of \lamEdd\ for our sources extends to both sides of $\lamEdd^{*}$, thus exhibiting Eddington ratios that are typical of the broader population of X-ray selected AGN.

We conclude that the highly luminous obscured AGN in our sample are neither extremely massive, nor do they accrete at extremely high Eddington ratios, compared to the general population of low-redshift AGN.
A forthcoming study by the BASS team (Weigel et al., in prep.) will address, in great detail, the BH mass and Eddington ratio functions for all BASS AGN (including those to appear in the upcoming DR2).

As mentioned in Section~\ref{sec:intro}, high radiative outputs from the accreting SMBH can have significant effects on the structure of the obscuring circumnuclear material, which is generally considered to be found in an axis-symmetric, toroidal configuration \cite[e.g.,][]{Antonucci1993,UrryPadovani1995_rev,Netzer2015_torus_rev}.
The inner boundary of this dusty torus is dictated by the radius at which the AGN radiation is sufficient for the temperature to exceed the sublimation temperature of the dust grains.

In the ``receding torus'' scenario \cite[][]{1991MNRAS.252..586L}, 
a higher AGN luminosity would result in a higher temperature at any given distance from the radiation source, thus sublimating the torus dust in increasingly large radii, and consequently lowering the covering factor (i.e., exposing a larger solid angle around the central AGN engine).
This scenario thus predicts a rather sharp decrease in the number (or fraction) of obscured sources at high AGN luminosities, which is indeed observed in large multi-wavelength AGN studies \cite[e.g.,][]{2003ApJ...598..886U,2005MNRAS.360..565S,Maiolino2007_IRS_S04,2008ApJ...679..140T}.
In terms of \lamEdd, however, we note that in this scenario high-\lamEdd\ sources may remain obscured, as long as they are not too luminous.
An alternative picture \cite[e.g.,][and references therein]{2009MNRAS.394L..89F}, suggests that the inner boundary of the torus, and indeed the overall amount of line-of-sight obscuring material, is dictated by radiation pressure, which in turn can be parametrized using $\Lbol/\mbh \propto \lamEdd$ (and not simply \Lbol). 
The obscuring material can be ``pushed away'' if \lamEdd\ is sufficiently high, where the exact threshold depends on the line-of-sight column density, \NH.
Consequently, one expects a ``blow-out'' region in the $\lamEdd-\NH$ parameter space, where long lived clouds of dusty gas cannot exist (see below).

The recent study by \citet{2017Natur.549..488R} investigated the two scenarios in detail, using the BASS/DR1 sample. 
We have reproduced the key $\log\lamEdd-\log\NH$ diagram from that study, shown here as Fig.~\ref{fig:Lum_NH_dists_LLEdd}.
The diagram shows 392 AGN for which \lamEdd\ and \NH\ were determined through BASS/DR1 (i.e., \citealp{2017ApJ...850...74K} and \citealp{2017ApJS..233...17R}, respectively).
For the purpose of Fig.~\ref{fig:Lum_NH_dists_LLEdd}, we have adopted \Lbol\ estimates that are based on the intrinsic $2-10$ keV luminosities, and a {\it fixed} bolometric correction of 20 -- that is, $\Lbol = 20 \times \Lhard$. 
This was done to be consistent with the analysis performed by \citet{2017Natur.549..488R}. 
The green shaded area in Fig.~\ref{fig:Lum_NH_dists_LLEdd} delineates the region where the radiation pressure is not high enough to expel the obscuring material, while the white wedge-shaped region highlights the ``blow-out'' region.
Essentially all our highly luminous obscured AGN (larger red symbols) are found outside the ``blow-out'' region, thus self-consistently explaining their obscured nature, and implying that this obscuration could be long-lived.
The positions of the two exceptions, that are found just within the wedge, are not conclusive given the large systematic uncertainties on \lamEdd\ (at least $\sim$0.3 dex; see the error-bars in the top-left corner of Fig.~\ref{fig:Lum_NH_dists_LLEdd}, and \citealp{2017ApJ...850...74K}). 
We have verified that using alternative estimates of \Lbol\ (and thus of \lamEdd), such as the \Lhard-based bolometric corrections of \cite{2004MNRAS.351..169M}, or the \Lbat-based estimates used throughout this work, does not significantly change the overall positioning of our sources with respect to the ``blow-out'' region in Fig.~\ref{fig:Lum_NH_dists_LLEdd}.

We conclude that our sample of extremely X-ray luminous obscured AGN is consistent with the radiation pressure driven unification framework, while their very existence is challenging the receding torus scenario.

The recently published study by \citet{KongHo2018_Q2} presented \mbh\ and \lamEdd\ estimates for a large, SDSS-based sample of luminous obscured AGN \cite[originally presented in][]{2008AJ....136.2373R}.
These AGN have bolometric luminosities in the range $45.5 \lesssim \log(\Lbol/\ergs) \lesssim 47.5$, estimated through scaling the luminosities of the strong, AGN-dominated \OIII\ emission line ($L(\oiii)$).
This range of \Lbol\ overlaps with the sample studied here, although our sample is somewhat concentrated towards the low-\Lbol\ end of this range (see right panel of Fig.~\ref{fig:Lum_NH_dists}). 
\citet{KongHo2018_Q2} present careful measurements of \sigs\ that -- when combined with the same $\mbh-\sigs$ relation we use here -- imply BH masses in the range $6.5 \lesssim \log(\mbh/\Msol) \lesssim 10$, and Eddington ratios in the range $-2.9 \lesssim \log\lamEdd \lesssim 1.8$, with $\sim$20\% of sources exceeding the Eddington limit. 
These high accretion rates obviously challenge the aforementioned ``radiative feedback'' unification framework. 
In particular, in terms of the $\lamEdd - \NH$ parameter space (Fig.~\ref{fig:Lum_NH_dists_LLEdd}), the only way to have long-term obscuring material at these high \lamEdd\ would be if these AGN were Compton-thick, with $\log(\NH/\cmii)\gtrsim 24$. 
This is, however, unlikely, as there is no reason to suspect that the (optical) selection of the large SDSS sample would include such a high fraction of rather elusive Compton thick AGN \cite[e.g.,][]{2015ApJ...815L..13R}.

As discussed in \cite{KongHo2018_Q2}, the broad range in \lamEdd\ they find, and indeed the high accretion rates found for many of their sources, are at least partially driven by the significant uncertainties in the \oiii-based bolometric corrections and thus \Lbol\ and \lamEdd. 
In this context, we recall that the BASS sample was used to demonstrate the large scatter between \oiii\ and (ultra-hard) X-ray emission \citep{Berney2015_BASS}, ranging over almost 2 dex. 
These outstanding issues could be mitigated perhaps by using higher-ionization lines, which were shown to be more tightly correlated with X-ray emission and thus, in principle, with \Lbol\ (e.g., [Ne~\textsc{iii}]~$\lambda$3869 - see \citealp{Berney2015_BASS}; and/or [Ne~\textsc{v}]~$\lambda$14.3 $\mu$m - see \citealp{2007ApJ...663L...9S,2012ApJ...753...38S}).

\section{Summary and conclusions}
\label{sec:conclusions}

We presented a multi-wavelength analysis of some of the most luminous ultra-hard X-ray selected, obscured AGN in the low-redshift Universe. 
We selected the 28 objects with the highest $14-195\,\kev$ (and thus, bolometric) luminosities from the \swift/BAT all-sky 70-month catalogue, focusing on narrow emission line sources located at least 6\deg\, off the galactic plane, and excluding beamed sources.
We analysed our sample in the optical, infrared, and radio regimes, with the general goal of determining whether the objects of our sample as a group have  distinctive properties. 
Our analysis mostly relied on data available through the first data release of the BAT AGN Spectroscopic Survey (BASS/DR1).
However, since black hole masses -- that are of crucial importance to our analysis -- were available for only 10 of our objects, we complemented the BASS/DR1 data with a dedicated observing campaign, using the VLT, Palomar, and Keck observatories, to obtain the missing black hole masses (and Eddington ratios).

\noindent
The results of our analysis can be summarized as follows:
\begin{itemize}

\item 
While selected purely based on their intense hard X-ray and narrow line emission, we find that the sources in our sample would also be robustly identified as luminous, obscured AGN, through other approaches:
their column densities classify them as X-ray obscured AGN ($22 < \log[\NH/\cmii] < 24.5$; Fig.~\ref{fig:Lum_NH_dists}); 
and their optical emission line ratios and mid-infrared colours agree with commonly used AGN selection criteria (Figs. \ref{fig:BPT} and \ref{fig:MIR_colors}, respectively).

\item 
The host galaxies of the objects of our sample for which we have SDSS or PanSTARRS images (54\%) all appear to have elliptical (or bulge-dominated) morphologies, in contrast to the trends reported by several previous studies for less luminous AGN, and to the expectation to find {\it some} spiral hosts (Section~\ref{subsec:Hosts}) based upon samples with similar K-band luminosities.

\item 
The hosts cover a wide range of stellar masses, but are concentrated on relatively high masses, $9.8 \lesssim \log(M_{*}/\Msol) \lesssim 11.7$ -- a narrower range than what is seen in the general BASS type-2 AGN population ($7.7 \lesssim \log[M_{*}/\Msol] \lesssim 13.7$).
The host luminosity and masses of our AGN coincide with the range commonly associated with the transformation of the galaxy population from (star forming) discs to (quiescent) spheroids (see right panel of Fig.~\ref{fig:morphology}).

\item 
The highly X-ray luminous, obscured AGN in our sample show a significantly higher fraction of radio-loud sources, and significantly higher occurrence rate of double radio lobes, compared to the total BASS/DR1 type-2 AGN population. 
Their radio luminosities spread over at least 3 dex, compared to about 1 dex in ultra-hard X-ray luminosity.
The great diversity in radio properties of our sources (i.e., radio luminosities, morphologies, and the presence of radio lobes), means there are no clear and robust defining characteristics for our type of sources. 

\item
The vast range in radio luminosities cannot be fully accounted for by the range in X-ray luminosities and BH masses, in the framework of the so-called ``fundamental plane'' (Fig.~\ref{fig:Fundplane}).

\item 
The BH masses of our sources cover $7.5 \lesssim \log(\mbh/\Msol) \lesssim 10.3$, and the accretion rates cover $-0.2 \lesssim \log\lamEdd \lesssim -2.2$ (Fig.~\ref{fig:Eddratio}). 
Thus, the most luminous obscured sources in the low redshift Universe are powered neither by particularly low-\mbh\ and high-\lamEdd\ SMBHs, nor by high-\mbh\ and low-\lamEdd\ ones.\footnote{The distribution of \lamEdd\ among our highly luminous sources {\it is} skewed towards higher \lamEdd\ than that of lower-luminosity BASS/DR1 AGN, which is expected given our high luminosity cut (see Section~\ref{sec:bhm}).}

\item
Based on the distribution of our sources in the $\lamEdd-\NH$ plane, we conclude that the most luminous obscured AGN in the local Universe are consistent with the radiative feedback driven unification scenario \citep{Fabian2008_rad_press,2017Natur.549..488R}, where AGN can remain obscured as long as their Eddington ratios are not high enough to expel the dusty obscuring circumnuclear gas, in contrast to the ``receding torus'' scenario (Fig. \ref{fig:Lum_NH_dists_LLEdd}).

\end{itemize}
 
We find that, as a group, our sample of some of the most luminous obscured AGN in BASS/DR1 does not exhibit any distinctive properties with respect to their black hole masses, Eddington ratios, and/or stellar masses of their host galaxies. 
Their host galaxies are all (or mostly) ellipticals, which is rather unexpected.
If this finding is corroborated by proper morphological decomposition of higher-quality, deeper multi-band imaging data, it may lend some indirect evidence in support of the popular idea that epochs of intense SMBH growth are linked to the transformation of galaxies from (star-forming) disks to (quenched) ellipticals (i.e., through major mergers).\\

The broad range of radio luminosities; the high fraction of radio-loud sources; and the high occurrence rate of double radio lobes among our sample of highly X-ray luminous, obscured AGN all suggest that intense, gas- and dust-rich SMBH growth may be linked to efficient launching of radio jets.
We speculate that this may be due to high SMBH spins enhancing the emission associated with both mechanisms.
However, the radio emission studied here probes timescales that are far longer than those associated with the nuclear X-ray emission (which is, in turn, linked to the ``instantenous'' accretion onto the SMBH).
Thus, to be able to test the possible links between intense (obscured) SMBH growth and jet launching in more detail, one would require to have a more comprehensive, homogeneous, and complete multi-frequency and high-resolution radio survey for the local AGN population, as represented in BASS.
In this context, we note the surprising dearth of (high-resolution) archival radio data for these extremely luminous and intriguing sources.

We indeed envision that such follow-up investigations will be undertaken as part of the ongoing BASS project. 
In particular, we are currently pursuing a large, arc-second resolution survey using the VLA, to probe the core radio emission of hundreds of BASS AGN (Smith et al., in prep.).

\section*{Acknowledgments} 

We thank the  anonymous referee, whose useful and detailed comments helped us to considerably improve this paper.
LFS, AKW and KS acknowledge support from the Swiss National Science Foundation (SNSF) Grants
PP00P2\textunderscore138979 and
PP00P2\textunderscore166159. 
MK acknowledges support from the SNSF through the Ambizione fellowship grant PZ00P2\textunderscore154799/1 and SNSF grant PP00P2\textunderscore138979/1, as well as support from NASA through ADAP award NNH16CT03C. 
KO and KS acknowledge support from the SNSF through Project grant 200021\textunderscore157021. 
KO acknowledges support from the Japan Society for the Promotion of Science (JSPS, ID: 17321). 
FR acknowledges support from FONDECYT Postdoctorado 3180506 and CONICYT project Basal AFB-170002.
FP acknowledges support from the NASA {\it Chandra} award No. AR8-19021A and from the Yale Keck program No. Y144.

This work made use of data products from the Wide-field Infrared Survey Explorer ({\it WISE}), which is a joint project of the University of Californfia, Los Angeles, and the Jet Propulsion Laboratory/California Institute of Technology, and {\it NEOWISE}, which is a project of the Jet Propulsion Laboratory/California Institute of Technology. {\it WISE} and {\it NEOWISE} are funded by the National Aeronautics and Space Administration.

This study was partially based on observations collected at the European Organisation for Astronomical Research in the Southern Hemisphere under ESO programmes 098.A-0635(B) and 099.A-0403(B).

This work has made use of the NASA/IPAC Extragalactic Database (NED) which is operated by the Jet Propulsion Laboratory, California Institute of Technology, under contract with the National Aeronautics and Space Administration. 
This research made use of the SIMBAD and VizieR \citep{2000A&AS..143...23O} online databases operated at CDS, Strasbourg, France. 
This research also made use of the data products from the Two Micron All Sky Survey (2MASS), which is a joint project of the University of Massachusetts and the Infrared Processing and Analysis Center/California Institute of Technology, funded by by NASA and the National Science Foundation.
This research made use of \texttt{Astropy}, a community-developed core Python package for Astronomy \citep{2013A&A...558A..33A}, \texttt{Matplotlib} \citep{Hunter:2007} and \texttt{NumPy} \citep{2011CSE....13b..22V}.
This research used the \texttt{TOPCAT} tool for catalogue cross-matching \citep{2005ASPC..347...29T}.

\clearpage

\bibliographystyle{mnras}
\bibliography{bibliography3}

\begin{thebibliography}{}
\makeatletter
\relax
\def\mn@urlcharsother{\let\do\@makeother \do\$\do\&\do\#\do\^\do\_\do\%\do\~}
\def\mn@doi{\begingroup\mn@urlcharsother \@ifnextchar [ {\mn@doi@}
  {\mn@doi@[]}}
\def\mn@doi@[#1]#2{\def\@tempa{#1}\ifx\@tempa\@empty \href
  {http://dx.doi.org/#2} {doi:#2}\else \href {http://dx.doi.org/#2} {#1}\fi
  \endgroup}
\def\mn@eprint#1#2{\mn@eprint@#1:#2::\@nil}
\def\mn@eprint@arXiv#1{\href {http://arxiv.org/abs/#1} {{\tt arXiv:#1}}}
\def\mn@eprint@dblp#1{\href {http://dblp.uni-trier.de/rec/bibtex/#1.xml}
  {dblp:#1}}
\def\mn@eprint@#1:#2:#3:#4\@nil{\def\@tempa {#1}\def\@tempb {#2}\def\@tempc
  {#3}\ifx \@tempc \@empty \let \@tempc \@tempb \let \@tempb \@tempa \fi \ifx
  \@tempb \@empty \def\@tempb {arXiv}\fi \@ifundefined
  {mn@eprint@\@tempb}{\@tempb:\@tempc}{\expandafter \expandafter \csname
  mn@eprint@\@tempb\endcsname \expandafter{\@tempc}}}

\bibitem[\protect\citeauthoryear{{Abazajian} et~al.,}{{Abazajian}
  et~al.}{2009}]{2009ApJS..182..543A}
{Abazajian} K.~N.,  et~al., 2009, \mn@doi [\apjs]
  {10.1088/0067-0049/182/2/543}, \href
  {http://adsabs.harvard.edu/abs/2009ApJS..182..543A} {182, 543}

\bibitem[\protect\citeauthoryear{{Abolfathi} et~al.,}{{Abolfathi}
  et~al.}{2018}]{2018ApJS..235...42A}
{Abolfathi} B.,  et~al., 2018, \mn@doi [\apjs] {10.3847/1538-4365/aa9e8a},
  \href {http://adsabs.harvard.edu/abs/2018ApJS..235...42A} {235, 42}

\bibitem[\protect\citeauthoryear{{Abramowicz} \& {Fragile}}{{Abramowicz} \&
  {Fragile}}{2013}]{Abramowicz2013}
{Abramowicz} M.~A.,  {Fragile} P.~C.,  2013, \mn@doi [Living Reviews in
  Relativity] {10.12942/lrr-2013-1}, \href
  {http://adsabs.harvard.edu/abs/2013LRR....16....1A} {16, 1}

\bibitem[\protect\citeauthoryear{{Acero} et~al.,}{{Acero}
  et~al.}{2015}]{2015ApJS..218...23A}
{Acero} F.,  et~al., 2015, \mn@doi [\apjs] {10.1088/0067-0049/218/2/23}, \href
  {http://adsabs.harvard.edu/abs/2015ApJS..218...23A} {218, 23}

\bibitem[\protect\citeauthoryear{{Adelman-McCarthy} et~al.,}{{Adelman-McCarthy}
  et~al.}{2008}]{2008ApJS..175..297A}
{Adelman-McCarthy} J.~K.,  et~al., 2008, \mn@doi [\apjs] {10.1086/524984},
  \href {http://adsabs.harvard.edu/abs/2008ApJS..175..297A} {175, 297}

\bibitem[\protect\citeauthoryear{{Aird}, {Coil}  \& {Georgakakis}}{{Aird}
  et~al.}{2018}]{2018MNRAS.474.1225A}
{Aird} J.,  {Coil} A.~L.,   {Georgakakis} A.,  2018, \mn@doi [\mnras]
  {10.1093/mnras/stx2700}, \href
  {http://adsabs.harvard.edu/abs/2018MNRAS.474.1225A} {474, 1225}

\bibitem[\protect\citeauthoryear{{Ajello} et~al.,}{{Ajello}
  et~al.}{2008}]{2008ApJ...673...96A}
{Ajello} M.,  et~al., 2008, \mn@doi [\apj] {10.1086/524104}, \href
  {http://adsabs.harvard.edu/abs/2008ApJ...673...96A} {673, 96}

\bibitem[\protect\citeauthoryear{{Akiyama} et~al.,}{{Akiyama}
  et~al.}{2018}]{2018PASJ...70S..34A}
{Akiyama} M.,  et~al., 2018, \mn@doi [\pasj] {10.1093/pasj/psx091}, \href
  {http://adsabs.harvard.edu/abs/2018PASJ...70S..34A} {70, S34}

\bibitem[\protect\citeauthoryear{{Akylas}, {Georgantopoulos}, {Ranalli},
  {Gkiokas}, {Corral}  \& {Lanzuisi}}{{Akylas}
  et~al.}{2016}]{2016A&A...594A..73A}
{Akylas} A.,  {Georgantopoulos} I.,  {Ranalli} P.,  {Gkiokas} E.,  {Corral} A.,
    {Lanzuisi} G.,  2016, \mn@doi [\aap] {10.1051/0004-6361/201628711}, \href
  {http://adsabs.harvard.edu/abs/2016A%26A...594A..73A} {594, A73}

\bibitem[\protect\citeauthoryear{{Alexandroff} et~al.,}{{Alexandroff}
  et~al.}{2013}]{2013MNRAS.435.3306A}
{Alexandroff} R.,  et~al., 2013, \mn@doi [\mnras] {10.1093/mnras/stt1500},
  \href {http://adsabs.harvard.edu/abs/2013MNRAS.435.3306A} {435, 3306}

\bibitem[\protect\citeauthoryear{{Anglada}, {Villuendas}, {Estalella},
  {Beltr{\'a}n}, {Rodr{\'{\i}}guez}, {Torrelles}  \& {Curiel}}{{Anglada}
  et~al.}{1998}]{1998AJ....116.2953A}
{Anglada} G.,  {Villuendas} E.,  {Estalella} R.,  {Beltr{\'a}n} M.~T.,
  {Rodr{\'{\i}}guez} L.~F.,  {Torrelles} J.~M.,   {Curiel} S.,  1998, \mn@doi
  [\aj] {10.1086/300637}, \href
  {http://adsabs.harvard.edu/abs/1998AJ....116.2953A} {116, 2953}

\bibitem[\protect\citeauthoryear{Antonucci}{Antonucci}{1993}]{Antonucci1993}
Antonucci R.,  1993, \mn@doi [\araa] {10.1146/annurev.aa.31.090193.002353}, 31,
  473

\bibitem[\protect\citeauthoryear{{Assef} et~al.,}{{Assef}
  et~al.}{2010}]{Assef2010_SED_templates}
{Assef} R.~J.,  et~al., 2010, \mn@doi [\apj] {10.1088/0004-637X/713/2/970},
  \href {http://adsabs.harvard.edu/abs/2010ApJ...713..970A} {713, 970}

\bibitem[\protect\citeauthoryear{Assef et~al.,}{Assef
  et~al.}{2015}]{Assef2015_HotDOGs}
Assef R.~J.,  et~al., 2015, \mn@doi [\apj] {10.1088/0004-637X/804/1/27}, 804,
  27

\bibitem[\protect\citeauthoryear{{Astropy Collaboration} et~al.,}{{Astropy
  Collaboration} et~al.}{2013}]{2013A&A...558A..33A}
{Astropy Collaboration} et~al., 2013, \mn@doi [\aap]
  {10.1051/0004-6361/201322068}, \href
  {http://adsabs.harvard.edu/abs/2013A%26A...558A..33A} {558, A33}

\bibitem[\protect\citeauthoryear{{Audibert}, {Riffel}, {Sales}, {Pastoriza}  \&
  {Ruschel-Dutra}}{{Audibert} et~al.}{2017}]{2017MNRAS.464.2139A}
{Audibert} A.,  {Riffel} R.,  {Sales} D.~A.,  {Pastoriza} M.~G.,
  {Ruschel-Dutra} D.,  2017, \mn@doi [\mnras] {10.1093/mnras/stw2477}, \href
  {http://adsabs.harvard.edu/abs/2017MNRAS.464.2139A} {464, 2139}

\bibitem[\protect\citeauthoryear{Bahcall, Kirhakos, Saxe  \& Schneider}{Bahcall
  et~al.}{1997}]{Bahcall1997}
Bahcall J.~N.,  Kirhakos S.,  Saxe D.~H.,   Schneider D.~P.,  1997, \mn@doi
  [\apj] {10.1086/303926}, 479, 642

\bibitem[\protect\citeauthoryear{Baldry, Glazebrook, Brinkmann, Ivezi{\'{c}},
  Lupton, Nichol  \& Szalay}{Baldry et~al.}{2004}]{Baldry2004_galaxy_bimod}
Baldry I.~K.,  Glazebrook K.,  Brinkmann J.,  Ivezi{\'{c}} {\v{Z}}.,  Lupton
  R.~H.,  Nichol R.~C.,   Szalay A.~S.,  2004, \mn@doi [\apj] {10.1086/380092},
  600, 681

\bibitem[\protect\citeauthoryear{{Baldwin}, {Phillips}  \&
  {Terlevich}}{{Baldwin} et~al.}{1981}]{1981PASP...93....5B}
{Baldwin} J.~A.,  {Phillips} M.~M.,   {Terlevich} R.,  1981, \mn@doi [\pasp]
  {10.1086/130766}, \href {http://adsabs.harvard.edu/abs/1981PASP...93....5B}
  {93, 5}

\bibitem[\protect\citeauthoryear{{Ballo}, {Heras}, {Barcons}  \&
  {Carrera}}{{Ballo} et~al.}{2012}]{2012A&A...545A..66B}
{Ballo} L.,  {Heras} F.~J.~H.,  {Barcons} X.,   {Carrera} F.~J.,  2012, \mn@doi
  [\aap] {10.1051/0004-6361/201117464}, \href
  {http://adsabs.harvard.edu/abs/2012A%26A...545A..66B} {545, A66}

\bibitem[\protect\citeauthoryear{{Balmaverde} et~al.,}{{Balmaverde}
  et~al.}{2012}]{2012A&A...545A.143B}
{Balmaverde} B.,  et~al., 2012, \mn@doi [\aap] {10.1051/0004-6361/201219561},
  \href {http://adsabs.harvard.edu/abs/2012A%26A...545A.143B} {545, A143}

\bibitem[\protect\citeauthoryear{{Banerji}, {McMahon}, {Hewett},
  {Alaghband-Zadeh}, {Gonzalez-Solares}, {Venemans}  \& {Hawthorn}}{{Banerji}
  et~al.}{2012}]{Banerji2012_red_QSO_UKIDSS}
{Banerji} M.,  {McMahon} R.~G.,  {Hewett} P.~C.,  {Alaghband-Zadeh} S.,
  {Gonzalez-Solares} E.,  {Venemans} B.~P.,   {Hawthorn} M.~J.,  2012, \mn@doi
  [\mnras] {10.1111/j.1365-2966.2012.22099.x}, \href
  {http://adsabs.harvard.edu/abs/2012MNRAS.427.2275B} {427, 2275}

\bibitem[\protect\citeauthoryear{{Banerji}, {Alaghband-Zadeh}, {Hewett}  \&
  {McMahon}}{{Banerji} et~al.}{2015}]{Banerji2015_red_QSO_MBH}
{Banerji} M.,  {Alaghband-Zadeh} S.,  {Hewett} P.~C.,   {McMahon} R.~G.,  2015,
  \mn@doi [\mnras] {10.1093/mnras/stu2649}, \href
  {http://adsabs.harvard.edu/abs/2015MNRAS.447.3368B} {447, 3368}

\bibitem[\protect\citeauthoryear{{Banerji}, {Carilli}, {Jones}, {Wagg},
  {McMahon}, {Hewett}, {Alaghband-Zadeh}  \& {Feruglio}}{{Banerji}
  et~al.}{2017}]{Banerji2017_red_QSO_ALMA}
{Banerji} M.,  {Carilli} C.~L.,  {Jones} G.,  {Wagg} J.,  {McMahon} R.~G.,
  {Hewett} P.~C.,  {Alaghband-Zadeh} S.,   {Feruglio} C.,  2017, \mn@doi
  [\mnras] {10.1093/mnras/stw3019}, \href
  {http://adsabs.harvard.edu/abs/2017MNRAS.465.4390B} {465, 4390}

\bibitem[\protect\citeauthoryear{{Barthelmy} et~al.,}{{Barthelmy}
  et~al.}{2005}]{2005SSRv..120..143B}
{Barthelmy} S.~D.,  et~al., 2005, \mn@doi [\ssr] {10.1007/s11214-005-5096-3},
  \href {http://adsabs.harvard.edu/abs/2005SSRv..120..143B} {120, 143}

\bibitem[\protect\citeauthoryear{Baumgartner, Tueller, Markwardt, Skinner,
  Barthelmy, Mushotzky, Evans  \& Gehrels}{Baumgartner
  et~al.}{2013}]{Baumgartner2013_SwiftBAT_70m}
Baumgartner W.~H.,  Tueller J.,  Markwardt C.~B.,  Skinner G.,  Barthelmy S.,
  Mushotzky R.~F.,  Evans P.~A.,   Gehrels N.,  2013, \mn@doi [\apjs]
  {10.1088/0067-0049/207/2/19}, 207, 19

\bibitem[\protect\citeauthoryear{{Becker}, {White}  \& {Helfand}}{{Becker}
  et~al.}{1995}]{1995ApJ...450..559B}
{Becker} R.~H.,  {White} R.~L.,   {Helfand} D.~J.,  1995, \mn@doi [\apj]
  {10.1086/176166}, \href {http://adsabs.harvard.edu/abs/1995ApJ...450..559B}
  {450, 559}

\bibitem[\protect\citeauthoryear{{Bell}}{{Bell}}{2003}]{2003ApJ...586..794B}
{Bell} E.~F.,  2003, \mn@doi [\apj] {10.1086/367829}, \href
  {http://adsabs.harvard.edu/abs/2003ApJ...586..794B} {586, 794}

\bibitem[\protect\citeauthoryear{{Bell}, {McIntosh}, {Katz}  \&
  {Weinberg}}{{Bell} et~al.}{2003a}]{2003ApJS..149..289B}
{Bell} E.~F.,  {McIntosh} D.~H.,  {Katz} N.,   {Weinberg} M.~D.,  2003a,
  \mn@doi [\apjs] {10.1086/378847}, \href
  {http://adsabs.harvard.edu/abs/2003ApJS..149..289B} {149, 289}

\bibitem[\protect\citeauthoryear{Bell, McIntosh, Katz  \& Weinberg}{Bell
  et~al.}{2003b}]{Bell2003_LF_MF}
Bell E.~F.,  McIntosh D.~H.,  Katz N.,   Weinberg M.~D.,  2003b, \mn@doi
  [\apjs] {10.1086/378847}, 149, 289

\bibitem[\protect\citeauthoryear{Berney et~al.,}{Berney
  et~al.}{2015}]{Berney2015_BASS}
Berney S.,  et~al., 2015, \mn@doi [\mnras] {10.1093/mnras/stv2181}, 454, 3622

\bibitem[\protect\citeauthoryear{{Best}}{{Best}}{2004}]{2004MNRAS.351...70B}
{Best} P.~N.,  2004, \mn@doi [\mnras] {10.1111/j.1365-2966.2004.07752.x}, \href
  {http://adsabs.harvard.edu/abs/2004MNRAS.351...70B} {351, 70}

\bibitem[\protect\citeauthoryear{{Best}, {Kauffmann}, {Heckman}  \&
  {Ivezi{\'c}}}{{Best} et~al.}{2005a}]{2005MNRAS.362....9B}
{Best} P.~N.,  {Kauffmann} G.,  {Heckman} T.~M.,   {Ivezi{\'c}} {\v Z}.,
  2005a, \mn@doi [\mnras] {10.1111/j.1365-2966.2005.09283.x}, \href
  {http://adsabs.harvard.edu/abs/2005MNRAS.362....9B} {362, 9}

\bibitem[\protect\citeauthoryear{{Best}, {Kauffmann}, {Heckman}, {Brinchmann},
  {Charlot}, {Ivezi{\'c}}  \& {White}}{{Best}
  et~al.}{2005b}]{2005MNRAS.362...25B}
{Best} P.~N.,  {Kauffmann} G.,  {Heckman} T.~M.,  {Brinchmann} J.,  {Charlot}
  S.,  {Ivezi{\'c}} {\v Z}.,   {White} S.~D.~M.,  2005b, \mn@doi [\mnras]
  {10.1111/j.1365-2966.2005.09192.x}, \href
  {http://adsabs.harvard.edu/abs/2005MNRAS.362...25B} {362, 25}

\bibitem[\protect\citeauthoryear{{Blandford} \& {K{\"o}nigl}}{{Blandford} \&
  {K{\"o}nigl}}{1979}]{1979ApJ...232...34B}
{Blandford} R.~D.,  {K{\"o}nigl} A.,  1979, \mn@doi [\apj] {10.1086/157262},
  \href {http://adsabs.harvard.edu/abs/1979ApJ...232...34B} {232, 34}

\bibitem[\protect\citeauthoryear{{Blandford} \& {Znajek}}{{Blandford} \&
  {Znajek}}{1977}]{BlandfordZnajek1977}
{Blandford} R.~D.,  {Znajek} R.~L.,  1977, \mn@doi [\mnras]
  {10.1093/mnras/179.3.433}, \href
  {https://ui.adsabs.harvard.edu/abs/1977MNRAS.179..433B} {179, 433}

\bibitem[\protect\citeauthoryear{{Blandford}, {Meier}  \&
  {Readhead}}{{Blandford} et~al.}{2018}]{2018arXiv181206025B}
{Blandford} R.,  {Meier} D.,   {Readhead} A.,  2018, \araa, \href
  {https://ui.adsabs.harvard.edu/abs/2018arXiv181206025B} {p. arXiv:1812.06025}

\bibitem[\protect\citeauthoryear{{Blanton} et~al.,}{{Blanton}
  et~al.}{2005}]{2005AJ....129.2562B}
{Blanton} M.~R.,  et~al., 2005, \mn@doi [\aj] {10.1086/429803}, \href
  {http://adsabs.harvard.edu/abs/2005AJ....129.2562B} {129, 2562}

\bibitem[\protect\citeauthoryear{{Bonchi}, {La Franca}, {Melini}, {Bongiorno}
  \& {Fiore}}{{Bonchi} et~al.}{2013}]{2013MNRAS.429.1970B}
{Bonchi} A.,  {La Franca} F.,  {Melini} G.,  {Bongiorno} A.,   {Fiore} F.,
  2013, \mn@doi [\mnras] {10.1093/mnras/sts456}, \href
  {http://adsabs.harvard.edu/abs/2013MNRAS.429.1970B} {429, 1970}

\bibitem[\protect\citeauthoryear{{Brandt} \& {Alexander}}{{Brandt} \&
  {Alexander}}{2015}]{2015A&ARv..23....1B}
{Brandt} W.~N.,  {Alexander} D.~M.,  2015, \mn@doi [\aapr]
  {10.1007/s00159-014-0081-z}, \href
  {http://adsabs.harvard.edu/abs/2015A%26ARv..23....1B} {23, 1}

\bibitem[\protect\citeauthoryear{{Burlon}, {Ghirlanda}, {Murphy}, {Chhetri},
  {Sadler}  \& {Ajello}}{{Burlon} et~al.}{2013}]{2013MNRAS.431.2471B}
{Burlon} D.,  {Ghirlanda} G.,  {Murphy} T.,  {Chhetri} R.,  {Sadler} E.,
  {Ajello} M.,  2013, \mn@doi [\mnras] {10.1093/mnras/stt343}, \href
  {http://adsabs.harvard.edu/abs/2013MNRAS.431.2471B} {431, 2471}

\bibitem[\protect\citeauthoryear{{Callingham} et~al.,}{{Callingham}
  et~al.}{2015}]{2015ApJ...809..168C}
{Callingham} J.~R.,  et~al., 2015, \mn@doi [\apj]
  {10.1088/0004-637X/809/2/168}, \href
  {http://adsabs.harvard.edu/abs/2015ApJ...809..168C} {809, 168}

\bibitem[\protect\citeauthoryear{Capellupo, Netzer, Lira  \&
  Trakhtenbrot}{Capellupo et~al.}{2015}]{Capellupo2015_ADs}
Capellupo D.~M.,  Netzer H.,  Lira P.,   Trakhtenbrot B.,  2015, \mn@doi
  [\mnras] {10.1093/mnras/stu2266}, 446, 3427

\bibitem[\protect\citeauthoryear{Caplar, Lilly  \& Trakhtenbrot}{Caplar
  et~al.}{2015}]{Caplar2015_coeval}
Caplar N.,  Lilly S.~J.,   Trakhtenbrot B.,  2015, \mn@doi [\apj]
  {10.1088/0004-637X/811/2/148}, 811, 148

\bibitem[\protect\citeauthoryear{{Cappellari} \& {Emsellem}}{{Cappellari} \&
  {Emsellem}}{2004}]{2004PASP..116..138C}
{Cappellari} M.,  {Emsellem} E.,  2004, \mn@doi [\pasp] {10.1086/381875}, \href
  {http://adsabs.harvard.edu/abs/2004PASP..116..138C} {116, 138}

\bibitem[\protect\citeauthoryear{{Carilli}, {Perley}, {Dreher}  \&
  {Leahy}}{{Carilli} et~al.}{1991}]{1991ApJ...383..554C}
{Carilli} C.~L.,  {Perley} R.~A.,  {Dreher} J.~W.,   {Leahy} J.~P.,  1991,
  \mn@doi [\apj] {10.1086/170813}, \href
  {http://adsabs.harvard.edu/abs/1991ApJ...383..554C} {383, 554}

\bibitem[\protect\citeauthoryear{{Castell{\'o}-Mor}, {Netzer}  \&
  {Kaspi}}{{Castell{\'o}-Mor} et~al.}{2016}]{2016MNRAS.458.1839C}
{Castell{\'o}-Mor} N.,  {Netzer} H.,   {Kaspi} S.,  2016, \mn@doi [\mnras]
  {10.1093/mnras/stw445}, \href
  {http://adsabs.harvard.edu/abs/2016MNRAS.458.1839C} {458, 1839}

\bibitem[\protect\citeauthoryear{{Condon}, {Cotton}, {Greisen}, {Yin},
  {Perley}, {Taylor}  \& {Broderick}}{{Condon}
  et~al.}{1998}]{1998AJ....115.1693C}
{Condon} J.~J.,  {Cotton} W.~D.,  {Greisen} E.~W.,  {Yin} Q.~F.,  {Perley}
  R.~A.,  {Taylor} G.~B.,   {Broderick} J.~J.,  1998, \mn@doi [\aj]
  {10.1086/300337}, \href {http://adsabs.harvard.edu/abs/1998AJ....115.1693C}
  {115, 1693}

\bibitem[\protect\citeauthoryear{{Croom}, {Smith}, {Boyle}, {Shanks}, {Miller},
  {Outram}  \& {Loaring}}{{Croom} et~al.}{2004}]{2004MNRAS.349.1397C}
{Croom} S.~M.,  {Smith} R.~J.,  {Boyle} B.~J.,  {Shanks} T.,  {Miller} L.,
  {Outram} P.~J.,   {Loaring} N.~S.,  2004, \mn@doi [\mnras]
  {10.1111/j.1365-2966.2004.07619.x}, \href
  {http://adsabs.harvard.edu/abs/2004MNRAS.349.1397C} {349, 1397}

\bibitem[\protect\citeauthoryear{{Daly}, {Stout}  \& {Mysliwiec}}{{Daly}
  et~al.}{2018}]{2018ApJ...863..117D}
{Daly} R.~A.,  {Stout} D.~A.,   {Mysliwiec} J.~N.,  2018, \mn@doi [\apj]
  {10.3847/1538-4357/aad08b}, \href
  {https://ui.adsabs.harvard.edu/abs/2018ApJ...863..117D} {863, 117}

\bibitem[\protect\citeauthoryear{{Davies} et~al.,}{{Davies}
  et~al.}{2017}]{2017MNRAS.466.4917D}
{Davies} R.~I.,  et~al., 2017, \mn@doi [\mnras] {10.1093/mnras/stx045}, \href
  {http://adsabs.harvard.edu/abs/2017MNRAS.466.4917D} {466, 4917}

\bibitem[\protect\citeauthoryear{{Deeley} et~al.,}{{Deeley}
  et~al.}{2017}]{2017MNRAS.467.3934D}
{Deeley} S.,  et~al., 2017, \mn@doi [\mnras] {10.1093/mnras/stx441}, \href
  {http://adsabs.harvard.edu/abs/2017MNRAS.467.3934D} {467, 3934}

\bibitem[\protect\citeauthoryear{{Fabian}, {Celotti}  \& {Erlund}}{{Fabian}
  et~al.}{2006}]{2006MNRAS.373L..16F}
{Fabian} A.~C.,  {Celotti} A.,   {Erlund} M.~C.,  2006, \mn@doi [\mnras]
  {10.1111/j.1745-3933.2006.00234.x}, \href
  {http://adsabs.harvard.edu/abs/2006MNRAS.373L..16F} {373, L16}

\bibitem[\protect\citeauthoryear{Fabian, Vasudevan  \& Gandhi}{Fabian
  et~al.}{2008}]{Fabian2008_rad_press}
Fabian A.~C.,  Vasudevan R.~V.,   Gandhi P.,  2008, \mn@doi [\mnras]
  {10.1111/j.1745-3933.2008.00430.x}, 385, L43

\bibitem[\protect\citeauthoryear{{Fabian}, {Vasudevan}, {Mushotzky}, {Winter}
  \& {Reynolds}}{{Fabian} et~al.}{2009}]{2009MNRAS.394L..89F}
{Fabian} A.~C.,  {Vasudevan} R.~V.,  {Mushotzky} R.~F.,  {Winter} L.~M.,
  {Reynolds} C.~S.,  2009, \mn@doi [\mnras] {10.1111/j.1745-3933.2009.00617.x},
  \href {http://adsabs.harvard.edu/abs/2009MNRAS.394L..89F} {394, L89}

\bibitem[\protect\citeauthoryear{{Falcke} \& {Biermann}}{{Falcke} \&
  {Biermann}}{1995}]{1995A&A...293..665F}
{Falcke} H.,  {Biermann} P.~L.,  1995, \aap, \href
  {http://adsabs.harvard.edu/abs/1995A%26A...293..665F} {293, 665}

\bibitem[\protect\citeauthoryear{{Falcke}, {K{\"o}rding}  \&
  {Markoff}}{{Falcke} et~al.}{2004}]{2004A&A...414..895F}
{Falcke} H.,  {K{\"o}rding} E.,   {Markoff} S.,  2004, \mn@doi [\aap]
  {10.1051/0004-6361:20031683}, \href
  {http://adsabs.harvard.edu/abs/2004A%26A...414..895F} {414, 895}

\bibitem[\protect\citeauthoryear{{Fanaroff} \& {Riley}}{{Fanaroff} \&
  {Riley}}{1974}]{1974MNRAS.167P..31F}
{Fanaroff} B.~L.,  {Riley} J.~M.,  1974, \mn@doi [\mnras]
  {10.1093/mnras/167.1.31P}, \href
  {http://adsabs.harvard.edu/abs/1974MNRAS.167P..31F} {167, 31P}

\bibitem[\protect\citeauthoryear{{Flesch}}{{Flesch}}{2016}]{2016PASA...33...52F}
{Flesch} E.~W.,  2016, \mn@doi [\pasa] {10.1017/pasa.2016.44}, \href
  {http://adsabs.harvard.edu/abs/2016PASA...33...52F} {33, e052}

\bibitem[\protect\citeauthoryear{{Flewelling} et~al.,}{{Flewelling}
  et~al.}{2016}]{2016arXiv161205243F}
{Flewelling} H.~A.,  et~al., 2016, preprint, \href
  {http://adsabs.harvard.edu/abs/2016arXiv161205243F} {} (\mn@eprint {arXiv}
  {1612.05243})

\bibitem[\protect\citeauthoryear{{Freudling}, {Romaniello}, {Bramich},
  {Ballester}, {Forchi}, {Garc{\'{\i}}a-Dabl{\'o}}, {Moehler}  \&
  {Neeser}}{{Freudling} et~al.}{2013}]{2013A&A...559A..96F}
{Freudling} W.,  {Romaniello} M.,  {Bramich} D.~M.,  {Ballester} P.,  {Forchi}
  V.,  {Garc{\'{\i}}a-Dabl{\'o}} C.~E.,  {Moehler} S.,   {Neeser} M.~J.,  2013,
  \mn@doi [\aap] {10.1051/0004-6361/201322494}, \href
  {http://adsabs.harvard.edu/abs/2013A%26A...559A..96F} {559, A96}

\bibitem[\protect\citeauthoryear{{Gehrels} et~al.,}{{Gehrels}
  et~al.}{2004}]{2004ApJ...611.1005G}
{Gehrels} N.,  et~al., 2004, \mn@doi [\apj] {10.1086/422091}, \href
  {http://adsabs.harvard.edu/abs/2004ApJ...611.1005G} {611, 1005}

\bibitem[\protect\citeauthoryear{{Ghisellini}, {Tavecchio}, {Maraschi},
  {Celotti}  \& {Sbarrato}}{{Ghisellini}
  et~al.}{2014}]{Ghisellini2014_jets_disks}
{Ghisellini} G.,  {Tavecchio} F.,  {Maraschi} L.,  {Celotti} A.,   {Sbarrato}
  T.,  2014, \mn@doi [\nat] {10.1038/nature13856}, \href
  {https://ui.adsabs.harvard.edu/abs/2014Natur.515..376G} {515, 376}

\bibitem[\protect\citeauthoryear{{Glikman}}{{Glikman}}{2017}]{Glikman2017_red_QSO_note}
{Glikman} E.,  2017, \mn@doi [Research Notes of the American Astronomical
  Society] {10.3847/2515-5172/aaa0c0}, \href
  {http://adsabs.harvard.edu/abs/2017RNAAS...1a..48G} {1, 48}

\bibitem[\protect\citeauthoryear{{Glikman}, {Helfand}, {White}, {Becker},
  {Gregg}  \& {Lacy}}{{Glikman} et~al.}{2007}]{Glikman2007_red_QSO_FIRST}
{Glikman} E.,  {Helfand} D.~J.,  {White} R.~L.,  {Becker} R.~H.,  {Gregg}
  M.~D.,   {Lacy} M.,  2007, \mn@doi [\apj] {10.1086/521073}, \href
  {http://adsabs.harvard.edu/abs/2007ApJ...667..673G} {667, 673}

\bibitem[\protect\citeauthoryear{{Glikman}, {Simmons}, {Mailly}, {Schawinski},
  {Urry}  \& {Lacy}}{{Glikman} et~al.}{2015}]{Glikman2015_red_QSO_mergers}
{Glikman} E.,  {Simmons} B.,  {Mailly} M.,  {Schawinski} K.,  {Urry} C.~M.,
  {Lacy} M.,  2015, \mn@doi [\apj] {10.1088/0004-637X/806/2/218}, \href
  {http://adsabs.harvard.edu/abs/2015ApJ...806..218G} {806, 218}

\bibitem[\protect\citeauthoryear{{Goulding} et~al.,}{{Goulding}
  et~al.}{2018}]{Goulding2018_red_QSO_xray}
{Goulding} A.~D.,  et~al., 2018, \mn@doi [\apj] {10.3847/1538-4357/aab040},
  \href {http://adsabs.harvard.edu/abs/2018ApJ...856....4G} {856, 4}

\bibitem[\protect\citeauthoryear{{Graham} \& {Scott}}{{Graham} \&
  {Scott}}{2013}]{2013ApJ...764..151G}
{Graham} A.~W.,  {Scott} N.,  2013, \mn@doi [\apj]
  {10.1088/0004-637X/764/2/151}, \href
  {http://adsabs.harvard.edu/abs/2013ApJ...764..151G} {764, 151}

\bibitem[\protect\citeauthoryear{Greene, Zakamska, Ho  \& Barth}{Greene
  et~al.}{2011}]{Greene2011_Q2_outflows}
Greene J.~E.,  Zakamska N.~L.,  Ho L.~C.,   Barth A.~J.,  2011, \mn@doi [\apj]
  {10.1088/0004-637X/732/1/9}, 732

\bibitem[\protect\citeauthoryear{Grier et~al.,}{Grier
  et~al.}{2013}]{Grier2013_sigs_PGs}
Grier C.~J.,  et~al., 2013, \mn@doi [\apj] {10.1088/0004-637X/773/2/90}, 773,
  90

\bibitem[\protect\citeauthoryear{G{\"{u}}ltekin et~al.,}{G{\"{u}}ltekin
  et~al.}{2009a}]{Gultekin2009}
G{\"{u}}ltekin K.,  et~al., 2009a, \mn@doi [\apj]
  {10.1088/0004-637X/698/1/198}, 698, 198

\bibitem[\protect\citeauthoryear{{G{\"u}ltekin}, {Cackett}, {Miller}, {Di
  Matteo}, {Markoff}  \& {Richstone}}{{G{\"u}ltekin}
  et~al.}{2009b}]{2009ApJ...706..404G}
{G{\"u}ltekin} K.,  {Cackett} E.~M.,  {Miller} J.~M.,  {Di Matteo} T.,
  {Markoff} S.,   {Richstone} D.~O.,  2009b, \mn@doi [\apj]
  {10.1088/0004-637X/706/1/404}, \href
  {http://adsabs.harvard.edu/abs/2009ApJ...706..404G} {706, 404}

\bibitem[\protect\citeauthoryear{Harrison}{Harrison}{2017}]{Harrison2017_BH_SF_review}
Harrison C.~M.,  2017, \mn@doi [Nature Astronomy] {10.1038/s41550-017-0165}, 1,
  0165

\bibitem[\protect\citeauthoryear{{Hart} et~al.,}{{Hart}
  et~al.}{2016}]{2016MNRAS.461.3663H}
{Hart} R.~E.,  et~al., 2016, \mn@doi [\mnras] {10.1093/mnras/stw1588}, \href
  {http://adsabs.harvard.edu/abs/2016MNRAS.461.3663H} {461, 3663}

\bibitem[\protect\citeauthoryear{{Hasinger}, {Miyaji}  \& {Schmidt}}{{Hasinger}
  et~al.}{2005}]{2005A&A...441..417H}
{Hasinger} G.,  {Miyaji} T.,   {Schmidt} M.,  2005, \mn@doi [\aap]
  {10.1051/0004-6361:20042134}, \href
  {http://adsabs.harvard.edu/abs/2005A%26A...441..417H} {441, 417}

\bibitem[\protect\citeauthoryear{{Heckman} \& {Best}}{{Heckman} \&
  {Best}}{2014}]{2014ARA&A..52..589H}
{Heckman} T.~M.,  {Best} P.~N.,  2014, \mn@doi [\araa]
  {10.1146/annurev-astro-081913-035722}, \href
  {http://adsabs.harvard.edu/abs/2014ARA%26A..52..589H} {52, 589}

\bibitem[\protect\citeauthoryear{{Heinz} \& {Sunyaev}}{{Heinz} \&
  {Sunyaev}}{2003}]{2003MNRAS.343L..59H}
{Heinz} S.,  {Sunyaev} R.~A.,  2003, \mn@doi [\mnras]
  {10.1046/j.1365-8711.2003.06918.x}, \href
  {http://adsabs.harvard.edu/abs/2003MNRAS.343L..59H} {343, L59}

\bibitem[\protect\citeauthoryear{Hickox \& Alexander}{Hickox \&
  Alexander}{2018}]{HickoxAlexander2018_obs_AGN_rev}
Hickox R.~C.,  Alexander D.~M.,  2018, \mn@doi [\araa]
  {10.1146/annurev-astro-081817-051803}, 56, 625

\bibitem[\protect\citeauthoryear{{Hickox}, {Mullaney}, {Alexander}, {Chen},
  {Civano}, {Goulding}  \& {Hainline}}{{Hickox}
  et~al.}{2014}]{2014ApJ...782....9H}
{Hickox} R.~C.,  {Mullaney} J.~R.,  {Alexander} D.~M.,  {Chen} C.-T.~J.,
  {Civano} F.~M.,  {Goulding} A.~D.,   {Hainline} K.~N.,  2014, \mn@doi [\apj]
  {10.1088/0004-637X/782/1/9}, \href
  {http://adsabs.harvard.edu/abs/2014ApJ...782....9H} {782, 9}

\bibitem[\protect\citeauthoryear{{Hickox}, {Lamassa}, {Silverman}  \&
  {Kolodzig}}{{Hickox} et~al.}{2016}]{2016IAUFM..29B.113H}
{Hickox} R.~C.,  {Lamassa} S.~M.,  {Silverman} J.~D.,   {Kolodzig} A.,  2016,
  \mn@doi [IAU Focus Meeting] {10.1017/S1743921316004592}, \href
  {http://adsabs.harvard.edu/abs/2016IAUFM..29B.113H} {29, 113}

\bibitem[\protect\citeauthoryear{Hunter}{Hunter}{2007}]{Hunter:2007}
Hunter J.~D.,  2007, \mn@doi [Computing In Science \& Engineering]
  {10.1109/MCSE.2007.55}, 9, 90

\bibitem[\protect\citeauthoryear{{Ichikawa}, {Ricci}, {Ueda}, {Matsuoka},
  {Toba}, {Kawamuro}, {Trakhtenbrot}  \& {Koss}}{{Ichikawa}
  et~al.}{2017}]{2017ApJ...835...74I}
{Ichikawa} K.,  {Ricci} C.,  {Ueda} Y.,  {Matsuoka} K.,  {Toba} Y.,  {Kawamuro}
  T.,  {Trakhtenbrot} B.,   {Koss} M.~J.,  2017, \mn@doi [\apj]
  {10.3847/1538-4357/835/1/74}, \href
  {http://adsabs.harvard.edu/abs/2017ApJ...835...74I} {835, 74}

\bibitem[\protect\citeauthoryear{{Ichikawa} et~al.,}{{Ichikawa}
  et~al.}{2019}]{Ichikawa2019}
{Ichikawa} K.,  et~al., 2019, \mn@doi [\apj] {10.3847/1538-4357/aaef8f}, \href
  {http://adsabs.harvard.edu/abs/2019ApJ...870...31I} {870, 31}

\bibitem[\protect\citeauthoryear{{Inayoshi} \& {Haiman}}{{Inayoshi} \&
  {Haiman}}{2016}]{Inayoshi2016_maxMBH}
{Inayoshi} K.,  {Haiman} Z.,  2016, \mn@doi [\apj]
  {10.3847/0004-637X/828/2/110}, \href
  {https://ui.adsabs.harvard.edu/abs/2016ApJ...828..110I} {828, 110}

\bibitem[\protect\citeauthoryear{{Jarrett} et~al.,}{{Jarrett}
  et~al.}{2011}]{2011ApJ...735..112J}
{Jarrett} T.~H.,  et~al., 2011, \mn@doi [\apj] {10.1088/0004-637X/735/2/112},
  \href {http://adsabs.harvard.edu/abs/2011ApJ...735..112J} {735, 112}

\bibitem[\protect\citeauthoryear{{Jarrett} et~al.,}{{Jarrett}
  et~al.}{2012}]{2012AJ....144...68J}
{Jarrett} T.~H.,  et~al., 2012, \mn@doi [\aj] {10.1088/0004-6256/144/2/68},
  \href {http://adsabs.harvard.edu/abs/2012AJ....144...68J} {144, 68}

\bibitem[\protect\citeauthoryear{Jin, Ward  \& Done}{Jin
  et~al.}{2012}]{Jin2012_Xop_3_SEDs}
Jin C.,  Ward M.,   Done C.,  2012, \mn@doi [\mnras]
  {10.1111/j.1365-2966.2012.21272.x}, 425, 907

\bibitem[\protect\citeauthoryear{{Kaneda} et~al.,}{{Kaneda}
  et~al.}{1995}]{1995ApJ...453L..13K}
{Kaneda} H.,  et~al., 1995, \mn@doi [\apjl] {10.1086/309742}, \href
  {http://adsabs.harvard.edu/abs/1995ApJ...453L..13K} {453, L13}

\bibitem[\protect\citeauthoryear{{Katz-Stone}, {Rudnick}  \&
  {Anderson}}{{Katz-Stone} et~al.}{1993}]{1993ApJ...407..549K}
{Katz-Stone} D.~M.,  {Rudnick} L.,   {Anderson} M.~C.,  1993, \mn@doi [\apj]
  {10.1086/172536}, \href {http://adsabs.harvard.edu/abs/1993ApJ...407..549K}
  {407, 549}

\bibitem[\protect\citeauthoryear{{Kauffmann} et~al.,}{{Kauffmann}
  et~al.}{2003}]{2003MNRAS.341...33K}
{Kauffmann} G.,  et~al., 2003, \mn@doi [\mnras]
  {10.1046/j.1365-8711.2003.06291.x}, \href
  {http://adsabs.harvard.edu/abs/2003MNRAS.341...33K} {341, 33}

\bibitem[\protect\citeauthoryear{{Kellermann}, {Sramek}, {Schmidt}, {Shaffer}
  \& {Green}}{{Kellermann} et~al.}{1989}]{1989AJ.....98.1195K}
{Kellermann} K.~I.,  {Sramek} R.,  {Schmidt} M.,  {Shaffer} D.~B.,   {Green}
  R.,  1989, \mn@doi [\aj] {10.1086/115207}, \href
  {http://adsabs.harvard.edu/abs/1989AJ.....98.1195K} {98, 1195}

\bibitem[\protect\citeauthoryear{{Kellermann}, {Condon}, {Kimball}, {Perley}
  \& {Ivezi{\'c}}}{{Kellermann} et~al.}{2016}]{2016ApJ...831..168K}
{Kellermann} K.~I.,  {Condon} J.~J.,  {Kimball} A.~E.,  {Perley} R.~A.,
  {Ivezi{\'c}} {\v Z}.,  2016, \mn@doi [\apj] {10.3847/0004-637X/831/2/168},
  \href {http://adsabs.harvard.edu/abs/2016ApJ...831..168K} {831, 168}

\bibitem[\protect\citeauthoryear{Kelly \& Shen}{Kelly \&
  Shen}{2013}]{KellyShen2013_BHMF}
Kelly B.~C.,  Shen Y.,  2013, \mn@doi [\apj] {10.1088/0004-637X/764/1/45}, 764,
  45

\bibitem[\protect\citeauthoryear{Kewley, Heisler, Dopita  \& Lumsden}{Kewley
  et~al.}{2001a}]{Kewley2001_BPT}
Kewley L.~J.,  Heisler C.~A.,  Dopita M.~A.,   Lumsden S.,  2001a, \mn@doi
  [\apjs] {10.1086/318944}, 132, 37

\bibitem[\protect\citeauthoryear{{Kewley}, {Dopita}, {Sutherland}, {Heisler}
  \& {Trevena}}{{Kewley} et~al.}{2001b}]{2001ApJ...556..121K}
{Kewley} L.~J.,  {Dopita} M.~A.,  {Sutherland} R.~S.,  {Heisler} C.~A.,
  {Trevena} J.,  2001b, \mn@doi [\apj] {10.1086/321545}, \href
  {http://adsabs.harvard.edu/abs/2001ApJ...556..121K} {556, 121}

\bibitem[\protect\citeauthoryear{King}{King}{2016}]{King2016_maxMBH}
King A.~R.,  2016, \mn@doi [\mnras] {10.1093/mnrasl/slv186}, 456, L109

\bibitem[\protect\citeauthoryear{{Komatsu} et~al.,}{{Komatsu}
  et~al.}{2011}]{2011ApJS..192...18K}
{Komatsu} E.,  et~al., 2011, \mn@doi [\apjs] {10.1088/0067-0049/192/2/18},
  \href {http://adsabs.harvard.edu/abs/2011ApJS..192...18K} {192, 18}

\bibitem[\protect\citeauthoryear{Kong \& Ho}{Kong \& Ho}{2018}]{KongHo2018_Q2}
Kong M.,  Ho L.~C.,  2018, \mn@doi [\apj] {arXiv:1804.09852v1}, 859, 116

\bibitem[\protect\citeauthoryear{K{\"{o}}rding, Falcke  \&
  Corbel}{K{\"{o}}rding et~al.}{2006}]{Kording2006_FP}
K{\"{o}}rding E.,  Falcke H.,   Corbel S.,  2006, \mn@doi [\aap]
  {10.1051/0004-6361:20054144}, 456, 439

\bibitem[\protect\citeauthoryear{{Kormendy}}{{Kormendy}}{2016}]{2016ASSL..418..431K}
{Kormendy} J.,  2016, in {Laurikainen} E.,  {Peletier} R.,   {Gadotti} D.,
  eds,  Astrophysics and Space Science Library Vol. 418, Galactic Bulges.
  p.~431 (\mn@eprint {arXiv} {1504.03330}),
  \mn@doi{10.1007/978-3-319-19378-6_16}

\bibitem[\protect\citeauthoryear{{Kormendy} \& {Ho}}{{Kormendy} \&
  {Ho}}{2013}]{2013ARA&A..51..511K}
{Kormendy} J.,  {Ho} L.~C.,  2013, \mn@doi [\araa]
  {10.1146/annurev-astro-082708-101811}, \href
  {http://adsabs.harvard.edu/abs/2013ARA%26A..51..511K} {51, 511}

\bibitem[\protect\citeauthoryear{{Koss}, {Mushotzky}, {Veilleux}  \&
  {Winter}}{{Koss} et~al.}{2010}]{2010ApJ...716L.125K}
{Koss} M.,  {Mushotzky} R.,  {Veilleux} S.,   {Winter} L.,  2010, \mn@doi
  [\apjl] {10.1088/2041-8205/716/2/L125}, \href
  {http://adsabs.harvard.edu/abs/2010ApJ...716L.125K} {716, L125}

\bibitem[\protect\citeauthoryear{{Koss}, {Mushotzky}, {Veilleux}, {Winter},
  {Baumgartner}, {Tueller}, {Gehrels}  \& {Valencic}}{{Koss}
  et~al.}{2011}]{2011ApJ...739...57K}
{Koss} M.,  {Mushotzky} R.,  {Veilleux} S.,  {Winter} L.~M.,  {Baumgartner} W.,
   {Tueller} J.,  {Gehrels} N.,   {Valencic} L.,  2011, \mn@doi [\apj]
  {10.1088/0004-637X/739/2/57}, \href
  {http://adsabs.harvard.edu/abs/2011ApJ...739...57K} {739, 57}

\bibitem[\protect\citeauthoryear{{Koss} et~al.,}{{Koss}
  et~al.}{2016}]{2016ApJ...825...85K}
{Koss} M.~J.,  et~al., 2016, \mn@doi [\apj] {10.3847/0004-637X/825/2/85}, \href
  {http://adsabs.harvard.edu/abs/2016ApJ...825...85K} {825, 85}

\bibitem[\protect\citeauthoryear{{Koss} et~al.,}{{Koss}
  et~al.}{2017}]{2017ApJ...850...74K}
{Koss} M.,  et~al., 2017, \mn@doi [\apj] {10.3847/1538-4357/aa8ec9}, \href
  {http://adsabs.harvard.edu/abs/2017ApJ...850...74K} {850, 74}

\bibitem[\protect\citeauthoryear{{Koss} et~al.,}{{Koss}
  et~al.}{2018}]{2018Natur.563..214K}
{Koss} M.~J.,  et~al., 2018, \mn@doi [\nat] {10.1038/s41586-018-0652-7}, \href
  {http://adsabs.harvard.edu/abs/2018Natur.563..214K} {563, 214}

\bibitem[\protect\citeauthoryear{{Kraft}, {Hardcastle}, {Worrall}  \&
  {Murray}}{{Kraft} et~al.}{2005}]{2005ApJ...622..149K}
{Kraft} R.~P.,  {Hardcastle} M.~J.,  {Worrall} D.~M.,   {Murray} S.~S.,  2005,
  \mn@doi [\apj] {10.1086/427822}, \href
  {http://adsabs.harvard.edu/abs/2005ApJ...622..149K} {622, 149}

\bibitem[\protect\citeauthoryear{{LaMassa} et~al.,}{{LaMassa}
  et~al.}{2017}]{LaMassa2017_S82X_red_QSO}
{LaMassa} S.~M.,  et~al., 2017, \mn@doi [\apj] {10.3847/1538-4357/aa87b5},
  \href {http://adsabs.harvard.edu/abs/2017ApJ...847..100L} {847, 100}

\bibitem[\protect\citeauthoryear{{Lawrence}}{{Lawrence}}{1991}]{1991MNRAS.252..586L}
{Lawrence} A.,  1991, \mn@doi [\mnras] {10.1093/mnras/252.4.586}, \href
  {http://adsabs.harvard.edu/abs/1991MNRAS.252..586L} {252, 586}

\bibitem[\protect\citeauthoryear{{Lintott} et~al.,}{{Lintott}
  et~al.}{2011}]{2011MNRAS.410..166L}
{Lintott} C.,  et~al., 2011, \mn@doi [\mnras]
  {10.1111/j.1365-2966.2010.17432.x}, \href
  {http://adsabs.harvard.edu/abs/2011MNRAS.410..166L} {410, 166}

\bibitem[\protect\citeauthoryear{{Liu}, {Pooley}  \& {Riley}}{{Liu}
  et~al.}{1992}]{1992MNRAS.257..545L}
{Liu} R.,  {Pooley} G.,   {Riley} J.~M.,  1992, \mn@doi [\mnras]
  {10.1093/mnras/257.4.545}, \href
  {http://adsabs.harvard.edu/abs/1992MNRAS.257..545L} {257, 545}

\bibitem[\protect\citeauthoryear{{Liu}, {Zakamska}, {Greene}, {Strauss},
  {Krolik}  \& {Heckman}}{{Liu} et~al.}{2009}]{2009ApJ...702.1098L}
{Liu} X.,  {Zakamska} N.~L.,  {Greene} J.~E.,  {Strauss} M.~A.,  {Krolik}
  J.~H.,   {Heckman} T.~M.,  2009, \mn@doi [\apj]
  {10.1088/0004-637X/702/2/1098}, \href
  {http://adsabs.harvard.edu/abs/2009ApJ...702.1098L} {702, 1098}

\bibitem[\protect\citeauthoryear{Lutz et~al.,}{Lutz
  et~al.}{2008}]{Lutz2008_SF_AGN}
Lutz D.,  et~al., 2008, \mn@doi [\apj] {10.1086/590367}, 684, 853

\bibitem[\protect\citeauthoryear{{Maia}, {Machado}  \& {Willmer}}{{Maia}
  et~al.}{2003}]{2003AJ....126.1750M}
{Maia} M.~A.~G.,  {Machado} R.~S.,   {Willmer} C.~N.~A.,  2003, \mn@doi [\aj]
  {10.1086/378360}, \href {http://adsabs.harvard.edu/abs/2003AJ....126.1750M}
  {126, 1750}

\bibitem[\protect\citeauthoryear{Maiolino, Shemmer, Imanishi, Netzer, Oliva,
  Lutz  \& Sturm}{Maiolino et~al.}{2007}]{Maiolino2007_IRS_S04}
Maiolino R.,  Shemmer O.,  Imanishi M.,  Netzer H.,  Oliva E.,  Lutz D.,
  Sturm E.,  2007, \mn@doi [\aap] {10.1051/0004-6361:20077252}, 468, 979

\bibitem[\protect\citeauthoryear{{Marconi}, {Risaliti}, {Gilli}, {Hunt},
  {Maiolino}  \& {Salvati}}{{Marconi} et~al.}{2004}]{2004MNRAS.351..169M}
{Marconi} A.,  {Risaliti} G.,  {Gilli} R.,  {Hunt} L.~K.,  {Maiolino} R.,
  {Salvati} M.,  2004, \mn@doi [\mnras] {10.1111/j.1365-2966.2004.07765.x},
  \href {http://adsabs.harvard.edu/abs/2004MNRAS.351..169M} {351, 169}

\bibitem[\protect\citeauthoryear{{Maselli} et~al.,}{{Maselli}
  et~al.}{2013}]{2013ApJS..206...17M}
{Maselli} A.,  et~al., 2013, \mn@doi [\apjs] {10.1088/0067-0049/206/2/17},
  \href {http://adsabs.harvard.edu/abs/2013ApJS..206...17M} {206, 17}

\bibitem[\protect\citeauthoryear{{Massaro}, {Maselli}, {Leto}, {Marchegiani},
  {Perri}, {Giommi}  \& {Piranomonte}}{{Massaro}
  et~al.}{2015}]{2015Ap&SS.357...75M}
{Massaro} E.,  {Maselli} A.,  {Leto} C.,  {Marchegiani} P.,  {Perri} M.,
  {Giommi} P.,   {Piranomonte} S.,  2015, \mn@doi [\apss]
  {10.1007/s10509-015-2254-2}, \href
  {https://ui.adsabs.harvard.edu/abs/2015Ap&SS.357...75M} {357, 75}

\bibitem[\protect\citeauthoryear{{Mateos} et~al.,}{{Mateos}
  et~al.}{2012}]{Mateos_2012}
{Mateos} S.,  et~al., 2012, \mn@doi [\mnras]
  {10.1111/j.1365-2966.2012.21843.x}, \href
  {http://adsabs.harvard.edu/abs/2012MNRAS.426.3271M} {426, 3271}

\bibitem[\protect\citeauthoryear{{Mateos} et~al.,}{{Mateos}
  et~al.}{2017}]{2017ApJ...841L..18M}
{Mateos} S.,  et~al., 2017, \mn@doi [\apjl] {10.3847/2041-8213/aa7268}, \href
  {http://adsabs.harvard.edu/abs/2017ApJ...841L..18M} {841, L18}

\bibitem[\protect\citeauthoryear{McConnell \& Ma}{McConnell \&
  Ma}{2013}]{McConnell2013_MM}
McConnell N.~J.,  Ma C.-P.,  2013, \mn@doi [\apj]
  {10.1088/0004-637X/764/2/184}, 764, 184

\bibitem[\protect\citeauthoryear{McConnell, Ma, Gebhardt, Wright, Murphy,
  Lauer, Graham  \& Richstone}{McConnell et~al.}{2011}]{McConnell2011_MBH_10}
McConnell N.~J.,  Ma C.-p.,  Gebhardt K.,  Wright S.~A.,  Murphy J.~D.,  Lauer
  T.~R.,  Graham J.~R.,   Richstone D.~O.,  2011, \mn@doi [\nat]
  {10.1038/nature10636}, 480, 215

\bibitem[\protect\citeauthoryear{McConnell, Ma, Murphy, Gebhardt, Lauer,
  Graham, Wright  \& Richstone}{McConnell et~al.}{2012}]{McConnell2012_BCGs}
McConnell N.~J.,  Ma C.-P.,  Murphy J.~D.,  Gebhardt K.,  Lauer T.~R.,  Graham
  J.~R.,  Wright S.~a.,   Richstone D.~O.,  2012, \mn@doi [\apj]
  {10.1088/0004-637X/756/2/179}, 756, 179

\bibitem[\protect\citeauthoryear{McDonald et~al.,}{McDonald
  et~al.}{2012}]{McDonald2012_Phoenix}
McDonald M.,  et~al., 2012, \mn@doi [\nat] {10.1038/nature11379}, 488, 349

\bibitem[\protect\citeauthoryear{{Merloni}, {Heinz}  \& {di Matteo}}{{Merloni}
  et~al.}{2003}]{2003MNRAS.345.1057M}
{Merloni} A.,  {Heinz} S.,   {di Matteo} T.,  2003, \mn@doi [\mnras]
  {10.1046/j.1365-2966.2003.07017.x}, \href
  {http://adsabs.harvard.edu/abs/2003MNRAS.345.1057M} {345, 1057}

\bibitem[\protect\citeauthoryear{{Mezcua}, {Hlavacek-Larrondo}, {Lucey},
  {Hogan}, {Edge}  \& {McNamara}}{{Mezcua} et~al.}{2018}]{2018MNRAS.474.1342M}
{Mezcua} M.,  {Hlavacek-Larrondo} J.,  {Lucey} J.~R.,  {Hogan} M.~T.,  {Edge}
  A.~C.,   {McNamara} B.~R.,  2018, \mn@doi [\mnras] {10.1093/mnras/stx2812},
  \href {http://adsabs.harvard.edu/abs/2018MNRAS.474.1342M} {474, 1342}

\bibitem[\protect\citeauthoryear{{Mineo}, {Gilfanov}, {Lehmer}, {Morrison}  \&
  {Sunyaev}}{{Mineo} et~al.}{2014}]{2014MNRAS.437.1698M}
{Mineo} S.,  {Gilfanov} M.,  {Lehmer} B.~D.,  {Morrison} G.~E.,   {Sunyaev} R.,
   2014, \mn@doi [\mnras] {10.1093/mnras/stt1999}, \href
  {http://adsabs.harvard.edu/abs/2014MNRAS.437.1698M} {437, 1698}

\bibitem[\protect\citeauthoryear{{Mingo} et~al.,}{{Mingo}
  et~al.}{2017}]{2017MNRAS.470.2762M}
{Mingo} B.,  et~al., 2017, \mn@doi [\mnras] {10.1093/mnras/stx1307}, \href
  {http://adsabs.harvard.edu/abs/2017MNRAS.470.2762M} {470, 2762}

\bibitem[\protect\citeauthoryear{Moffett et~al.,}{Moffett
  et~al.}{2016}]{Moffett2016_SMF_morph}
Moffett A.~J.,  et~al., 2016, \mn@doi [\mnras] {10.1093/mnras/stw1861}, 462,
  4336

\bibitem[\protect\citeauthoryear{{Mountrichas} et~al.,}{{Mountrichas}
  et~al.}{2017}]{2017MNRAS.468.3042M}
{Mountrichas} G.,  et~al., 2017, \mn@doi [\mnras] {10.1093/mnras/stx572}, \href
  {http://adsabs.harvard.edu/abs/2017MNRAS.468.3042M} {468, 3042}

\bibitem[\protect\citeauthoryear{{Murgia}, {Markevitch}, {Govoni}, {Parma},
  {Fanti}, {de Ruiter}  \& {Mack}}{{Murgia} et~al.}{2012}]{2012A&A...548A..75M}
{Murgia} M.,  {Markevitch} M.,  {Govoni} F.,  {Parma} P.,  {Fanti} R.,  {de
  Ruiter} H.~R.,   {Mack} K.-H.,  2012, \mn@doi [\aap]
  {10.1051/0004-6361/201219702}, \href
  {http://adsabs.harvard.edu/abs/2012A%26A...548A..75M} {548, A75}

\bibitem[\protect\citeauthoryear{Netzer}{Netzer}{2009}]{Netzer2009_SF_AGN}
Netzer H.,  2009, \mn@doi [\mnras] {10.1111/j.1365-2966.2009.15434.x}, 399,
  1907

\bibitem[\protect\citeauthoryear{Netzer}{Netzer}{2015}]{Netzer2015_torus_rev}
Netzer H.,  2015, \mn@doi [\araa] {10.1146/annurev-astro-082214-122302}, 53,
  365

\bibitem[\protect\citeauthoryear{{Nisbet} \& {Best}}{{Nisbet} \&
  {Best}}{2016}]{2016MNRAS.455.2551N}
{Nisbet} D.~M.,  {Best} P.~N.,  2016, \mn@doi [\mnras] {10.1093/mnras/stv2450},
  \href {http://adsabs.harvard.edu/abs/2016MNRAS.455.2551N} {455, 2551}

\bibitem[\protect\citeauthoryear{{Nyland} et~al.,}{{Nyland}
  et~al.}{2018}]{2018ApJ...859...23N}
{Nyland} K.,  et~al., 2018, \mn@doi [\apj] {10.3847/1538-4357/aab3d1}, \href
  {http://adsabs.harvard.edu/abs/2018ApJ...859...23N} {859, 23}

\bibitem[\protect\citeauthoryear{{Ochsenbein}, {Bauer}  \&
  {Marcout}}{{Ochsenbein} et~al.}{2000}]{2000A&AS..143...23O}
{Ochsenbein} F.,  {Bauer} P.,   {Marcout} J.,  2000, \mn@doi [\aaps]
  {10.1051/aas:2000169}, \href
  {http://adsabs.harvard.edu/abs/2000A%26AS..143...23O} {143, 23}

\bibitem[\protect\citeauthoryear{{Oh}, {Sarzi}, {Schawinski}  \& {Yi}}{{Oh}
  et~al.}{2011}]{2011ApJS..195...13O}
{Oh} K.,  {Sarzi} M.,  {Schawinski} K.,   {Yi} S.~K.,  2011, \mn@doi [\apjs]
  {10.1088/0067-0049/195/2/13}, \href
  {http://adsabs.harvard.edu/abs/2011ApJS..195...13O} {195, 13}

\bibitem[\protect\citeauthoryear{{Oh}, {Yi}, {Schawinski}, {Koss},
  {Trakhtenbrot}  \& {Soto}}{{Oh} et~al.}{2015}]{2015ApJS..219....1O}
{Oh} K.,  {Yi} S.~K.,  {Schawinski} K.,  {Koss} M.,  {Trakhtenbrot} B.,
  {Soto} K.,  2015, \mn@doi [\apjs] {10.1088/0067-0049/219/1/1}, \href
  {http://adsabs.harvard.edu/abs/2015ApJS..219....1O} {219, 1}

\bibitem[\protect\citeauthoryear{{Oh} et~al.,}{{Oh}
  et~al.}{2017}]{2017MNRAS.464.1466O}
{Oh} K.,  et~al., 2017, \mn@doi [\mnras] {10.1093/mnras/stw2467}, \href
  {http://adsabs.harvard.edu/abs/2017MNRAS.464.1466O} {464, 1466}

\bibitem[\protect\citeauthoryear{{Oh} et~al.,}{{Oh}
  et~al.}{2018}]{2018ApJS..235....4O}
{Oh} K.,  et~al., 2018, \mn@doi [\apjs] {10.3847/1538-4365/aaa7fd}, \href
  {http://adsabs.harvard.edu/abs/2018ApJS..235....4O} {235, 4}

\bibitem[\protect\citeauthoryear{{Osterbrock}}{{Osterbrock}}{1981}]{1981ApJ...249..462O}
{Osterbrock} D.~E.,  1981, \mn@doi [\apj] {10.1086/159306}, \href
  {http://adsabs.harvard.edu/abs/1981ApJ...249..462O} {249, 462}

\bibitem[\protect\citeauthoryear{Pacucci, Natarajan  \& Ferrara}{Pacucci
  et~al.}{2017}]{Pacucci2017_maxMBH_feedback}
Pacucci F.,  Natarajan P.,   Ferrara A.,  2017, \mn@doi [\apjl]
  {10.3847/2041-8213/835/2/L36}, 835, L36

\bibitem[\protect\citeauthoryear{{Padmanabhan} et~al.,}{{Padmanabhan}
  et~al.}{2008}]{2008ApJ...674.1217P}
{Padmanabhan} N.,  et~al., 2008, \mn@doi [\apj] {10.1086/524677}, \href
  {http://adsabs.harvard.edu/abs/2008ApJ...674.1217P} {674, 1217}

\bibitem[\protect\citeauthoryear{{Padovani}}{{Padovani}}{1993}]{1993MNRAS.263..461P}
{Padovani} P.,  1993, \mn@doi [\mnras] {10.1093/mnras/263.2.461}, \href
  {http://adsabs.harvard.edu/abs/1993MNRAS.263..461P} {263, 461}

\bibitem[\protect\citeauthoryear{{Padovani}}{{Padovani}}{2017}]{2017NatAs...1E.194P}
{Padovani} P.,  2017, \mn@doi [Nature Astronomy] {10.1038/s41550-017-0194},
  \href {http://adsabs.harvard.edu/abs/2017NatAs...1E.194P} {1, 0194}

\bibitem[\protect\citeauthoryear{{Panessa} et~al.,}{{Panessa}
  et~al.}{2015}]{2015MNRAS.447.1289P}
{Panessa} F.,  et~al., 2015, \mn@doi [\mnras] {10.1093/mnras/stu2455}, \href
  {http://adsabs.harvard.edu/abs/2015MNRAS.447.1289P} {447, 1289}

\bibitem[\protect\citeauthoryear{{Plotkin}, {Markoff}, {Kelly}, {K{\"o}rding}
  \& {Anderson}}{{Plotkin} et~al.}{2012}]{2012MNRAS.419..267P}
{Plotkin} R.~M.,  {Markoff} S.,  {Kelly} B.~C.,  {K{\"o}rding} E.,   {Anderson}
  S.~F.,  2012, \mn@doi [\mnras] {10.1111/j.1365-2966.2011.19689.x}, \href
  {http://adsabs.harvard.edu/abs/2012MNRAS.419..267P} {419, 267}

\bibitem[\protect\citeauthoryear{{Powell} et~al.,}{{Powell}
  et~al.}{2018}]{2018ApJ...858..110P}
{Powell} M.~C.,  et~al., 2018, \mn@doi [\apj] {10.3847/1538-4357/aabd7f}, \href
  {http://adsabs.harvard.edu/abs/2018ApJ...858..110P} {858, 110}

\bibitem[\protect\citeauthoryear{{Rafter}, {Crenshaw}  \& {Wiita}}{{Rafter}
  et~al.}{2009}]{2009AJ....137...42R}
{Rafter} S.~E.,  {Crenshaw} D.~M.,   {Wiita} P.~J.,  2009, \mn@doi [\aj]
  {10.1088/0004-6256/137/1/42}, \href
  {http://adsabs.harvard.edu/abs/2009AJ....137...42R} {137, 42}

\bibitem[\protect\citeauthoryear{{Ramos Almeida}}{{Ramos
  Almeida}}{2014}]{2014cosp...40E2695R}
{Ramos Almeida} C.,  2014, in 40th COSPAR Scientific Assembly. pp E1.19--6--14

\bibitem[\protect\citeauthoryear{{Ramos Almeida} \& {Ricci}}{{Ramos Almeida} \&
  {Ricci}}{2017}]{2017NatAs...1..679R}
{Ramos Almeida} C.,  {Ricci} C.,  2017, \mn@doi [Nature Astronomy]
  {10.1038/s41550-017-0232-z}, \href
  {http://adsabs.harvard.edu/abs/2017NatAs...1..679R} {1, 679}

\bibitem[\protect\citeauthoryear{{Randall}, {Hopkins}, {Norris}  \&
  {Edwards}}{{Randall} et~al.}{2011}]{2011MNRAS.416.1135R}
{Randall} K.~E.,  {Hopkins} A.~M.,  {Norris} R.~P.,   {Edwards} P.~G.,  2011,
  \mn@doi [\mnras] {10.1111/j.1365-2966.2011.19116.x}, \href
  {http://adsabs.harvard.edu/abs/2011MNRAS.416.1135R} {416, 1135}

\bibitem[\protect\citeauthoryear{Reines \& Volonteri}{Reines \&
  Volonteri}{2015}]{ReinesVolonteri2015_MM_AGN}
Reines A.~E.,  Volonteri M.,  2015, \mn@doi [\apj]
  {10.1088/0004-637X/813/2/82}, 813, 82

\bibitem[\protect\citeauthoryear{{Reyes} et~al.,}{{Reyes}
  et~al.}{2008}]{2008AJ....136.2373R}
{Reyes} R.,  et~al., 2008, \mn@doi [\aj] {10.1088/0004-6256/136/6/2373}, \href
  {http://adsabs.harvard.edu/abs/2008AJ....136.2373R} {136, 2373}

\bibitem[\protect\citeauthoryear{{Ricci}, {Ueda}, {Koss}, {Trakhtenbrot},
  {Bauer}  \& {Gandhi}}{{Ricci} et~al.}{2015}]{2015ApJ...815L..13R}
{Ricci} C.,  {Ueda} Y.,  {Koss} M.~J.,  {Trakhtenbrot} B.,  {Bauer} F.~E.,
  {Gandhi} P.,  2015, \mn@doi [\apjl] {10.1088/2041-8205/815/1/L13}, \href
  {http://adsabs.harvard.edu/abs/2015ApJ...815L..13R} {815, L13}

\bibitem[\protect\citeauthoryear{{Ricci} et~al.,}{{Ricci}
  et~al.}{2017a}]{2017ApJS..233...17R}
{Ricci} C.,  et~al., 2017a, \mn@doi [\apjs] {10.3847/1538-4365/aa96ad}, \href
  {http://adsabs.harvard.edu/abs/2017ApJS..233...17R} {233, 17}

\bibitem[\protect\citeauthoryear{{Ricci} et~al.,}{{Ricci}
  et~al.}{2017b}]{Ricci2017_mergers}
{Ricci} C.,  et~al., 2017b, \mn@doi [\mnras] {10.1093/mnras/stx173}, \href
  {http://adsabs.harvard.edu/abs/2017MNRAS.468.1273R} {468, 1273}

\bibitem[\protect\citeauthoryear{{Ricci} et~al.,}{{Ricci}
  et~al.}{2017c}]{2017Natur.549..488R}
{Ricci} C.,  et~al., 2017c, \mn@doi [\nat] {10.1038/nature23906}, \href
  {http://adsabs.harvard.edu/abs/2017Natur.549..488R} {549, 488}

\bibitem[\protect\citeauthoryear{{Richards} et~al.,}{{Richards}
  et~al.}{2006}]{2006AJ....131.2766R}
{Richards} G.~T.,  et~al., 2006, \mn@doi [\aj] {10.1086/503559}, \href
  {http://adsabs.harvard.edu/abs/2006AJ....131.2766R} {131, 2766}

\bibitem[\protect\citeauthoryear{{Rigby}, {Diamond-Stanic}  \&
  {Aniano}}{{Rigby} et~al.}{2009}]{2009ApJ...700.1878R}
{Rigby} J.~R.,  {Diamond-Stanic} A.~M.,   {Aniano} G.,  2009, \mn@doi [\apj]
  {10.1088/0004-637X/700/2/1878}, \href
  {http://adsabs.harvard.edu/abs/2009ApJ...700.1878R} {700, 1878}

\bibitem[\protect\citeauthoryear{{Ross} et~al.,}{{Ross}
  et~al.}{2013}]{2013ApJ...773...14R}
{Ross} N.~P.,  et~al., 2013, \mn@doi [\apj] {10.1088/0004-637X/773/1/14}, \href
  {http://adsabs.harvard.edu/abs/2013ApJ...773...14R} {773, 14}

\bibitem[\protect\citeauthoryear{{Ross} et~al.,}{{Ross}
  et~al.}{2015}]{Ross2015_red_QSO_BOSS}
{Ross} N.~P.,  et~al., 2015, \mn@doi [\mnras] {10.1093/mnras/stv1710}, \href
  {http://adsabs.harvard.edu/abs/2015MNRAS.453.3932R} {453, 3932}

\bibitem[\protect\citeauthoryear{{Satyapal}, {Vega}, {Heckman}, {O'Halloran}
  \& {Dudik}}{{Satyapal} et~al.}{2007}]{2007ApJ...663L...9S}
{Satyapal} S.,  {Vega} D.,  {Heckman} T.,  {O'Halloran} B.,   {Dudik} R.,
  2007, \mn@doi [\apjl] {10.1086/519995}, \href
  {http://adsabs.harvard.edu/abs/2007ApJ...663L...9S} {663, L9}

\bibitem[\protect\citeauthoryear{{Sazonov}, {Churazov}  \&
  {Krivonos}}{{Sazonov} et~al.}{2015}]{2015MNRAS.454.1202S}
{Sazonov} S.,  {Churazov} E.,   {Krivonos} R.,  2015, \mn@doi [\mnras]
  {10.1093/mnras/stv2069}, \href
  {http://adsabs.harvard.edu/abs/2015MNRAS.454.1202S} {454, 1202}

\bibitem[\protect\citeauthoryear{{Schawinski}, {Thomas}, {Sarzi}, {Maraston},
  {Kaviraj}, {Joo}, {Yi}  \& {Silk}}{{Schawinski}
  et~al.}{2007}]{2007MNRAS.382.1415S}
{Schawinski} K.,  {Thomas} D.,  {Sarzi} M.,  {Maraston} C.,  {Kaviraj} S.,
  {Joo} S.-J.,  {Yi} S.~K.,   {Silk} J.,  2007, \mn@doi [\mnras]
  {10.1111/j.1365-2966.2007.12487.x}, \href
  {http://adsabs.harvard.edu/abs/2007MNRAS.382.1415S} {382, 1415}

\bibitem[\protect\citeauthoryear{{Schawinski} et~al.,}{{Schawinski}
  et~al.}{2009}]{2009MNRAS.396..818S}
{Schawinski} K.,  et~al., 2009, \mn@doi [\mnras]
  {10.1111/j.1365-2966.2009.14793.x}, \href
  {http://adsabs.harvard.edu/abs/2009MNRAS.396..818S} {396, 818}

\bibitem[\protect\citeauthoryear{Schulze \& Wisotzki}{Schulze \&
  Wisotzki}{2010}]{Schulze2010_BHMF}
Schulze A.,  Wisotzki L.,  2010, \mn@doi [\aap] {10.1051/0004-6361/201014193},
  516, A87

\bibitem[\protect\citeauthoryear{Schulze et~al.,}{Schulze
  et~al.}{2015}]{Schulze2015_BHMF}
Schulze A.,  et~al., 2015, \mn@doi [\mnras] {10.1093/mnras/stu2549}, 447, 2085

\bibitem[\protect\citeauthoryear{{Secrest}, {Satyapal}, {Gliozzi}, {Cheung},
  {Seth}  \& {B{\"o}ker}}{{Secrest} et~al.}{2012}]{2012ApJ...753...38S}
{Secrest} N.~J.,  {Satyapal} S.,  {Gliozzi} M.,  {Cheung} C.~C.,  {Seth} A.~C.,
    {B{\"o}ker} T.,  2012, \mn@doi [\apj] {10.1088/0004-637X/753/1/38}, \href
  {http://adsabs.harvard.edu/abs/2012ApJ...753...38S} {753, 38}

\bibitem[\protect\citeauthoryear{{Secrest}, {Dudik}, {Dorland}, {Zacharias},
  {Makarov}, {Fey}, {Frouard}  \& {Finch}}{{Secrest}
  et~al.}{2015}]{2015ApJS..221...12S}
{Secrest} N.~J.,  {Dudik} R.~P.,  {Dorland} B.~N.,  {Zacharias} N.,  {Makarov}
  V.,  {Fey} A.,  {Frouard} J.,   {Finch} C.,  2015, \mn@doi [\apjs]
  {10.1088/0067-0049/221/1/12}, \href
  {http://adsabs.harvard.edu/abs/2015ApJS..221...12S} {221, 12}

\bibitem[\protect\citeauthoryear{Shemmer, Netzer, Maiolino, Oliva, Croom,
  Corbett  \& di Fabrizio}{Shemmer et~al.}{2004}]{Shemmer2004}
Shemmer O.,  Netzer H.,  Maiolino R.,  Oliva E.,  Croom S.~M.,  Corbett E.~A.,
   di Fabrizio L.,  2004, \mn@doi [\apj] {10.1086/423607}, 614, 547

\bibitem[\protect\citeauthoryear{{Sikora}, {Stawarz}  \& {Lasota}}{{Sikora}
  et~al.}{2007}]{2007ApJ...658..815S}
{Sikora} M.,  {Stawarz} {\L}.,   {Lasota} J.-P.,  2007, \mn@doi [\apj]
  {10.1086/511972}, \href {http://adsabs.harvard.edu/abs/2007ApJ...658..815S}
  {658, 815}

\bibitem[\protect\citeauthoryear{{Simpson}}{{Simpson}}{2005}]{2005MNRAS.360..565S}
{Simpson} C.,  2005, \mn@doi [\mnras] {10.1111/j.1365-2966.2005.09043.x}, \href
  {http://adsabs.harvard.edu/abs/2005MNRAS.360..565S} {360, 565}

\bibitem[\protect\citeauthoryear{{Skrutskie} et~al.,}{{Skrutskie}
  et~al.}{2006}]{2006AJ....131.1163S}
{Skrutskie} M.~F.,  et~al., 2006, \mn@doi [\aj] {10.1086/498708}, \href
  {http://adsabs.harvard.edu/abs/2006AJ....131.1163S} {131, 1163}

\bibitem[\protect\citeauthoryear{Stern \& Laor}{Stern \&
  Laor}{2013}]{Stern2013_BPT}
Stern J.,  Laor A.,  2013, \mn@doi [\mnras] {10.1093/mnras/stt211}, 431, 836

\bibitem[\protect\citeauthoryear{{Stern} et~al.,}{{Stern}
  et~al.}{2012}]{2012ApJ...753...30S}
{Stern} D.,  et~al., 2012, \mn@doi [\apj] {10.1088/0004-637X/753/1/30}, \href
  {http://adsabs.harvard.edu/abs/2012ApJ...753...30S} {753, 30}

\bibitem[\protect\citeauthoryear{{Stern} et~al.,}{{Stern}
  et~al.}{2014}]{Stern2014_HotDOGs_Xray}
{Stern} D.,  et~al., 2014, \mn@doi [\apj] {10.1088/0004-637X/794/2/102}, \href
  {http://adsabs.harvard.edu/abs/2014ApJ...794..102S} {794, 102}

\bibitem[\protect\citeauthoryear{{Tadhunter}}{{Tadhunter}}{2016}]{2016A&ARv..24...10T}
{Tadhunter} C.,  2016, \mn@doi [\aapr] {10.1007/s00159-016-0094-x}, \href
  {http://adsabs.harvard.edu/abs/2016A%26ARv..24...10T} {24, 10}

\bibitem[\protect\citeauthoryear{{Taylor}}{{Taylor}}{2005}]{2005ASPC..347...29T}
{Taylor} M.~B.,  2005, in {Shopbell} P.,  {Britton} M.,   {Ebert} R.,  eds,
  Astronomical Society of the Pacific Conference Series Vol. 347, Astronomical
  Data Analysis Software and Systems XIV. p.~29

\bibitem[\protect\citeauthoryear{Trakhtenbrot \& Netzer}{Trakhtenbrot \&
  Netzer}{2012}]{TrakhtNetzer2012_Mg2}
Trakhtenbrot B.,  Netzer H.,  2012, \mn@doi [\mnras]
  {10.1111/j.1365-2966.2012.22056.x}, 427, 3081

\bibitem[\protect\citeauthoryear{Trakhtenbrot et~al.,}{Trakhtenbrot
  et~al.}{2016}]{Trakhtenbrot2016_COSMOSFIRE_MBH}
Trakhtenbrot B.,  et~al., 2016, \mn@doi [\apj] {10.3847/0004-637X/825/1/4},
  825, 4

\bibitem[\protect\citeauthoryear{{Trakhtenbrot} et~al.,}{{Trakhtenbrot}
  et~al.}{2017}]{2017MNRAS.470..800T}
{Trakhtenbrot} B.,  et~al., 2017, \mn@doi [\mnras] {10.1093/mnras/stx1117},
  \href {http://adsabs.harvard.edu/abs/2017MNRAS.470..800T} {470, 800}

\bibitem[\protect\citeauthoryear{{Treister}, {Krolik}  \&
  {Dullemond}}{{Treister} et~al.}{2008}]{2008ApJ...679..140T}
{Treister} E.,  {Krolik} J.~H.,   {Dullemond} C.,  2008, \mn@doi [\apj]
  {10.1086/586698}, \href {http://adsabs.harvard.edu/abs/2008ApJ...679..140T}
  {679, 140}

\bibitem[\protect\citeauthoryear{{Treister}, {Schawinski}, {Urry}  \&
  {Simmons}}{{Treister} et~al.}{2012}]{2012ApJ...758L..39T}
{Treister} E.,  {Schawinski} K.,  {Urry} C.~M.,   {Simmons} B.~D.,  2012,
  \mn@doi [\apjl] {10.1088/2041-8205/758/2/L39}, \href
  {http://adsabs.harvard.edu/abs/2012ApJ...758L..39T} {758, L39}

\bibitem[\protect\citeauthoryear{{Ueda}, {Akiyama}, {Ohta}  \& {Miyaji}}{{Ueda}
  et~al.}{2003}]{2003ApJ...598..886U}
{Ueda} Y.,  {Akiyama} M.,  {Ohta} K.,   {Miyaji} T.,  2003, \mn@doi [\apj]
  {10.1086/378940}, \href {http://adsabs.harvard.edu/abs/2003ApJ...598..886U}
  {598, 886}

\bibitem[\protect\citeauthoryear{{Ueda}, {Hayashida}, {Anabuki}, {Nakajima},
  {Koyama}  \& {Tsunemi}}{{Ueda} et~al.}{2013}]{2013ApJ...778...33U}
{Ueda} S.,  {Hayashida} K.,  {Anabuki} N.,  {Nakajima} H.,  {Koyama} K.,
  {Tsunemi} H.,  2013, \mn@doi [\apj] {10.1088/0004-637X/778/1/33}, \href
  {http://adsabs.harvard.edu/abs/2013ApJ...778...33U} {778, 33}

\bibitem[\protect\citeauthoryear{{Ueda}, {Akiyama}, {Hasinger}, {Miyaji}  \&
  {Watson}}{{Ueda} et~al.}{2014}]{2014ApJ...786..104U}
{Ueda} Y.,  {Akiyama} M.,  {Hasinger} G.,  {Miyaji} T.,   {Watson} M.~G.,
  2014, \mn@doi [\apj] {10.1088/0004-637X/786/2/104}, \href
  {http://adsabs.harvard.edu/abs/2014ApJ...786..104U} {786, 104}

\bibitem[\protect\citeauthoryear{Urry \& Padovani}{Urry \&
  Padovani}{1995}]{UrryPadovani1995_rev}
Urry C.~M.,  Padovani P.,  1995, \mn@doi [\pasp] {10.1086/133630}, 107, 803

\bibitem[\protect\citeauthoryear{Vasudevan \& Fabian}{Vasudevan \&
  Fabian}{2007}]{VasudevanFabian2007_BC}
Vasudevan R.~V.,  Fabian A.~C.,  2007, \mn@doi [\mnras]
  {10.1111/j.1365-2966.2007.12328.x}, 381, 1235

\bibitem[\protect\citeauthoryear{{Vasudevan}, {Mushotzky}, {Winter}  \&
  {Fabian}}{{Vasudevan} et~al.}{2009}]{2009MNRAS.399.1553V}
{Vasudevan} R.~V.,  {Mushotzky} R.~F.,  {Winter} L.~M.,   {Fabian} A.~C.,
  2009, \mn@doi [\mnras] {10.1111/j.1365-2966.2009.15371.x}, \href
  {http://adsabs.harvard.edu/abs/2009MNRAS.399.1553V} {399, 1553}

\bibitem[\protect\citeauthoryear{{Veilleux} \& {Osterbrock}}{{Veilleux} \&
  {Osterbrock}}{1987}]{1987ApJS...63..295V}
{Veilleux} S.,  {Osterbrock} D.~E.,  1987, \mn@doi [\apjs] {10.1086/191166},
  \href {http://adsabs.harvard.edu/abs/1987ApJS...63..295V} {63, 295}

\bibitem[\protect\citeauthoryear{{Veilleux} et~al.,}{{Veilleux}
  et~al.}{2009}]{2009ApJ...701..587V}
{Veilleux} S.,  et~al., 2009, \mn@doi [\apj] {10.1088/0004-637X/701/1/587},
  \href {http://adsabs.harvard.edu/abs/2009ApJ...701..587V} {701, 587}

\bibitem[\protect\citeauthoryear{{Vernet} et~al.,}{{Vernet}
  et~al.}{2011}]{2011A&A...536A.105V}
{Vernet} J.,  et~al., 2011, \mn@doi [\aap] {10.1051/0004-6361/201117752}, \href
  {http://adsabs.harvard.edu/abs/2011A%26A...536A.105V} {536, A105}

\bibitem[\protect\citeauthoryear{Vestergaard \& Osmer}{Vestergaard \&
  Osmer}{2009}]{VestergaardOsmer2009_BHMF}
Vestergaard M.,  Osmer P.~S.,  2009, \mn@doi [\apj]
  {10.1088/0004-637X/699/1/800}, 699, 800

\bibitem[\protect\citeauthoryear{Villar-Mart{\'{i}}n, Humphrey, Delgado, Colina
   \& Arribas}{Villar-Mart{\'{i}}n
  et~al.}{2011}]{Villar-Martin2011_Q2_outflows}
Villar-Mart{\'{i}}n M.,  Humphrey A.,  Delgado R.~G.,  Colina L.,   Arribas S.,
   2011, \mn@doi [\mnras] {10.1111/j.1365-2966.2011.19622.x}, 418, 2032

\bibitem[\protect\citeauthoryear{Villar-Mart{\'{i}}n, Arribas, Emonts,
  Humphrey, Tadhunter, Bessiere, {Cabrera Lavers}  \& {Ramos
  Almeida}}{Villar-Mart{\'{i}}n et~al.}{2016}]{Villar-Martin2016_Q2_outflows}
Villar-Mart{\'{i}}n M.,  Arribas S.,  Emonts B.,  Humphrey A.,  Tadhunter C.,
  Bessiere P.,  {Cabrera Lavers} A.,   {Ramos Almeida} C.,  2016, \mn@doi
  [\mnras] {10.1093/mnras/stw901}, 460, 130

\bibitem[\protect\citeauthoryear{{Villarroel}, {Nyholm}, {Karlsson},
  {Comer{\'o}n}, {Korn}, {Sollerman}  \& {Zackrisson}}{{Villarroel}
  et~al.}{2017}]{2017ApJ...837..110V}
{Villarroel} B.,  {Nyholm} A.,  {Karlsson} T.,  {Comer{\'o}n} S.,  {Korn}
  A.~J.,  {Sollerman} J.,   {Zackrisson} E.,  2017, \mn@doi [\apj]
  {10.3847/1538-4357/aa5d5a}, \href
  {http://adsabs.harvard.edu/abs/2017ApJ...837..110V} {837, 110}

\bibitem[\protect\citeauthoryear{{Vito} et~al.,}{{Vito}
  et~al.}{2018}]{Vito2018_HotDOGs_Xray}
{Vito} F.,  et~al., 2018, \mn@doi [\mnras] {10.1093/mnras/stx3120}, \href
  {http://adsabs.harvard.edu/abs/2018MNRAS.474.4528V} {474, 4528}

\bibitem[\protect\citeauthoryear{{Vollmer} et~al.,}{{Vollmer}
  et~al.}{2010}]{2010A&A...511A..53V}
{Vollmer} B.,  et~al., 2010, \mn@doi [\aap] {10.1051/0004-6361/200913460},
  \href {http://adsabs.harvard.edu/abs/2010A%26A...511A..53V} {511, A53}

\bibitem[\protect\citeauthoryear{Wang et~al.,}{Wang
  et~al.}{2015}]{Wang2015_z5_hiM}
Wang F.,  et~al., 2015, \mn@doi [\apj] {10.1088/2041-8205/807/1/L9}, 807, L9

\bibitem[\protect\citeauthoryear{{Watabe}, {Kawakatu}, {Imanishi}  \&
  {Takeuchi}}{{Watabe} et~al.}{2009}]{2009MNRAS.400.1803W}
{Watabe} Y.,  {Kawakatu} N.,  {Imanishi} M.,   {Takeuchi} T.~T.,  2009, \mn@doi
  [\mnras] {10.1111/j.1365-2966.2009.15345.x}, \href
  {http://adsabs.harvard.edu/abs/2009MNRAS.400.1803W} {400, 1803}

\bibitem[\protect\citeauthoryear{Weigel, Schawinski  \& Bruderer}{Weigel
  et~al.}{2016}]{Weigel2016_SMF}
Weigel A.~K.,  Schawinski K.,   Bruderer C.,  2016, \mn@doi [\mnras]
  {10.1093/mnras/stw756}, 459, 2150

\bibitem[\protect\citeauthoryear{{Weigel}, {Schawinski}, {Caplar}, {Wong},
  {Treister}  \& {Trakhtenbrot}}{{Weigel} et~al.}{2017}]{2017ApJ...845..134W}
{Weigel} A.~K.,  {Schawinski} K.,  {Caplar} N.,  {Wong} O.~I.,  {Treister} E.,
   {Trakhtenbrot} B.,  2017, \mn@doi [\apj] {10.3847/1538-4357/aa803b}, \href
  {http://adsabs.harvard.edu/abs/2017ApJ...845..134W} {845, 134}

\bibitem[\protect\citeauthoryear{{Weigel}, {Schawinski}, {Treister},
  {Trakhtenbrot}  \& {Sanders}}{{Weigel} et~al.}{2018}]{2018MNRAS.476.2308W}
{Weigel} A.~K.,  {Schawinski} K.,  {Treister} E.,  {Trakhtenbrot} B.,
  {Sanders} D.~B.,  2018, \mn@doi [\mnras] {10.1093/mnras/sty383}, \href
  {http://adsabs.harvard.edu/abs/2018MNRAS.476.2308W} {476, 2308}

\bibitem[\protect\citeauthoryear{{White} \& {Becker}}{{White} \&
  {Becker}}{1992}]{1992ApJS...79..331W}
{White} R.~L.,  {Becker} R.~H.,  1992, \mn@doi [\apjs] {10.1086/191656}, \href
  {http://adsabs.harvard.edu/abs/1992ApJS...79..331W} {79, 331}

\bibitem[\protect\citeauthoryear{{Willett} et~al.,}{{Willett}
  et~al.}{2013}]{2013MNRAS.435.2835W}
{Willett} K.~W.,  et~al., 2013, \mn@doi [\mnras] {10.1093/mnras/stt1458}, \href
  {http://adsabs.harvard.edu/abs/2013MNRAS.435.2835W} {435, 2835}

\bibitem[\protect\citeauthoryear{{Wong} et~al.,}{{Wong}
  et~al.}{2016}]{2016MNRAS.460.1588W}
{Wong} O.~I.,  et~al., 2016, \mn@doi [\mnras] {10.1093/mnras/stw957}, \href
  {http://adsabs.harvard.edu/abs/2016MNRAS.460.1588W} {460, 1588}

\bibitem[\protect\citeauthoryear{Woo, Schulze, Park, Kang, Kim  \&
  Riechers}{Woo et~al.}{2013}]{Woo2013_RM_Msig}
Woo J.-H.,  Schulze A.,  Park D.,  Kang W.-R.,  Kim S.~C.,   Riechers D.~a.,
  2013, \mn@doi [\apj] {10.1088/0004-637X/772/1/49}, 772, 49

\bibitem[\protect\citeauthoryear{Wright et~al.,}{Wright
  et~al.}{2010}]{Wright2010_WISE}
Wright E.~L.,  et~al., 2010, \mn@doi [\aj] {10.1088/0004-6256/140/6/1868}, 140,
  1868

\bibitem[\protect\citeauthoryear{Wu et~al.,}{Wu
  et~al.}{2015}]{Wu2015_z6_nature}
Wu X.-B.,  et~al., 2015, \mn@doi [\nat] {10.1038/nature14241}, 518, 512

\bibitem[\protect\citeauthoryear{{Wu} et~al.,}{{Wu}
  et~al.}{2018}]{Wu2018_HotDOGs_MBH}
{Wu} J.,  et~al., 2018, \mn@doi [\apj] {10.3847/1538-4357/aa9ff3}, \href
  {http://adsabs.harvard.edu/abs/2018ApJ...852...96W} {852, 96}

\bibitem[\protect\citeauthoryear{{York} et~al.,}{{York}
  et~al.}{2000}]{2000AJ....120.1579Y}
{York} D.~G.,  et~al., 2000, \mn@doi [\aj] {10.1086/301513}, \href
  {http://adsabs.harvard.edu/abs/2000AJ....120.1579Y} {120, 1579}

\bibitem[\protect\citeauthoryear{{Yuan}, {Yu}  \& {Ho}}{{Yuan}
  et~al.}{2009}]{2009ApJ...703.1034Y}
{Yuan} F.,  {Yu} Z.,   {Ho} L.~C.,  2009, \mn@doi [\apj]
  {10.1088/0004-637X/703/1/1034}, \href
  {http://adsabs.harvard.edu/abs/2009ApJ...703.1034Y} {703, 1034}

\bibitem[\protect\citeauthoryear{{Yuan}, {Strauss}  \& {Zakamska}}{{Yuan}
  et~al.}{2016}]{2016MNRAS.462.1603Y}
{Yuan} S.,  {Strauss} M.~A.,   {Zakamska} N.~L.,  2016, \mn@doi [\mnras]
  {10.1093/mnras/stw1747}, \href
  {http://adsabs.harvard.edu/abs/2016MNRAS.462.1603Y} {462, 1603}

\bibitem[\protect\citeauthoryear{{Zakamska} et~al.,}{{Zakamska}
  et~al.}{2003}]{2003AJ....126.2125Z}
{Zakamska} N.~L.,  et~al., 2003, \mn@doi [\aj] {10.1086/378610}, \href
  {http://adsabs.harvard.edu/abs/2003AJ....126.2125Z} {126, 2125}

\bibitem[\protect\citeauthoryear{{Zakamska}, {Strauss}, {Heckman}, {Ivezi{\'c}}
   \& {Krolik}}{{Zakamska} et~al.}{2004}]{2004AJ....128.1002Z}
{Zakamska} N.~L.,  {Strauss} M.~A.,  {Heckman} T.~M.,  {Ivezi{\'c}} {\v Z}.,
  {Krolik} J.~H.,  2004, \mn@doi [\aj] {10.1086/423220}, \href
  {http://adsabs.harvard.edu/abs/2004AJ....128.1002Z} {128, 1002}

\bibitem[\protect\citeauthoryear{{Zakamska} et~al.,}{{Zakamska}
  et~al.}{2006}]{2006AJ....132.1496Z}
{Zakamska} N.~L.,  et~al., 2006, \mn@doi [\aj] {10.1086/506986}, \href
  {http://adsabs.harvard.edu/abs/2006AJ....132.1496Z} {132, 1496}

\bibitem[\protect\citeauthoryear{{de Vries}, {Becker}  \& {White}}{{de Vries}
  et~al.}{2006}]{2006AJ....131..666D}
{de Vries} W.~H.,  {Becker} R.~H.,   {White} R.~L.,  2006, \mn@doi [\aj]
  {10.1086/499303}, \href {http://adsabs.harvard.edu/abs/2006AJ....131..666D}
  {131, 666}

\bibitem[\protect\citeauthoryear{{van der Walt}, {Colbert}  \&
  {Varoquaux}}{{van der Walt} et~al.}{2011}]{2011CSE....13b..22V}
{van der Walt} S.,  {Colbert} S.~C.,   {Varoquaux} G.,  2011, \mn@doi
  [Computing in Science and Engineering] {10.1109/MCSE.2011.37}, \href
  {https://ui.adsabs.harvard.edu/abs/2011CSE....13b..22V} {13, 22}

\makeatother
\end{thebibliography}
\bsp

\label{lastpage}

\end{document}